\documentclass[twocolumn,notitlepage,prb,superscriptaddress,longbibliography]{revtex4-2}

\usepackage{times}
\usepackage{graphicx}
\usepackage{float}
\usepackage{latexsym,amsmath,amssymb,bm,euscript}
\usepackage{color}
\usepackage{epstopdf}
\usepackage[colorlinks=true,linkcolor=blue,citecolor=blue,urlcolor=blue]{hyperref}
\usepackage{hyperref}

\usepackage{tikz}
\usepackage{txfonts}
\usepackage{ulem}
\usepackage{xspace}
\usepackage{xfrac}

\begin{document}

\title{Nematic chiral spin liquid in a Kitaev magnet under external magnetic field}

\author{Shun-Yao Yu}
\affiliation{Department of Physics, Beihang University, Beijing 100191, China}

\author{Han Li}
\affiliation{Kavli Institute for Theoretical Sciences, University of Chinese Academy of Sciences, Beijing 100190, China}

\author{Qi-Rong Zhao}
\affiliation{Department of Physics, Renmin University of China, Beijing 100872, China}

\author{Yuan Gao}
\affiliation{Department of Physics, Beihang University, Beijing 100191, China}
\affiliation{CAS Key Laboratory of Theoretical Physics, Institute of Theoretical Physics, Chinese Academy of Sciences, Beijing 100190, China}

\author{Xiao-Yu Dong}
\affiliation{Hefei National Laboratory, Hefei 230088, China}

\author{Zheng-Xin Liu}
\email{liuzxphys@ruc.edu.cn}
\affiliation{Department of Physics, Renmin University of China, Beijing 100872, China}

\author{Wei Li}
\email{w.li@itp.ac.cn}
\affiliation{CAS Key Laboratory of Theoretical Physics, Institute of Theoretical Physics, Chinese Academy of Sciences, Beijing 100190, China}
\affiliation{CAS Center of Excellence in Topological Quantum Computation, University of Chinese Academy of Sciences, Beijing 100190, China}
\affiliation{Peng Huanwu Collaborative Center for Research and Education, Beihang University, Beijing 100191, China}

\author{Shou-Shu Gong}
\email{shoushu.gong@buaa.edu.cn}
\affiliation{Department of Physics, Beihang University, Beijing 100191, China}
\affiliation{School of Physical Sciences, Great Bay University, Dongguan 523000, China}

\date{\today}

\begin{abstract}
The possible existence of a quantum spin liquid (QSL) phase, an exotic state of matter with long-range quantum entanglement and fractionalized excitations, in $\alpha$-RuCl$_3$ has sparked widespread interests in exploring QSLs in various Kitaev models under magnetic fields. Recently, a $K$-$J$-$\Gamma$-$\Gamma'$ model has been proposed to accurately describe the compound by fitting the measured thermodynamic data [Nat. Commun. {\bf 12}, 4007 (2021)], where $J$ is Heisenberg interaction, and $\Gamma$, $\Gamma'$ are off-diagonal exchanges on top of the dominant Kitaev coupling $K$. Based on this effective model, an intermediate QSL phase in presence of an out-of-plane fields along the $[1 1 1]$ direction, between the low-field zigzag order and high-field polarized phase, has been predicted. By combining density matrix renormalization group (DMRG), exponential tensor renormalization group (XTRG), and variational Monte Carlo (VMC) calculations, we address the nature of this QSL phase in the honeycomb $K$-$J$-$\Gamma$-$\Gamma'$ model under the $[1 1 1]$-direction field.
Our DMRG calculations find the algebraic-like decay of spin correlation function, the finite spin scalar chiral order, and lattice nematic order. Together with the XTRG results of power-law specific heat at low temperature, our findings naturally suggest a gapless nematic chiral spin liquid. On the other hand, our VMC study finds a gapped nematic chiral spin liquid with the variational energy very close to that obtained in DMRG. As VMC finds a very small gap that is beyond the current resolution of both ground-state DMRG and finite-temperature XTRG calculations on finite-width cylinders, we resort the full clarification for the nature of the intermediate QSL to future studies. Lastly, we discuss the implications of our results to the recent experiment on $\alpha$-RuCl$_3$ and the QSL-like phase in a similar $K$-$\Gamma$-$\Gamma'$ model.
\end{abstract}

\maketitle

\section{Introduction}

Quantum spin liquid (QSL) is an exotic quantum state of matter where the spins of electrons remain disordered even at absolute zero temperature~\cite{Savary2016,Zhou2017}. 
Unlike conventional phases of matter, QSL states do not fit the Landau-Ginzburg-Wilson framework, and exhibit long-range quantum entanglement and fractionalized excitations~\cite{Wen1991,Senthil2000,Senthil2001}.
The exactly soluble Kitaev honeycomb model plays the milestone role for understanding the physics in QSL.
In particular, a non-Abelian chiral spin liquid (CSL) can be induced by breaking time-reversal symmetry in the Kitaev model, which possesses the non-Abelian anyon that has potential application in topological quantum computing~\cite{Kitaev2003,Kitaev2006}. Experimentalists have been searching for the Kitaev QSL candidate materials in various compounds with $4d$ and $5d$ transition metal-based honeycomb lattice. 
These compounds have strong spin-orbit coupling that may induce the bond-dependent Kitaev interaction~\cite{Jackeli2009}. 
Some examples of these compounds are A$_2$IrO$_3$ with A = Li, Na~\cite{Singh2012}, H$_3$LiIr$_2$O$_6$, which is a hydrogen intercalated modification of A$_2$IrO$_3$~\cite{Kitagawa2018}, and $\alpha$-RuCl$_3$~\cite{Sears2015,Johnson2015,Banerjee2017,Do2017}.
Recently, the cobalt-based $3d^7$ honeycomb magnets such as A$_2$Co$_2$TeO$_6$~\cite{Lin2021NC, Yao2022PRL}, A$_3$Co$_2$SbO$_6$~\cite{Sano2018,Songvilay2020}, and BaCo$_2$(AsO$_4$)$_2$~\cite{Zhong2020,Zhang2023} are also thought to be candidates of Kitaev materials.

Among these compounds, $\alpha$-RuCl$_3$ has been widely recognized as a prominent Kitaev material.
Due to the non-Kitaev interactions in the material, the compound exhibits a zigzag magnetic order at low temperature~\cite{Sears2015,Johnson2015}. By suppressing the zigzag order with external magnetic fields~\cite{Sears2017,Zheng2017,Banerjee2018}, in the intermediate fields experimental evidences for a QSL have been reported, e.g., the excitation continuum in the inelastic neutron scattering spectrum~\cite{Banerjee2017,
Do2017,Banerjee2016}, the half-quantized thermal Hall conductivity~\cite{Kasahara2018,Yokoi2021,Bruin2022} 
suggesting the presence of gapless Majorana edge modes~\cite{Aviv2018,Ye2018}, and the quantum oscillations 
in the low-temperature longitudinal heat conductivity possibly due to a spinon Fermi surface~\cite{Czajka2021}. 
Despite these intriguing progresses, nature of this state are still under debate~\cite{Janssen2019,
Baek2017,Wolter2017,Gass2020,Hentrich2018,Hentrich2019,Hentrich2020,Balz2019,Balz2021,Bachus2020,Bachus2021,
Leahy2017,Jansa2018,Winter2018,Kaib2019,Chern2021}.

Motivated by the exciting experimental progress, there have been a plenty of theoretical studies in an effort to find QSL in related spin models. Amongst others, the $K$-$\Gamma$-$\Gamma'$ model with the Kitaev interaction $K$ and off-diagonal couplings $\Gamma, \Gamma'$ has been considered as a minimal model to describe a broad class of Kitaev materials including $\alpha$-RuCl$_3$~\cite{Winter2017nc,Cookmeyer2018,Kim2016,Suzuki2019,Ran2017,Ozel2019,Kim2015}. 
Besides the well known Kitaev spin liquids, other promising QSL candidates have also been proposed in extended Kitaev models, including the QSL in the antiferromagnetic (AFM) Kitaev model under an intermediate magnetic field~\cite{Zhu2018,Gohlke2018dynamic,Jiang2018,Patel2019,Hickey2019,Ronquillo2019,Zou2020}, different QSL phases in the $K$-$\Gamma$ model with or without a magnetic field~\cite{Gohlke2018,Liu2018,Yamada2021}, the multinode QSL~\cite{Wang2020} and nematic QSL in the $K$-$\Gamma$-$\Gamma'$ model~\cite{Gordon2019,Lee2020,Gohlke2020}.
However, it is hard to accurately estimate the microscopic parameters that can match the experimental data and thus it is unclear how to connect the found QSLs in model studies to the compounds in experiment.

Recently, a $K$-$J$-$\Gamma$-$\Gamma'$ model has been proposed for $\alpha$-RuCl$_3$. Multiple accurate many-body simulations have been combined to examine this model, which reproduces the major experimental features in equilibrium and dynamical measurements, giving the dominant ferromagnetic Kitaev interaction, additional Heisenberg coupling $J$, and other parameters~\cite{Li2021}.
Interestingly, this model under the $[1 1 1]$-direction magnetic field shows an emergent QSL phase between the low-field zigzag order and high-field polarized phase~\cite{Li2021}.
While the variational Monte Calro (VMC) study proposes a gapped CSL, the exact diagonalization results seem to suggest a gapless state~\cite{Li2021}.
Remarkably, the theoretically predicted two-transition scenario in Ref.~\cite{Li2021} has been witnessed by recent high magnetic field experiment on $\alpha$-RuCl$_3$ at $\sim 35$~T and $\sim 83$~T, respectively, with the field along the out-of-plane $c^*$ axis~\cite{Zhou2022arXiv}. 
A field-induced intermediate phase is found experimentally in the field-angle phase diagram~\cite{Zhou2022arXiv}.
The experimental findings and the good agreements with model study open a new route to search for QSL in $\alpha$-RuCl$_3$ and call for further understanding on the QSL phase in the $K$-$J$-$\Gamma$-$\Gamma'$ model.

In this work, we study the nature of the intermediate QSL phase in the $K$-$J$-$\Gamma$-$\Gamma'$ model using three methods: density matrix renormalization group (DMRG), VMC, and exponential tensor renormalization group (XTRG) for finite-temperature calculations.
The DMRG calculations on different cylinder geometries up to the circumference of $6$ unit cells identify the existence of an intermediate phase, with a robust nonzero scalar spin chiral order $\langle \mathbf{S}_i \cdot (\mathbf{S}_j \times \mathbf{S}_k) \rangle$ and a lattice nematic order characterizing a $C_3$ symmetry breaking.
In addition, the algebraic-like spin correlation decay observed in the ground state and the power-law behavior of the specific heat $C_m \sim T^2$ obtained at low temperature strongly suggest a gapless QSL state.

We further investigate the intermediate phase using the VMC with extended variational parameters, which unveils that allowing the $C_3$ symmetry breaking can further lower the variational energy of the previously identified gapped CSL~\cite{Li2021} and enlarge the spinon gap, giving a variational energy very close to that obtained in DMRG. 
Using quantitative analysis, we examine how the mean field dispersion varies and find that the spinon gap is very small compared to the total band width ($\sim 0.02$), which may explain why the spin correlation decay looks like algebraic on our studied cylinders.
The spinon density of states which is approximately proportional to energy above the spinon gap can agree with the $C_m \sim T^2$ behavior at low temperature.
The Van Hove singularity at the higher energy $E/|K| \sim 0.4$ also supports the double peak structure found in the specific heat.
The gapless or gapped nature of the QSL may require further simulations on larger sizes, but the predicted low-temperature properties above $T \approx 2$K, which are less affected by finite-size effects, could be tested in high-field experiments on $\alpha$-RuCl$_3$.

This paper is organized as follows. In Sec.~\ref{sec:model}, we introduce the $K$-$J$-$\Gamma$-$\Gamma'$ model and numerical methods employed in this study. The DMRG simulations are demonstrated in Sec.~\ref{sec:dmrg}, and the XTRG thermodynamics results are presented in Sec.~\ref{sec:thermal}.
In Sec.~\ref{sec:VMC}, we show the nematic order obtained in the VMC calculations and compare the results with the DMRG and XTRG results. We present our summary and discussion in Sec.~\ref{sec:summary}.

\begin{figure}
\centering
\includegraphics[width = 0.9\linewidth]{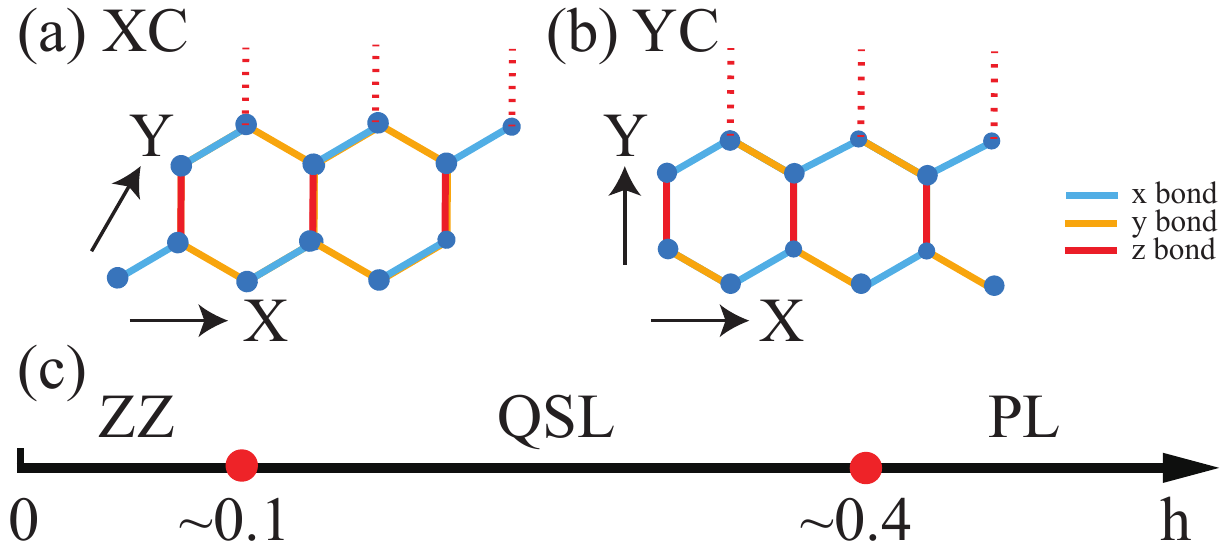}
\includegraphics[width = 1\linewidth]{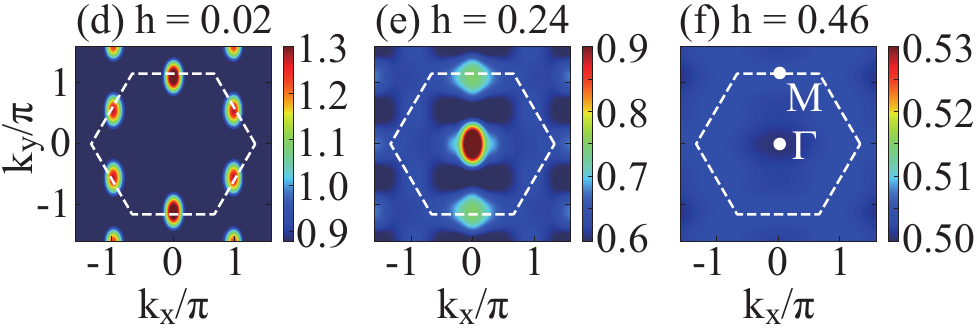}
\caption{Lattice geometries and the 1D phase diagram obtained by DMRG. (a) and (b) show the XC and YC cylinders used in our calculations. The cylinders have the open boundary condition along the $X$ direction and periodic boundary condition along the $Y$ direction. The three different colors in the solid bonds denote the different dominant Kitaev couplings. The dashed bonds mark the periodic connection. These two cylinders have the size $L_x = 3, L_y = 2$ and are denoted as XC2-3 and YC2-3. The external magnetic field is applied perpendicular to the honeycomb plane. (c) Quantum phase diagram of the model Eq.~\eqref{eq:model}. As the magnetic field increases, the system shows a zigzag (ZZ) magnetic order phase for $h \lesssim 0.1$, a polarized (PL) phase for $h \gtrsim 0.4$, and an intermediate quantum spin liquid (QSL) phase with spin scalar chiral order and lattice nematic order. (d)-(f) show the characteristic intra-sublattice spin structure factor $\Tilde{S}(\bf k)$ defined in Eq.~\eqref{eq:sq2} for the three phases, which are obtained on the YC4-18 cylinder for $h = 0.02, 0.24, 0.46$, respectively. The dashed white hexagon denotes the Brillouin zone of the sublattice, and the ${\bf \Gamma}$ and ${\bf M}$ points are marked in (f).
}
\label{fig:model}
\end{figure}

\section{Model and Methods}
\label{sec:model}

We study the honeycomb-lattice $K$-$J$-$\Gamma$-$\Gamma'$ model in an out-of-plane magnetic field along the $[1 1 1]$ direction, which has the Hamiltonian defined as
\begin{equation}\label{eq:model}
\begin{aligned}
        &H = \sum_{\langle i,j \rangle_\gamma} [K S_i^\gamma S_j^\gamma +J \mathbf{S}_i \cdot \mathbf{S}_j +\Gamma(S_i^\alpha S_j^\beta +S_i^\beta S_j^\alpha)  \\
            &+\Gamma'(S_i^\gamma S_j^\alpha+S_i^\gamma S_j^\beta +S_i^\alpha S_j^\gamma +S_i^\beta S_j^\gamma)]+h \sum_{i,\alpha} S^{\alpha}_i,
\end{aligned}
\end{equation}
where $\langle i,j\rangle_{\gamma}$ denotes the nearest-neighbor (NN) $\gamma$ bond with $\{ \alpha, \beta, \gamma\}$ being $\{x,y,z\}$ under a cyclic permutation, $\mathbf{S}_i = \{S_i^x, S_i^y, S_i^z \}$ are the pseudo spin-$1/2$ operators at the site $i$. $K$ denotes the FM Kitaev interaction, $J$ is the FM Heisenberg interaction, $\Gamma $ and $\Gamma' $ are the off-diagonal spin couplings. 
Following the microscopic model determined in Ref.~\cite{Li2021}, we choose the Kitaev coupling as the energy unit $K = -1.0$, $J = -0.1$, $\Gamma = 0.3$ and $\Gamma' = -0.02$.
Without external magnetic field, this system has the finite magnetic point group $D_{3d} \times Z_2^T$, where $Z_2^T = \{ E, T \}$ is the time-reversal symmetry group,  and each element in $D_{3d}$ stands for a combination of spin rotation and lattice rotation owing to the spin-orbit coupling.
In this paper, we focus on the system under an external magnetic field along the $[1 1 1]$ direction, which gives the Zeeman coupling $h\sum_{i,\alpha}S^{\alpha}_i$ in Eq.~\eqref{eq:model}. 

In DMRG calculation of the ground state~\cite{White1992}, we study the system on a cylinder geometry with the periodic boundary condition along the circumference direction ($Y$ direction) and the open boundary condition along the axis direction ($X$ direction). 
To avoid possible bias from the lattice geometry, we consider two geometries, namely the XC and YC cylinders shown in Figs.~\ref{fig:model}(a) and \ref{fig:model}(b). For the XC cylinder, there is a $\pi/3$ angle between the $Y$ and $X$ direction. 
For the YC cylinder, the $Y$ direction is perpendicular to the $X$ direction.
We denote the system size as XC$L_y$-$L_x$ and YC$L_y$-$L_x$, where $L_y$ and $L_x$ denote the numbers of unit cells along the two directions, respectively.
When we show the results on both cylinders together, we use the length along the circumference direction to demonstrate different system sizes, which is denoted as $W_y$ with the lattice constant as the unit.
In our DMRG simulation, we keep up to $D=2000$ bond states, which allow us to obtain accurate results on the $L_y = 6$ cylinder with the truncation error at about $10^{-6}$.
To reduce the finite-size effect, we also implement the infinite-DMRG (iDMRG)~\cite{Mcculloch2008} by using the TeNPy library~\cite{Hauschild2018} to check our results, where the system size along the $X$ direction can be taken as long enough to eliminate the size effect in this direction. In the iDMRG simulations, we denote the lattice as XC$L_y$ and YC$L_y$.

To compute the thermodynamic properties, we employ the exponential tensor renormalization group (XTRG) method~\cite{Chen2018,Li2022} to simulate the model on the YC cylinder with $L_y = 4$ and $L_x$ up to $12$. 
The retained bond dimension $D$ is up to $1000$, which ensures the truncation error smaller than $10^{-4}$ down to the lowest temperature $T/|K| = 0.008$.

\section{DMRG results for the ground state}
\label{sec:dmrg}

In this section, we demonstrate our DMRG simulation results, focusing on the QSL phase.
Since the phase boundaries have been investigated in a recent study~\cite{Li2021}, we leave our results on the determination of phase transition points in Appendix~\ref{app-sec:transition}, which identify the transitions at $h \simeq 0.1$ and $h \simeq 0.4$ as shown in Fig.~\ref{fig:model}(c), in consistent with the previous finding~\cite{Li2021}.
With increased magnetic field, an intermediate QSL emerges between a zigzag magnetic order at low field and a polarized phase at high field.  

The three phases have different features of spin structure factor. In Figs.~\ref{fig:model}(d-f),
we show the intra-sublattice structure factor defined as
\begin{equation} \label{eq:sq2}
    \tilde{S}({\bf k}) = \frac{1}{N} \sum_{i,j} e^{-i {\bf k} \cdot {\bf r}_{ij}} [\langle {\bf S}_i \cdot {\bf S}_j \rangle - \langle {\bf S}_i \rangle \cdot \langle {\bf S}_j \rangle],
\end{equation}
where the summation is for the sites in the same sublattice. The obtained results of $\tilde{S}({\bf k})$ are the same for the two sublattices of the honeycomb lattice.
In the zigzag phase [Fig.~\ref{fig:model}(d)], the prominent peaks at the ${\bf M}$ points characterize the zigzag order.
In the QSL phase [Fig.~\ref{fig:model}(e)], a dominant peak appears at the ${\bf \Gamma}$ point with a broad peak at a ${\bf M}$ point.
In the polarized phase [Fig.~\ref{fig:model}(f)], the featureless structure factor reflects the very short-range spin correlation. 
In the following, we focus on the characterization of the QSL phase.

\begin{figure*}
    \centering
    \includegraphics[width = 0.9\linewidth]{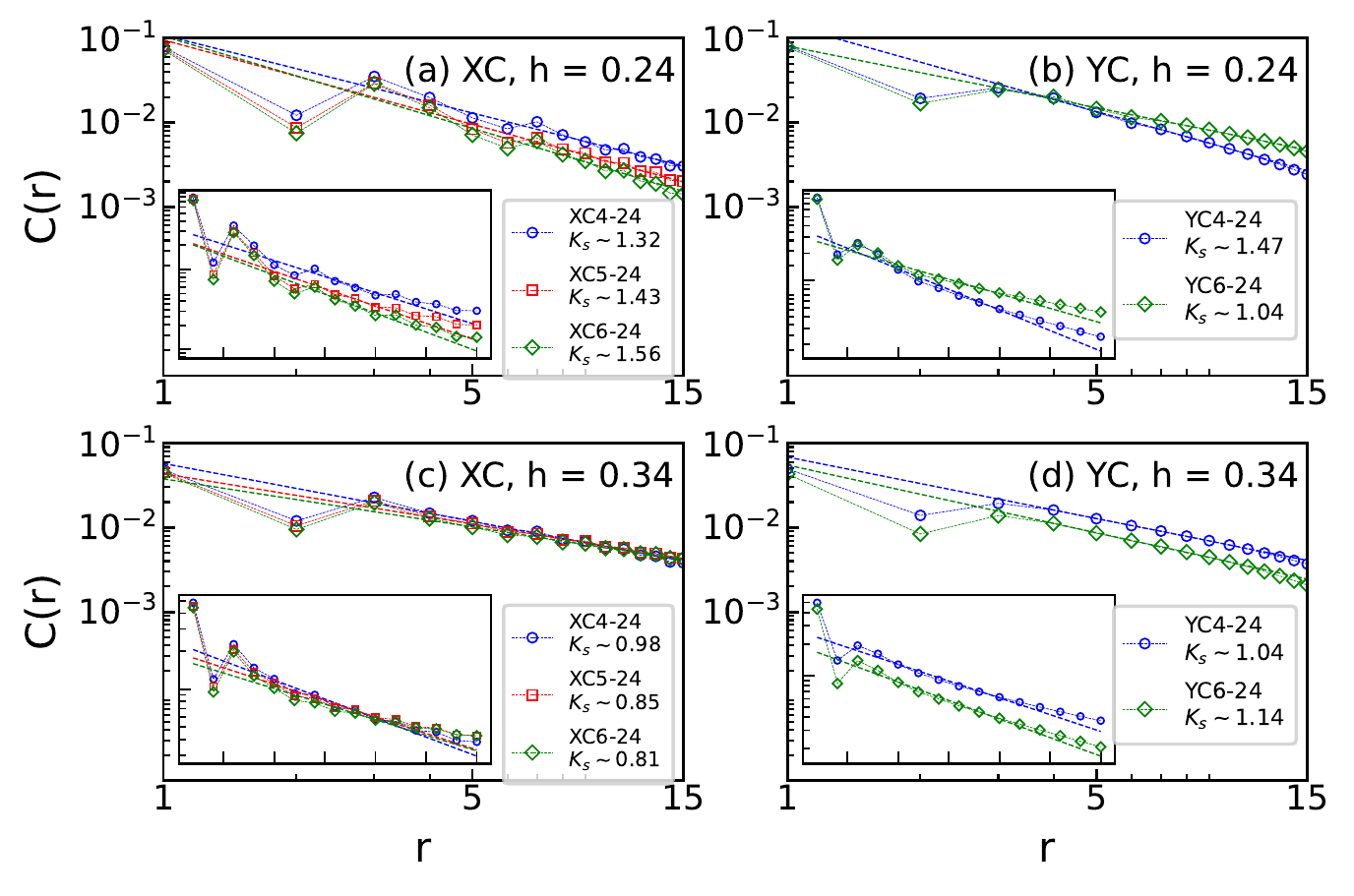}
    \caption{Absolute spin correlation function $C(r)$ along the $X$ direction in the QSL phase. $r$ is the distance of the sites to the reference site in the same row. (a) and (b) are the double-logarithmic plots of the spin correlations for $h = 0.24$ on the XC and YC cylinders with $L_x = 24$. $K_s$ is the fitted power exponent of algebraic decay. The insets are the semi-logarithmic plots for the same data. (c) and (d) are the similar figures for $h = 0.34$. The reference site is chosen on the fifth column of the cylinder.}
    \label{fig:spincor}
\end{figure*}

\subsection{Algebraic-like spin correlation function and static spin structure factor}

\begin{figure*}
    \centering
    \includegraphics[width = 1\linewidth]{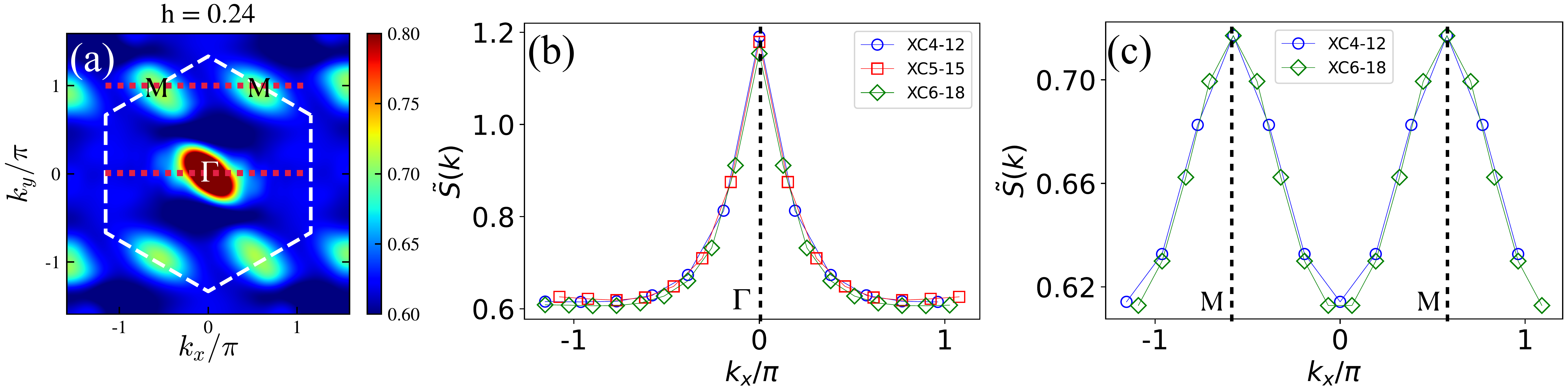}
    \caption{Static intra-sublattice spin structure factor at $h = 0.24$ in the QSL phase. (a) shows the typical results of $\tilde{S}({\bf k})$ obtained on the XC4-24 cylinder, with the white dashed hexagon indicating the Brillouin zone. The red dashed lines denote the two momentum cuts crossing the ${\bf \Gamma}$ and ${\bf M}$ points, which represent the momentum used in (b) and (c). (b) and (c) show the $\tilde{S}({\bf k})$ along the momentum cuts in (a), in which $\tilde{S}({\bf k})$ are obtained from the Fourier transform of the middle $L_y \times 3L_y$ sites.}
    \label{fig:sq}
\end{figure*}

We first study the spin-spin correlation function $C(r)$ defined as
\begin{equation}
    C(r) = |\langle \mathbf{S}_i \cdot \mathbf{S}_j \rangle -\langle \mathbf{S}_i \rangle \cdot \langle \mathbf{S}_j \rangle |,
\end{equation}
where $r$ is the distance between the two sites $i, j$ in the same row along the $X$ direction.
In Fig.~\ref{fig:spincor}, we show the double-logarithmic plots of the spin correlations on different geometries and system sizes.
In all the cases, the spin correlations drop markedly with about an order of magnitude in the distance of $r \simeq 1$, which is similar to the observation in the spin systems with the dominant bond-directional Kitaev interaction~\cite{Kitaev2006}.
Remarkably, the spin correlations decay slowly at long distance, which can be fitted quite well using the algebraic behavior $C(r) \sim 1/r^{K_{s}}$ with small power exponents $K_s \lesssim 1.5$.
Although the spin correlations decay slowly, one can find in Fig.~\ref{fig:spincor} that with growing $L_y$, the spin correlations only change slightly, which agrees with the vanished magnetic order in two-dimensional limit (also see Fig.~\ref{fig:sq} below).
With increasing the field, the power exponent $K_s$ remains small in the QSL phase until the transition to the polarized phase with spin correlations decaying exponentially.   

As a comparison, we also demonstrate the semi-logarithmic plots of the same data in the insets of Fig.~\ref{fig:spincor}.
We fit the most data as shown by the dashed lines, showing the deviation of the longer-distance data from the exponential 
decay. These analyses strongly suggest the algebraic-like decay of spin correlation, which supports a gapless QSL. 

We further discuss the static spin structure factor. For the sake of convenience, we stick to the intra-sublattice structure factor defined in Eq.~\eqref{eq:sq2}. As shown in Fig.~\ref{fig:sq} (for XC) and Fig.~\ref{fig:model}(e) (for YC), $\tilde{S}({\bf k})$ exhibits a dominant peak at the ${\bf \Gamma}$ point, which appears to be singular [Fig.~\ref{fig:sq}(b)] and is consistent with the small power exponent ($K_s < 2$) of spin correlation. In Figs.~\ref{fig:sq}(b) and \ref{fig:sq}(c), we plot $\tilde{S}({\bf k})$ versus $k_x$ following the two cuts in Fig.~\ref{fig:sq}(a), crossing the ${\bf \Gamma}$ and ${\bf M}$ points, respectively.
For a direct comparison, we plot the $\tilde{S}({\bf k})$ obtained from the middle $L_y \times 3L_y$ sites.
With increased $L_y$, the peak at $\tilde{S}({\bf \Gamma})$ becomes more diffusive, which agrees with the absent magnetic order.
The round peak at the ${\bf M}$ point may characterize the residual fluctuation of the neighboring zigzag order.

\subsection{Spin scalar chiral order}

\begin{figure}
    \centering
    \includegraphics[width = 1.0\linewidth]{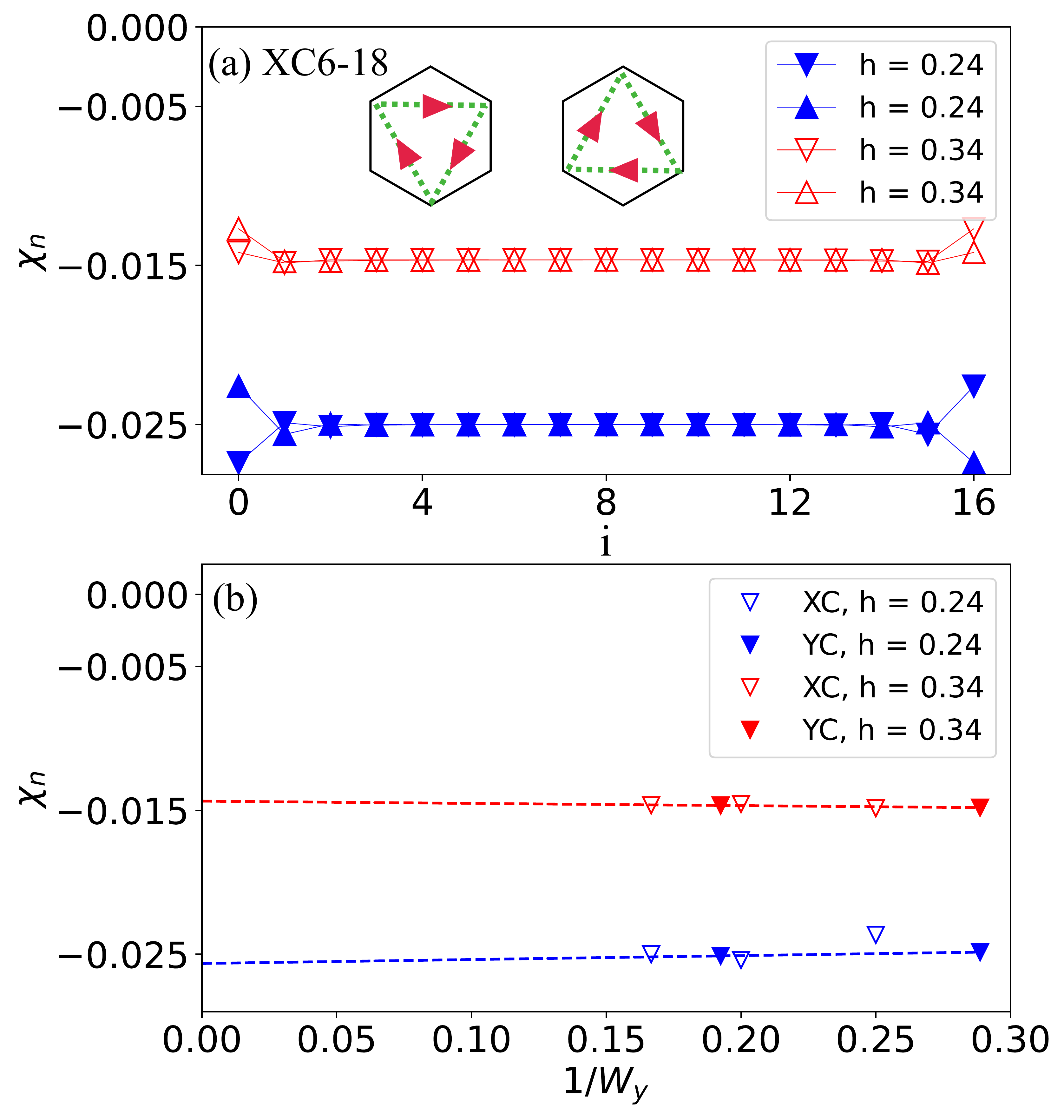}
    \caption{Scalar spin chiral order in the QSL phase. (a) The spatial dependence of the chiral orders computed for the equilateral triangles defined following the clockwise direction. $i$ denotes the column number and the data for $h = 0.24, 0.34$ are obtained on the XC6-18 cylinder. (b) System width dependence of the chiral order obtained on various cylinders. $W_y$ is the width of the cylinder along the circumference direction, with the lattice constant as the unit. The dashed lines are linear extrapolations of the data for the guide to the eye.}
    \label{fig:chiral}
\end{figure}

While the magnetic order is absent in the QSL phase, the CSL found in the previous VMC study~\cite{Li2021} motivates us to examine the spin chiral order. We calculate the spin scalar chiral order $\chi_{ijk}$ defined as
\begin{equation}
\label{eq:chiral}
\chi_{ijk} = \langle \mathbf{S}_i \cdot (\mathbf{S}_j \times \mathbf{S}_k) \rangle,
\end{equation}
where the sites $i, j, k$ from the same sublattice form a triangle in the clockwise direction as shown in the inset of Fig.~\ref{fig:chiral}(a). For a time-reversal symmetric system, the spontaneously emergent chiral order could either be left-chiral or right-chiral, which is chosen randomly in DMRG simulations~\cite{gong2014}. For our studied system with an external magnetic field, we always obtain a certain chirality. We have checked the chiral orders of different types of triangles allowed on the honeycomb lattice, and we find that the chiral orders of the equilateral triangles shown in the inset of Fig.~\ref{fig:chiral}(a) have the maximal magnitudes (see the results for other types of triangles in Appendix~\ref{app-sec:chiral}). Thus, we focus on the equilateral triangles inside the hexagon. In each hexagon, there are two types of equilateral triangles, and both of them have the same chirality denoted as $\chi_n$.

We first check the spatial dependence of the chiral order $\chi_n$. 
Two examples at $h = 0.24, 0.34$ on the XC6-18 cylinder are demonstrated in Fig.~\ref{fig:chiral}(a).
Indeed, the two types of triangles have the same chirality and both of them are quite uniform in the bulk of system, revealing a small finite-size effect along the $X$ direction.
We have also checked the chiral orders using iDMRG simulations, which obtain the results in agreement with the calculations on the finite-length cylinders. 

Furthermore, we examine the $L_y$ dependence of the chiral orders on both the XC and YC cylinders [Fig.~\ref{fig:chiral}(b)].
Clearly, the chiral orders only change slightly with growing $L_y$, strongly indicating the robust scalar chiral order in the thermodynamic limit.
We remark that this finite chiral order is not from a trivial product of local magnetic moments.
We have checked the result of $\langle \mathbf{S}_i \rangle \cdot (\langle \mathbf{S}_j \rangle \times \langle \mathbf{S}_k \rangle)$, which turns out to be vanishingly small in the QSL phase.

The nonzero scalar chiral order has also been found in some other Kitaev systems. 
One model is a similar $K$-$J$-$\Gamma$-$\Gamma'$ model either without or with a small magnetic field~\cite{Luo2022, Luo2022prb}.
Near the dominant $\Gamma$ interaction region, a QSL-like phase with chiral order has been found.
Different from the studied model, in this $\Gamma$-dominant phase the two equilateral triangles inside the hexagon have the opposite chiralities~\cite{Luo2022, Luo2022prb}.  
The other system is the $K$-$\Gamma$ model in the $[1 1 1]$-direction magnetic field, which has both the uniform chiral and staggered chiral phases on two-leg ladder, with tuning coupling ratio and field strength~\cite{Sorensen2021}. 
The spin chiral order can emerge in magnetically disordered phases of various Kitaev models due to the rich interplay among different interactions and magnetic field.

\subsection{Lattice nematic order}

\begin{figure}
    \centering
    \includegraphics[scale = 0.38]{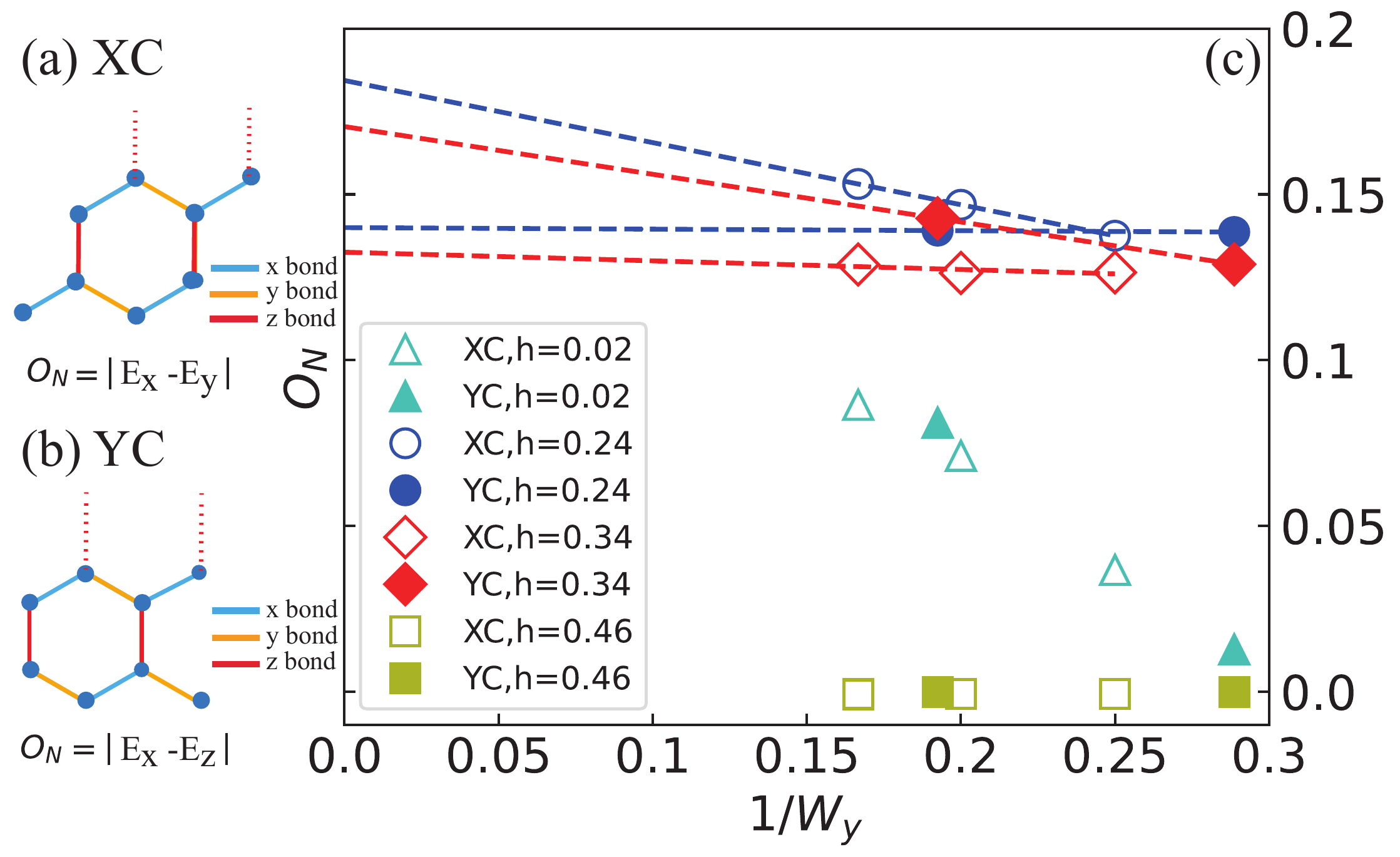}
    \caption{Size dependence of the lattice nematic order in the QSL phase. 
    (a) and (b) show the definitions of the lattice nematic order parameters $O_N$ on the XC and YC cylinders, respectively. 
    The nematic order is the bond-energy difference between the bond along the axis $(X)$ and the circumference $(Y)$ direction.
     (c) Size dependence of the nematic order on the XC and YC cylinders with different widths $W_y$ at the field $h=0.24$ and $0.34$.
      For a comparison, the nematic orders in the zigzag phase ($h = 0.02$) and the polarized phase ($h = 0.46$) are also shown. 
     }
    \label{fig:nematic}
\end{figure}

To rule out the valence bond order in the QSL phase, we have measured the bond energy to detect the translation symmetry breaking. In the bulk of the cylinder, we find a uniform bond energy distribution that agrees with the absent valence bond order. Furthermore, we find the bond energies along the axis ($X$) direction and the circumference ($Y$) direction show a clear difference, indicating a possible lattice nematic order associated with $C_3$ lattice rotational symmetry breaking. To detect such a nematic order, we can measure the difference of the bond energies as the nematic order parameter $O_N$.
As shown in Figs.~\ref{fig:nematic}(a) and \ref{fig:nematic}(b), we define the bond energies along the three bond directions as $E_{\gamma}$ (including all the energy contributions to the NN $\gamma$ bond), where $\gamma \in \{x, y, z\}$. On the XC cylinder, $E_x = E_z$ is guaranteed due to the symmetry along the $Y$ direction, and the nematic order can be described by $O_N \equiv |E_x - E_y|$. Similarly, on the YC cylinder $O_N$ can be defined as $O_N \equiv |E_x - E_z|$.

\begin{figure*}
    \centering
    \includegraphics[width = 0.75\linewidth]{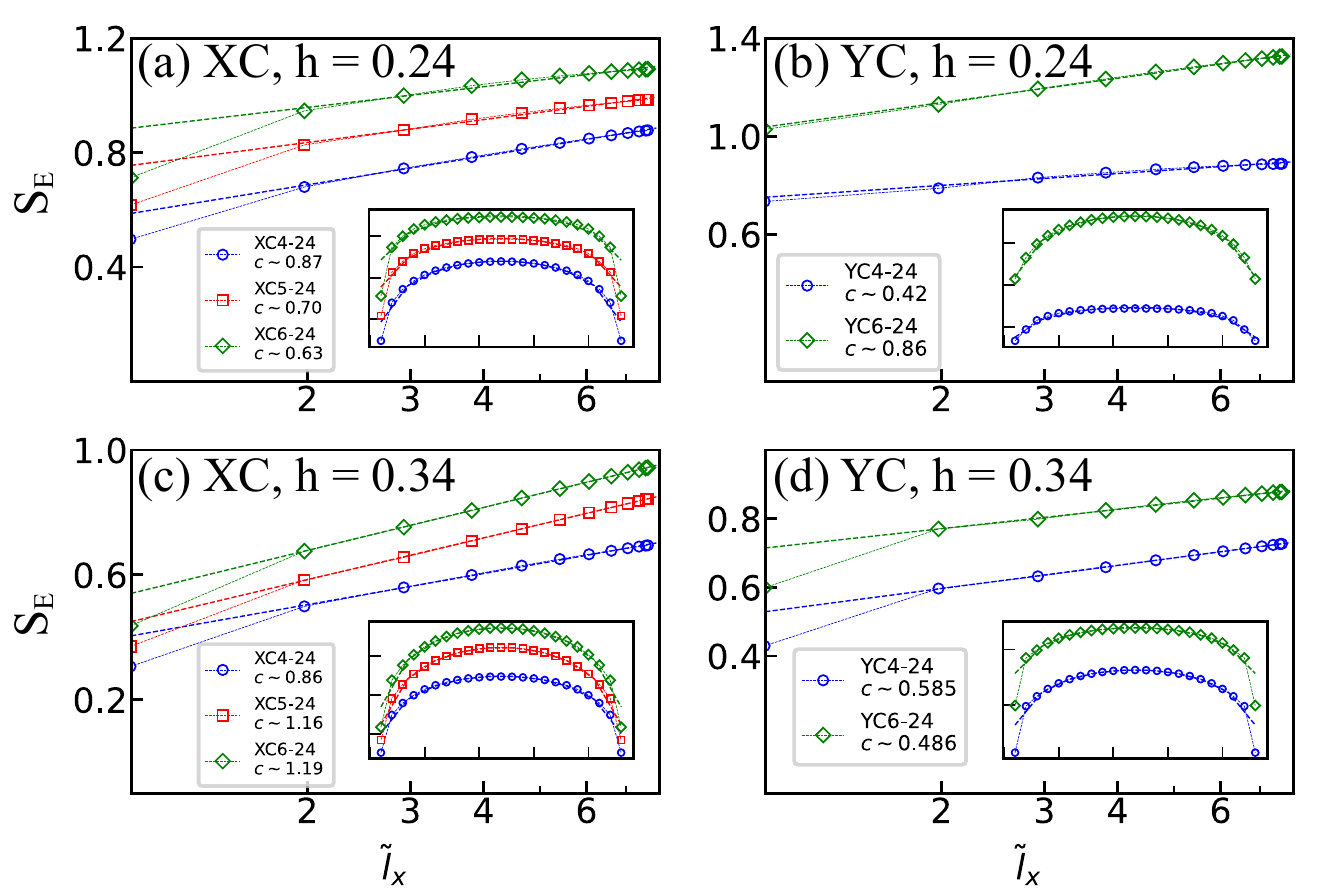}
    \caption{Entanglement entropy versus the conformal distance $\tilde{l}_x$ in the QSL phase. The conformal distance is defined as $\tilde{l}_x = \frac{L_x}{\pi}\sin(\frac{\pi l_x }{L_x})$. (a) and (b) are the entropy for $h = 0.24$ obtained on the $L_x = 24$ XC and YC cylinders, respectively. The dashed lines denote the fitting of entropy using the formula $S(\tilde{l}_x) = \frac{c}{6}\ln(\tilde{l}_x)+g $, giving the fitted central charge $c$. The insets show the entropy versus the subsystem length $l_x$. (c) and (d) are the similar plots for $h = 0.34$.}
    \label{entanglement}
\end{figure*}

We emphasize that for our studied systems, a nonzero $O_N$ always exists in the bulk of the system due to the quasi-one-dimensional cylindrical geometry that has already broken the lattice $C_3$ rotational symmetry, but the $L_y$ dependence of $O_N$ can determine whether the ground state has an intrinsic nematic order in two-dimensional limit.
Such a scheme for determining nematic order has been adopted in previous DMRG studies on various spin systems~\cite{hu2019,hu2020prb,hu2020prr}: for the quantum states with lattice rotational symmetry breaking, $O_N$ will remain finite after the extrapolation to $L_y \to \infty$; otherwise $O_N$ decreases rapidly with $L_y$ and appears to approach zero in the thermodynamic limit.

Following this strategy, we first check the nematic order $O_N$ in the bulk of the long cylinders for different $L_x$ to ensure the converged $O_N$ with system length (not shown here). 
Then, we examine the $L_y$ dependence of $O_N$ as shown in Fig.~\ref{fig:nematic}(c).
For a comparison, one can see that in the polarized phase ($h = 0.46$) $O_N$ is vanishing small and decreases rapidly to zero, which agrees with the preserved lattice rotational symmetry in this state.
On the other hand, in the zigzag phase ($h = 0.02$), $O_N$ increases with $L_y$ and approaches a finite value, which is consistent with the spacial anisotropy associated with the zigzag spin order.
Interestingly, in the QSL phase, $O_N$ shows relatively large values and the results slightly increase with growing $L_y$, strongly indicating a finite nematic order in the thermodynamic limit.
We remark that although this method can identify an intrinsic nematic order, the finite-size cylinder calculation will not allow the symmetries along the $Y$ direction to break.
Thus, our results in Fig.~\ref{fig:nematic} cannot determine whether the $C_3$ symmetry breaking state has $E_x \neq E_y \neq E_z$ or two of the three bonds have the same energy. 

In the recent studies on a similar $K$-$\Gamma$-$\Gamma'$ model in the $[1 1 1]$-direction magnetic field, a possible nematic QSL phase has also been found~\cite{Gordon2019,Lee2020,Gohlke2020}. The iDMRG simulation can obtain the ground state with $E_x \neq E_y \neq E_z$ in some field regime of the nematic phase~\cite{Gohlke2020}, which may be owing to the infinite system length that allows a complete $C_3$ symmetry breaking.
We have also tested the nematic order of our model using iDMRG (see the details in Appendix~\ref{app-sec:nematic}), which can obtain the complete $C_3$ symmetry breaking in some range of the QSL phase as well, indicating that such a nematic order with $E_x \neq E_y \neq E_z$ is highly possible in the thermodynamic limit. 
While our results can identify the nematic order, the explicit nature of the nematicity still needs further studies on larger system sizes, e.g., by tensor network simulations~\cite{Lee2020}.

\subsection{Entanglement entropy}

In this subsection, we calculate and analyze the entanglement entropy versus subsystem length $l_x$ by cutting the cylinder into two parts. For gapless quasi-one-dimensional systems, the entropy will have a logarithmic correction to the area law and follow the formula~\cite{Calabrese2004} 
\begin{equation} \label{eq:entropy}
    S(l_x) = \frac{c}{6} \ln{[(\frac{L_x}{\pi})\sin (\frac{\pi l_x}{L_x})]}+g,
\end{equation}
where $S(l_x)$ is the bipartite entanglement entropy, $L_x$ is the length of system, $g$ is a non-universal constant, and $c$ is the central charge characterizing the number of gapless modes. In principle, the fitting of the central charge from Eq.~\eqref{eq:entropy} should be applied for a large $L_x$. In our study, we study the entanglement entropy behavior with the system length up to $L_x = 24$, and we find the obtained central charge results are consistent at different system lengths (see Appendix~\ref{app-sec:entropy}).

In Fig.~\ref{entanglement}, we show the entanglement entropy for $h=0.24,0.34$ on both the XC and YC cylinders with $L_x = 24$, including the entanglement entropy versus the conformal distance $\tilde{l}_x = (L_x/\pi)\sin(\pi l_x / L_x)$ and the $l_x$ dependence of the entanglement entropy in the insets. 
On both geometries, the entanglement entropy can be fitted well using Eq.~\eqref{eq:entropy}, giving nonzero central charge.
Although the obtained central charge dependents on lattice geometry and magnetic field, the nonzero values support the gapless nature of the phase within present resolution. 
One inspiring observation is that with growing $L_y$ for each case, the obtained central charge seems to approach either $c \sim 0.5$ or $c \sim 1.0$, which may suggest emergent gapless Majorana excitations.

\section{Gapless nature identified from thermodynamics}
\label{sec:thermal}

To further explore the nature of the intermediate QSL phase, we turn to study the specific heat $C_{\rm m}$ and thermal entropy $S$. In a previous work by some of the authors~\cite{Li2021}, a prominent double-peak/hump feature has been revealed in the curve of the specific heat $C_{\rm m}$ versus temperature, for the intermediate field regime, which characterizes two temperature scales $T_{\rm H}$ (the higher-temperature scale) and $T''_{\rm L}$ (the lower-temperature scale). Below $T''_{\rm L}$, the calculated specific heat data seem to fall into a power-law decay~\cite{Li2021}.

\begin{figure}
  \includegraphics[width=1\linewidth]{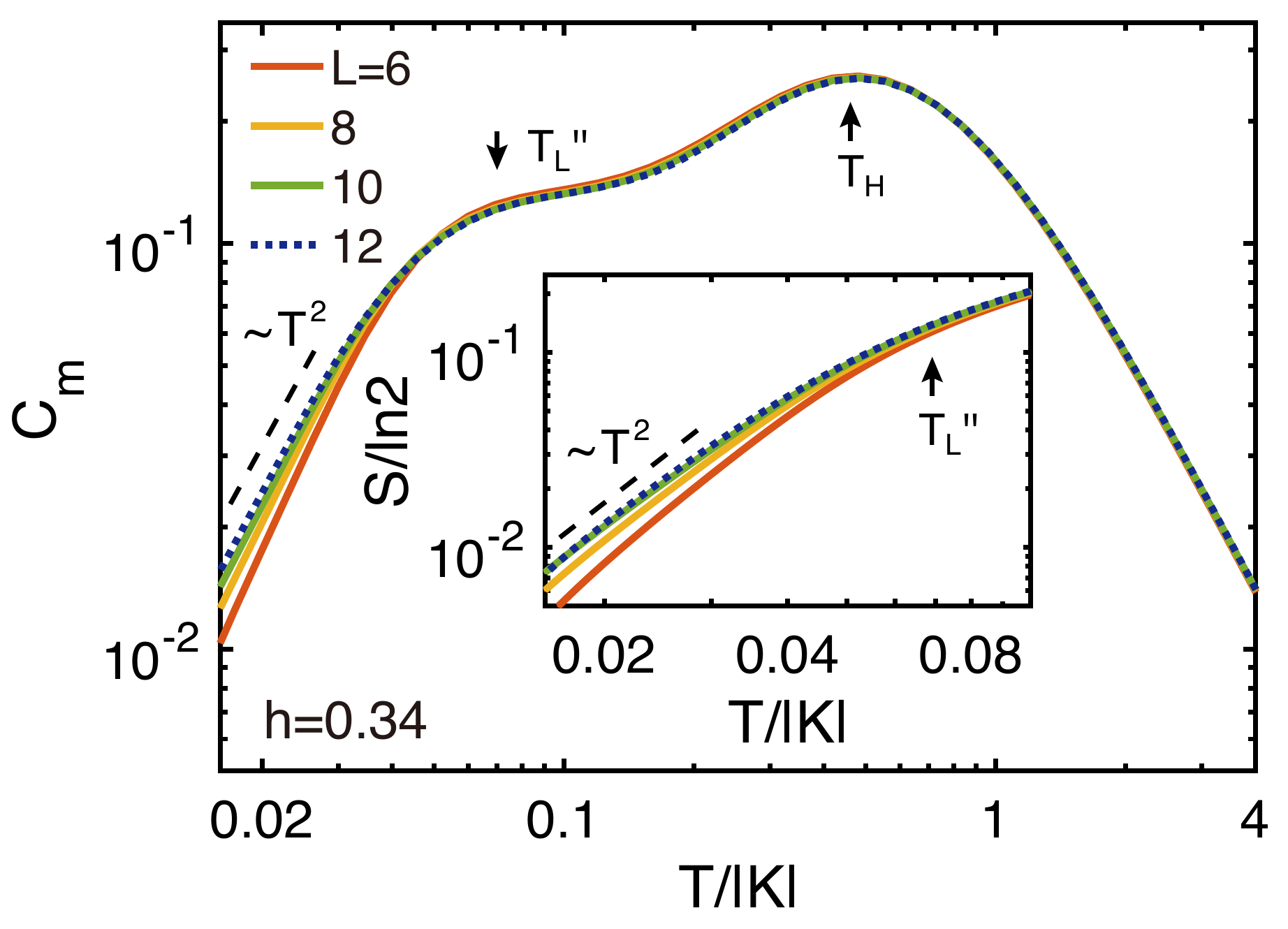}
  \caption{The low-temperature specific heat $C_{\rm m}$ curves under the out-of-plane field $h=0.34$ on the YC4-$L$ cylinders with $L$ up to $12$. The corresponding thermal entropy $S$ are shown in the inset.
  Below the lower-temperature scale $T''_{\rm L}\simeq 0.07$ as marked by the black arrow in both the main plot and the inset, the dashed line indicates a power-law decay ($\sim T^2$) and serves as a guide to the eye.
}
\label{fig:FiniteT}
\end{figure}

To clarify this possible power-law behavior, here we improve the XTRG calculation to larger system sizes, namely YC4-$L_x$ cylinders ($L_x=6,8,10,12$) and down to lower temperature. 
As shown in Fig.~\ref{fig:FiniteT}, we find that both the position and height of the $C_{\rm m}$ hump at $T''_{\rm L} \simeq 0.07$ (under the magnetic field $h=0.34$) are quite robust with growing system size.
Below $T''_{L}$, the $C_{\rm m}$ curves exhibit a power-law decay $C_{\rm m} \sim T^{\alpha}$ with the power exponent approaching $\alpha \simeq 2$, reflecting the gapless excitations (or with a very small excitation gap) in this low-energy scale.

Besides, we also show the thermal entropy $S$ in the inset of Fig.~\ref{fig:FiniteT}, where the low-temperature data has well converged versus bond dimension $D$ and also follows a power-law decay close to $S\sim T^2$. It is noteworthy that there are considerable amount of entropies below $T''_{L}$, indicating that the low-temperature states are strongly fluctuating
and have large density of states in low-energy excitations.

\section{Variational Monte Carlo results}
\label{sec:VMC}

\begin{figure}
    \includegraphics[width=1\linewidth]{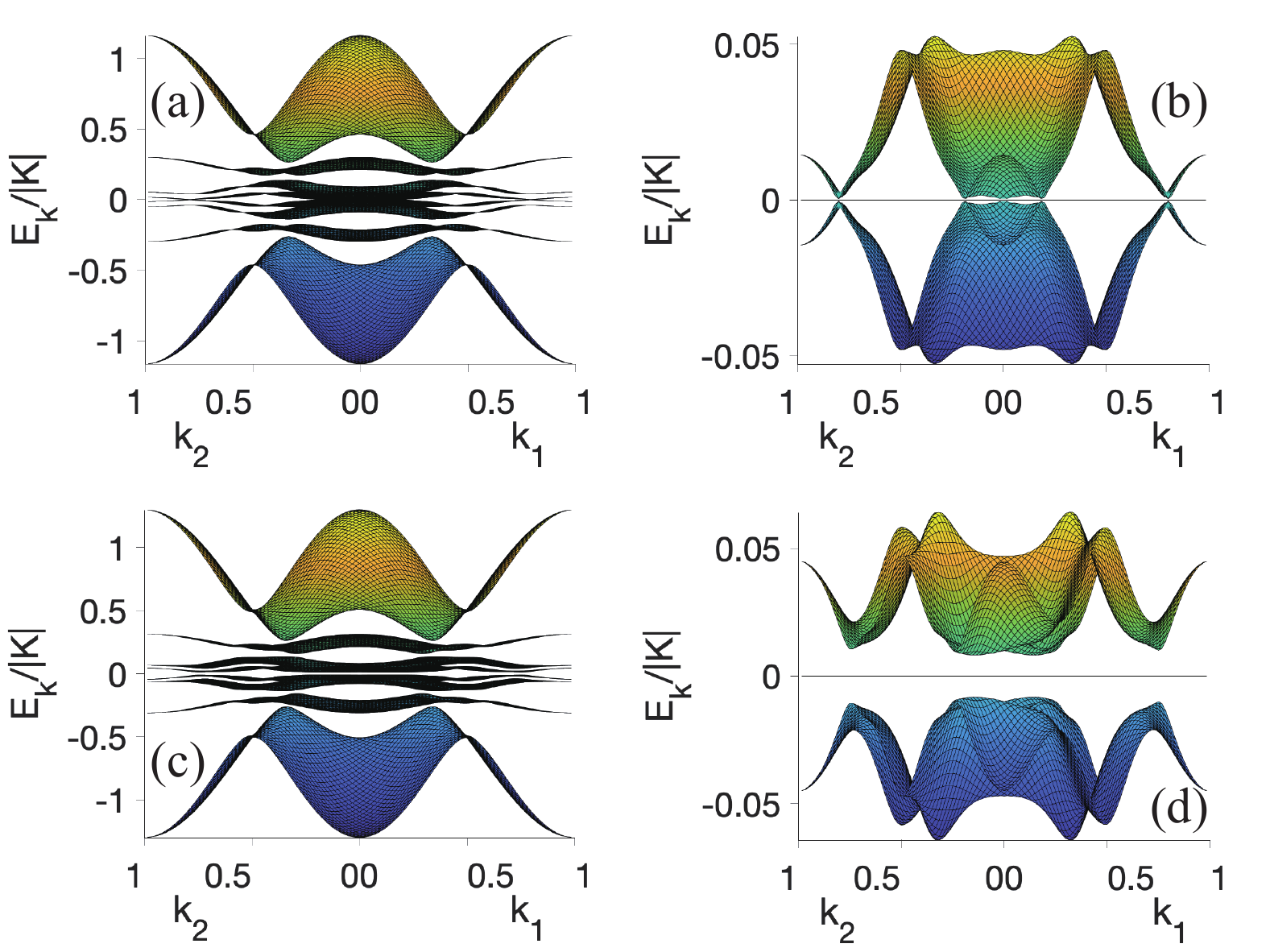}
    \caption{Mean-field band dispersion of the CSL state either without or with the nematic order. $k_1$ and $k_2$ are the two reciprocal lattice basis vectors. The bands are shown as 3D surface plots and viewed from the 45$^\circ$ angle. (a) The band dispersion of the CSL with the $C_3$ symmetry. (b) Zoom in of the lowest-energy band in (a). (c) The band dispersion of the CSL breaking the $C_3$ symmetry. (d) Zoom in of the lowest-energy band in (c). These band dispersions are obtained at $h = 0.3$.}
    \label{fig:band}
\end{figure}

Finally, we investigate the intermediate phase of the system using VMC, and try to provide a physical picture to help understand part of the above numerical results. 
 
We introduce the fermionic slave-particle representation for spin operator as $S_i^{\alpha} = \frac{1}{2}C_i^\dagger \sigma^{\alpha} C_i$ with $C_i^\dagger = (c_{i\uparrow}^\dagger, c_{i\downarrow}^\dagger)$, $\alpha \equiv x,y,z,$ and $\sigma^{\alpha}$ the Pauli matrices. To ensure that the Hilbert space of the fermions faithfully represents that of the physical spin, a constraint for particle number $\hat{N}_i \equiv c_{i \uparrow}^\dagger c_{i \uparrow} + c_{i \downarrow}^\dagger c_{i \downarrow} = 1$ is imposed for every site. This complex fermion representation is equivalent to the Majorana representation using $b_x,b_y,b_z, c$ fermions~\cite{Kitaev2006}. For the model described by Eq.~\eqref{eq:model}, the most general expression of the mean-field Hamiltonian reads~\cite{Wang2019} 
\begin{eqnarray}\label{eq:MF}
H^{\rm SL}_{\rm mf} &=& \sum_{\langle i,j \rangle \in \alpha \beta (\gamma)} \Big(\operatorname{Tr}[U_{ji}^{(0)} \Psi_i^\dagger \Psi_j] + \operatorname{Tr}[U_{ji}^{(1)} \Psi_i^\dagger (i R^{\gamma}_{\alpha \beta})\Psi_j] \nonumber \\
&+&\operatorname{Tr}[U_{ji}^{(2)} \Psi_i^\dagger \sigma^\gamma \Psi_j] + \operatorname{Tr}[U_{ji}^{(3)} \Psi_i^\dagger \sigma^\gamma R^{\gamma}_{\alpha \beta}\Psi_j] + {\rm H.c.} \Big)  \nonumber \\
&+& \sum_i \Big[\operatorname{Tr}({\pmb \lambda_i} \cdot \Psi_i {\pmb \tau} \Psi^{\dagger}_i) + {h\over4}\sum_\alpha {\rm Tr} (\Psi_i^\dag \sigma_\alpha \Psi_i)\Big],
\end{eqnarray}
where $\Psi_i = (C_i, \bar{C}_i)$, $\bar{C}_i = (c_{i \downarrow}^\dagger, -c_{i\uparrow}^\dagger)^T$, and $R^{\gamma
}_{\alpha \beta} = -i (\sigma^\alpha + \sigma^\beta)/\sqrt{2}$ is a rotation matrix. Considering the SU(2) gauge symmetry in this fermionic representation, the particle-number constraint $\hat{N}_i = 1$ is extended to the SU(2) gauge-invariant three-component form $\operatorname{Tr}(\Psi_i \tau \Psi^{\dagger}_i) = 0$, which is implemented by the Lagrangian multipliers $\lambda^{x,y,z}$ with $\tau^{x,y,z}$ being the generators of the SU(2) gauge group.
The mean-field matrices $U^{0,1,2,3}_{ji}$ can be expanded with the identity matrix and $\tau^{x,y,z}$. For the studied model, the SU(2) gauge symmetry breaks down to its $Z_2$ subgroup named the invariant gauge group (IGG)~\cite{Wen2002}.

While a QSL state preserves the space group symmetry, the symmetry group of its mean-field Hamiltonian is an enlarged one called the projective symmetry group (PSG), whose group elements are the space group operations followed by certain SU(2) gauge transformation. We adopt the Kitaev PSG~\cite{Song2016,You2012}, namely the PSG of the mean-field description of the pure Kitaev model, then the general form of the coefficients $U_{ji}^{(m)}$ can be expressed as
\begin{eqnarray}
&&U^{(0)^z}_{ji}\ = i \theta (\eta_0 + \rho_a + \rho_c) \tau^0, \nonumber \\
&&U^{(0)^{x,y}}_{ji} = i (\eta_0 + \rho_a + \rho_c) \tau^0, \nonumber \\
&&U^{(1)^z}_{ji}\ = i \theta \eta_3 \tau^c  + i \theta (\rho_a - \rho_c + \rho_d + 2 \rho_f) (\tau^x + \tau^y), \nonumber \\
&&U^{(1)^{x,y}}_{ji} = i \eta_3 \tau^c + i (\rho_a - \rho_c + \rho_d + 2 \rho_f) (\tau^{y,x} + \tau^z), \nonumber \\
&&U^{(2)^z}_{ji}\ = i \theta \eta_5 \tau^c + i \theta (\rho_a + \rho_c) \tau^z + i \theta \rho_f (\tau^x + \tau^y), \nonumber 
\end{eqnarray}
\begin{eqnarray}
&&U^{(2)^{x,y}}_{ji} = i \eta_5 \tau^c + i (\rho_a + \rho_c) \tau^{x,y} + i \rho_f (\tau^{y,x} + \tau^z), \nonumber \\
&&U^{(3)^{z}}_{ji}\ = i (\eta^z_7 + \theta \rho_c - \theta \rho_a - \theta \rho_d) (\tau^x - \tau^y), \nonumber \\
&&U^{(3)^{x,y}}_{ji} = \pm i \eta^x_7 \tau^c \pm i (\rho_c - \rho_a - \rho_d) (\tau^{y,x} - \tau^z).\nonumber 
\end{eqnarray}
with $\tau^c ={1\over\sqrt3}(\tau^x + \tau^y + \tau^z)$. Hence we have the complete set of variational parameters $\{ \rho_a, \rho_c, \rho_d, \rho_f, \eta_0, \eta_3, \eta_5, \eta^{x}_7, \eta^{z}_7, \theta, \lambda \}$, in which the subset $\{ \rho_a, \rho_c, \rho_d, \rho_f, \eta_0, \eta_3, \eta_5 \}$ preserves all the symmetries of the original Hamiltonian Eq.~\eqref{eq:model}. To make the VMC compatible with the spacial nematic order, we introduce three additional parameters $\{ \eta^{x}_7, \eta^{z}_7, \theta \}$ in the mean-field Hamiltonian to break the $C_3$ symmetry but preserve a residual mirror symmetry. The mirror plane is parallel to the $z$ bond and perpendicular to the lattice plane. Then Gutzwiller projection is performed to the mean-field ground state to enforce the particle number constraint. Thus we obtain the trial wave function $|\Psi(\{x\})\rangle = P_G |\Psi_{\rm mf} (\{x\}) \rangle = \sum_\alpha f(\alpha) |\alpha\rangle$, where $\{x\}$ denote the variational parameters and $\alpha$ stands for the Ising bases in the many-body Hilbert space. The energy of the trial state $E(\{x\}) = \langle \Psi(\{x\}) |H| \Psi(\{x\})\rangle / \langle \Psi(\{x\}) |\Psi(\{x\})\rangle = \sum_\alpha \frac{|f(\alpha)|^2}{\sum_\gamma |f(\gamma)|^2}(\sum_\beta \langle \beta | H | \alpha \rangle \frac{f(\beta)*}{f(\alpha)*})$ is computed using Monte Carlo sampling. Then the optimal parameters $\{x\}$ are determined by minimizing the energy $E(\{x\})$. Our VMC calculations are performed on the torus geometry up to the size $9 \times 9 \times 2$ with $162$ sites.

In the absence of the variational parameters $\{ \eta^x_7, \eta^z_7, \theta \}$, an intermediate CSL phase between the zigzag phase and the polarized trivial phase is identified with Chern number $\nu = 2$~\cite{Li2021}. This CSL is an Abelian QSL with the quasiparticle excitations $a$, $\bar{a}$, $\epsilon$, $1$, where $1$ denotes the vacuum, $\epsilon$ is the fermion, $a$ and $\bar{a}$ are two types of vortices with topological spin $e^{i \pi /4}$~\cite{Kitaev2006}. These excitations follow the fusion rules $\epsilon \times \epsilon = 1$, $a \times \bar{a} = 1$, $a \times \epsilon = \bar{a}$, $\bar{a} \times \epsilon = a$, and $\bar{a} \times \bar{a} = \epsilon$~\cite{Kitaev2006}. The gapless edge of this CSL contains $2$ branches of chiral Majorana excitations with total chiral central charge $c_-=1$.

By considering the additional parameters $\{ \eta^{x}_7, \eta^{z}_7, \theta \}$, our VMC calculations still support a CSL state, but with a nematic order which further lowers the variational energy. For example, on the $8 \times 8 \times 2$ size at the magnetic field $h = 0.3$, the variational energy of the CSL state with the $C_3$ symmetry is $-0.3691$. By allowing the $C_3$ symmetry breaking, the energy is lowered to $-0.3746$ with an emergent nematic order. This optimized variational energy is very close to the energies found in the DMRG simulation, e.g., $-0.3790$ in the bulk of the XC6 cylinder. 
The details of the VMC identification of the nematic order can be found in Appendix~\ref{app:VMC}. Besides the spontaneous nematic order in the new ground state, the obtained low-lying energies also have a drastic change. In Fig.~\ref{fig:band}, we show the dispersion of the fermion bands in the optimized mean-field Hamiltonian Eq.~\eqref{eq:MF} with $h = 0.3$, which indicates that with the $C_3$ symmetry breaking a small excitation gap is opened. 

The mean field dispersion is plotted for a system with $120\times 120\times 2$ sites. There are totally eight bands due to the two sublattices, two spin components and the particle-hole redundancy. A remarkable feature of the spinon dispersion is that six of the eight bands are quit flat and are close to zero energy, and the rest two bands are very dispersive. The existence of the nearly flat bands is very similar to the mean field spectrum of the pure Kitaev model, where the dispersive band mainly come from the $c$ fermions and the flat bands result from the $b_{x,y,z}$ fermions. Our variational calculation indicates that four parameters are dominating in the intermediate QSL phase, namely $\rho_a, \rho_c, \eta_0$ and $\eta_3$, where $\rho_a,\rho_c$ come from the Kitaev interactions, $\eta_0$ comes from the Heisenberg interaction and $\eta_3$ comes from the $\Gamma$ interaction.

It will be illustrative to compare the VMC results with those from DMRG and XTRG. The size of the spinon gap with respect to the total band width is $0.02|K|$ (we have set the band width of the spinons to be $2|K|$) which is very small. Practically, it is not easy to detect such a small gap using correlation functions. This qualitatively interprets the algebraic-like spin correlation decay observed in DMRG results. Furthermore, above the energy $0.02|K|$, the fermionic spinons start to show density of states which is proximately proportional to energy in a small energy window. This agrees with the $C_m\propto T^2$ law of the specific heat at the temperature region $0.02|K|\sim0.04|K|$. With further increasing energy, a Van Hove singularity appears in the dispersive band at $E\sim 0.5|K|$ which yields a large density of states. Hence a peak (or shoulder) is expected to appear in the specific heat $C_m$ at around $0.5|K|$. We notice that the existence of two-temperature scale structure in the specific heat is indeed observed in the XTRG results shown in Fig.~\ref{fig:FiniteT}.

\section{Summary and discussion}
\label{sec:summary}

By combining the DMRG, XTRG, and VMC calculations, we investigate the nature of the intermediate QSL phase of a honeycomb $K$-$J$-$\Gamma$-$\Gamma'$ model in an out-of-plane magnetic field along the $[1 1 1]$ direction, which is sandwiched between a low-field zigzag order phase and a high-field polarized phase.

{By using DMRG simulations on different cylinder geometries, we confirm the absent magnetic order and unveil the algebraic-like spin correlation function decay $C(r) \sim 1/r^{K_s}$ with small power exponent $K_s \lesssim 1.5$. 
Besides, we identify a robust spin scalar chiral order $\langle \mathbf{S}_i \cdot (\mathbf{S}_j \times \mathbf{S}_k) \rangle$ and a lattice nematic order characterizing the $C_3$ symmetry breaking.
We also extend the XTRG simulation to longer 4-leg cylinders and down to lower temperature.
Below the lower-temperature scale $T''_{\rm L} \simeq 0.07$, we observe a robust power-law behavior of the specific heat $C_{\rm m} \sim T^{2}$, as well as considerable amount of thermal entropy. 
By allowing $C_3$ symmetry breaking in the VMC calculation, we obtain the same gapped CSL reported in Ref.~\cite{Li2021} but with an additional nematic order that can further lower the variational energy.

While our DMRG results also find the chiral order and nematic order that can agree with the nematic CSL found by VMC, the ground-state and thermal measurements suggest a nearly gapless QSL.
Yet, the VMC results provide another picture to understand the finite-size simulation results.
The small ratio of the gap with respect to the spinon band width ($\sim 0.02$) indicates a large correlation length that may go beyond the resolution of present DMRG calculation. 
The finite density of states for spinons, which are nearly proportional to energy in a small energy window above the gap, also agrees with the behavior of the specific heat $C_m \sim T^2$.
Identifying the nature of gapped or gapless may need further studies on larger system size.

It is noteworthy that the QSL predicted here for $\alpha$-RuCl$_3$ with relatively large low-energy excitation could be further studied experimentally in high-field NMR~\cite{Meier2016}, ESR spectroscopy~\cite{Akaki2018}, and specific heat measurements. Besides, this intermediate QSL may also relate to other recently proposed field-induced states in the Kitaev candidates, e.g., the broad magnetic continuum observed by time-domain terahertz spectroscopy measurements on BaCo$_2$(AsO$_4$)$_2$~\cite{Zhang2023}.

We also notice that this QSL phase might be related to the nematic QSL-like phase in the $K$-$\Gamma$-$\Gamma'$ model with dominant FM Kitaev interaction and off-diagonal exchanges $\Gamma > 0$, $\Gamma' < 0$~\cite{Gordon2019,Lee2020,Gohlke2020}.
While this phase was found not connected to the Kitaev spin liquid, many-body simulations found the $C_3$ symmetry breaking~\cite{Lee2020,Gohlke2020} and gapless-like excitations~\cite{Gohlke2020}. 
It would be interesting to explore the potential connection of the two phases and whether the phase in the $K$-$\Gamma$-$\Gamma'$ model may also be a nematic CSL.}

\begin{acknowledgments}
We acknowledge stimulating discussions with Lei Wang, Xing-Yu Zhang, and Qiang Luo. This work was supported by the National Natural Science Foundation of China Grants No. 11874078, No. 11834014, No. 12222412, No. 11974036, N0. 11974421, No. 12047503 and No.12134020; the Innovation Program for Quantum Science and Technology (2021ZD0301900); CAS Project for Young Scientists in Basic Research (Grant No. YSBR-057); and China National Postdoctoral Program for Innovative Talents (Grant No. BX20220291). 
\end{acknowledgments}

\appendix

\section{Quantum phase transitions}
\label{app-sec:transition}

\begin{figure}
    \centering
    \includegraphics[width = 1.0\linewidth]{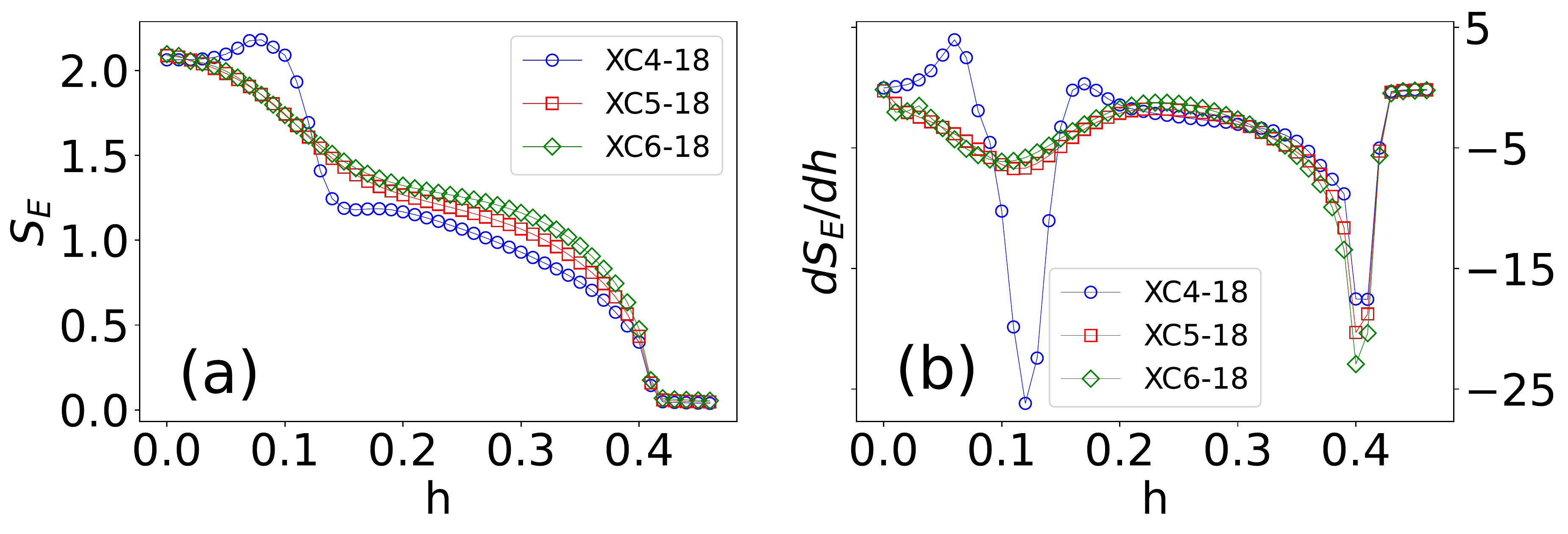}
    \caption{Field dependence of entanglement entropy and its derivative. (a) shows the entropy $S_E$ versus the field $h$ on the XC4-18, XC5-18, and XC6-18 cylinders. (b) shows the derivative of $S_E$ to the field $dS_E/dh$.}
    \label{sp-entropy}
\end{figure}

In Ref.~\cite{Li2021}, the two phase transitions have been identified from the results of magnetization and susceptibility. Here we calculate entanglement entropy to characterize the phase transitions.
As shown in Fig.~\ref{sp-entropy} on the XC cylinders, the entropy exhibits quick drops near the transition points. 
Although the entropy on the XC4 cylinder behaves differently from the wider sizes near the lower transition point, which may be owing to the finite-size effect, the estimated transition points are very close on different system sizes. 
Our determined transition points also agree with those found in Ref.~\cite{Li2021}.

\section{Spin scalar chiral orders in different triangles}
\label{app-sec:chiral}

\begin{figure}
    \centering
    \includegraphics[scale = 0.16]{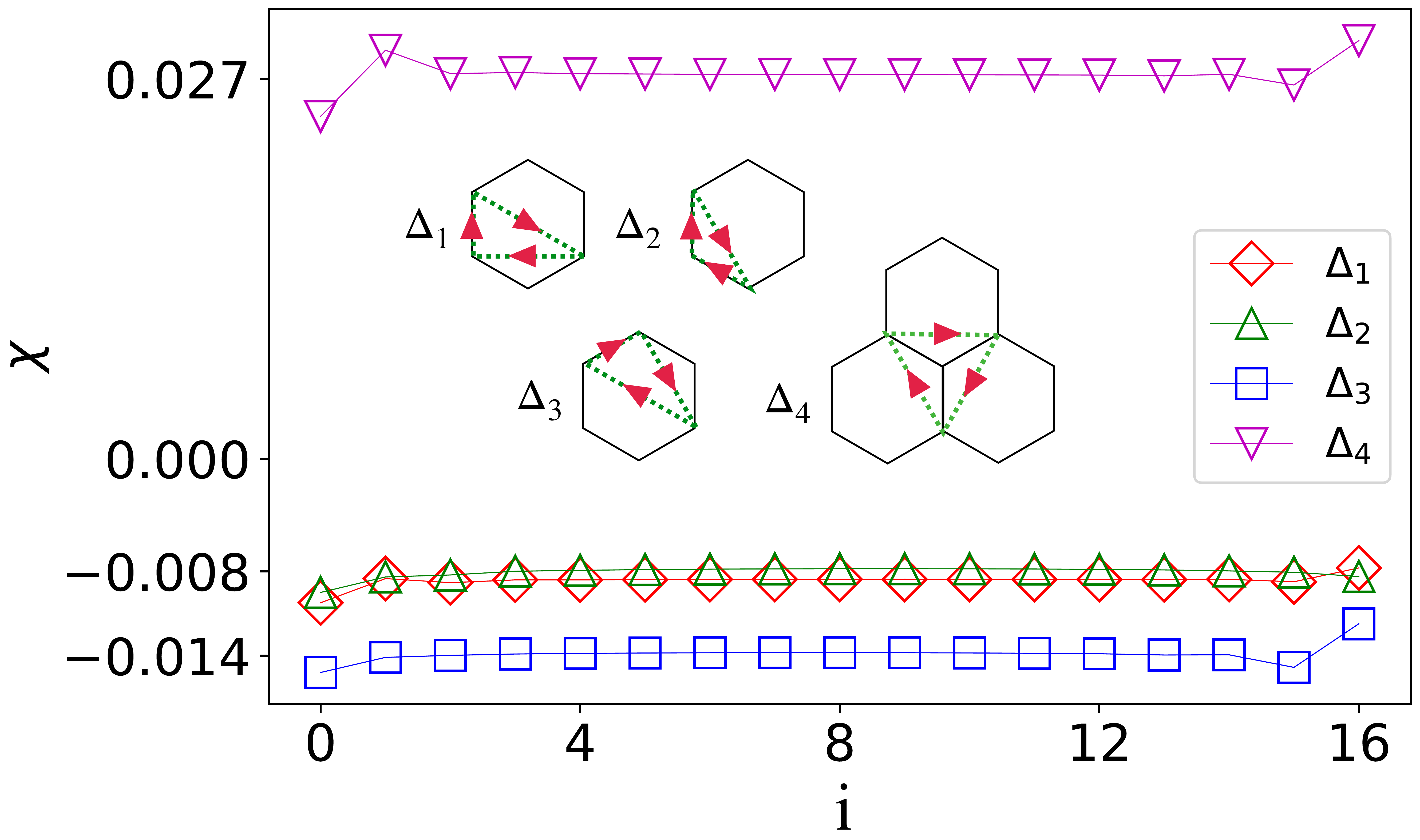}
    \caption{The scalar spin chiral orders in different types of triangles. The data are obtained for $h = 0.24$ on a XC6-18 cylinder. The inset shows the definitions of the four kinds of triangles. $i$ denotes the column number.}
    \label{sp-SC}
\end{figure}

In the main text, we have shown the spin scalar chiral order of the equilateral triangles inside the hexagon. 
Here we demonstrate the chiral orders for other types of triangles. 
Four typical triangles are shown in Fig.~\ref{sp-SC}, where the chiral order of the triangle $\Delta_4$ has the maximum magnitude and has the opposite chirality to the chiral order shown in the main text (inside the hexagon).

\section{Nematic order from iDMRG simulation}
\label{app-sec:nematic}

\begin{figure}
    \centering
    \includegraphics[scale = 0.55]{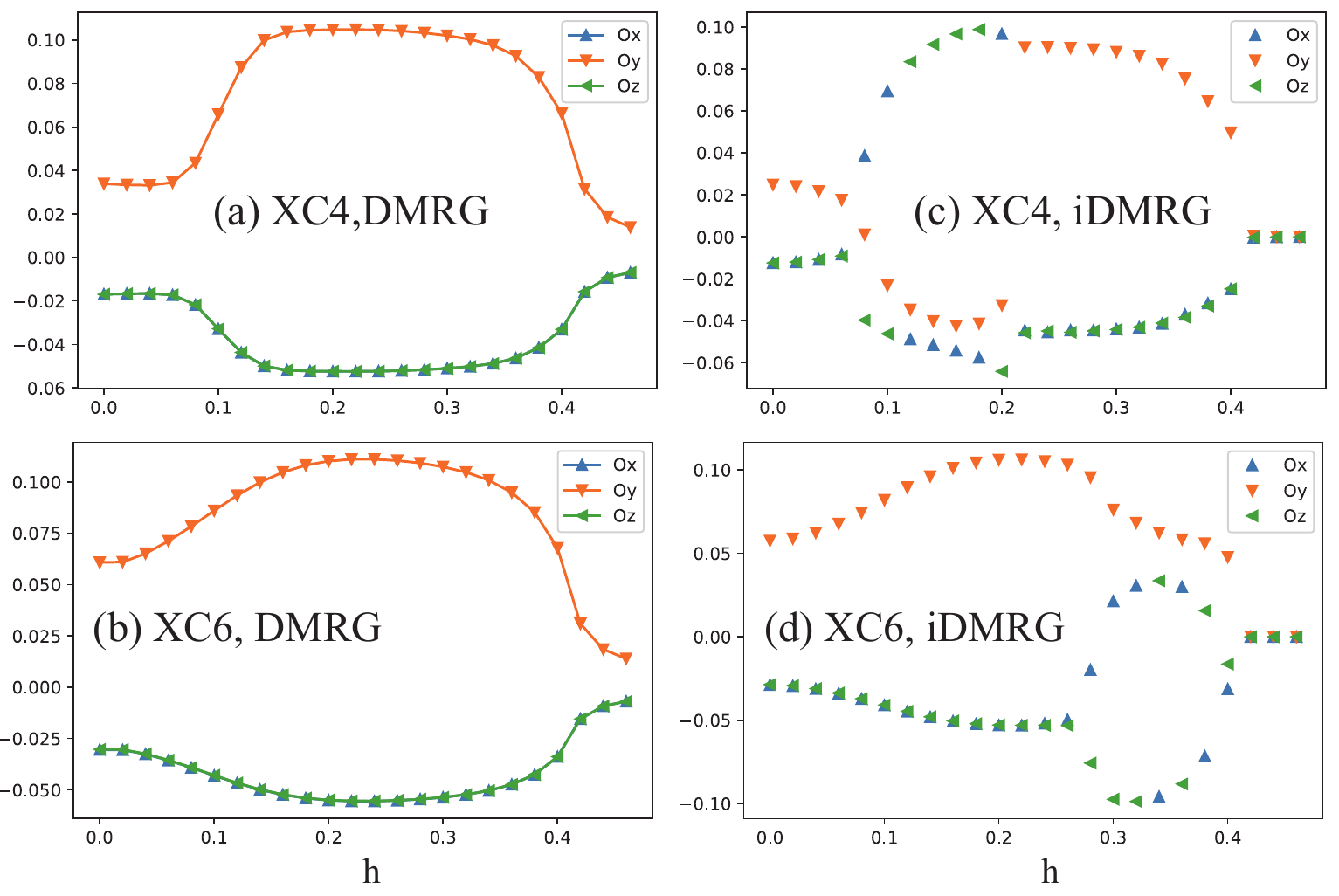}
    \caption{Lattice nematic orders versus the magnetic field. (a) and (b) are the results obtained from DMRG on the XC4 and XC6 cylinders. (c) and (d) are the results obtained from iDMRG on the XC4 and XC6 cylinders. $O_{\gamma} = E_{\gamma} - (E_x + E_y + E_z)/3$, where $\gamma = x, y, z$ denote the three bond directions.}
    \label{sp-nematic}
\end{figure}

In the matin text, we have shown the nematic order obtained by finite DMRG calculation. Here, we show more results from iDMRG calculation.
We define $O_\gamma = E_\gamma - (E_x + E_y + E_z)/3$, where $\gamma = x, y, z$ denote the three bond directions.
In Fig.~\ref{sp-nematic}, we show $O_\gamma$ obtained by DMRG and iDMRG.
In DMRG results, two bond energies are always the same, which is required by the symmetries along the circumference direction.
In iDMRG simulation, probably due to the infinite system length allows the complete $C_3$ symmetry breaking, one can obtain the ground state with $E_x \neq E_y \neq E_z$ in a regime of the intermediate phase, indicating that the complete $C_3$ symmetry breaking is possible in two-dimensional limit.

\section{Fitting of the central charge}
\label{app-sec:entropy}

\begin{figure}
    \centering
    \includegraphics[width = 1\linewidth]{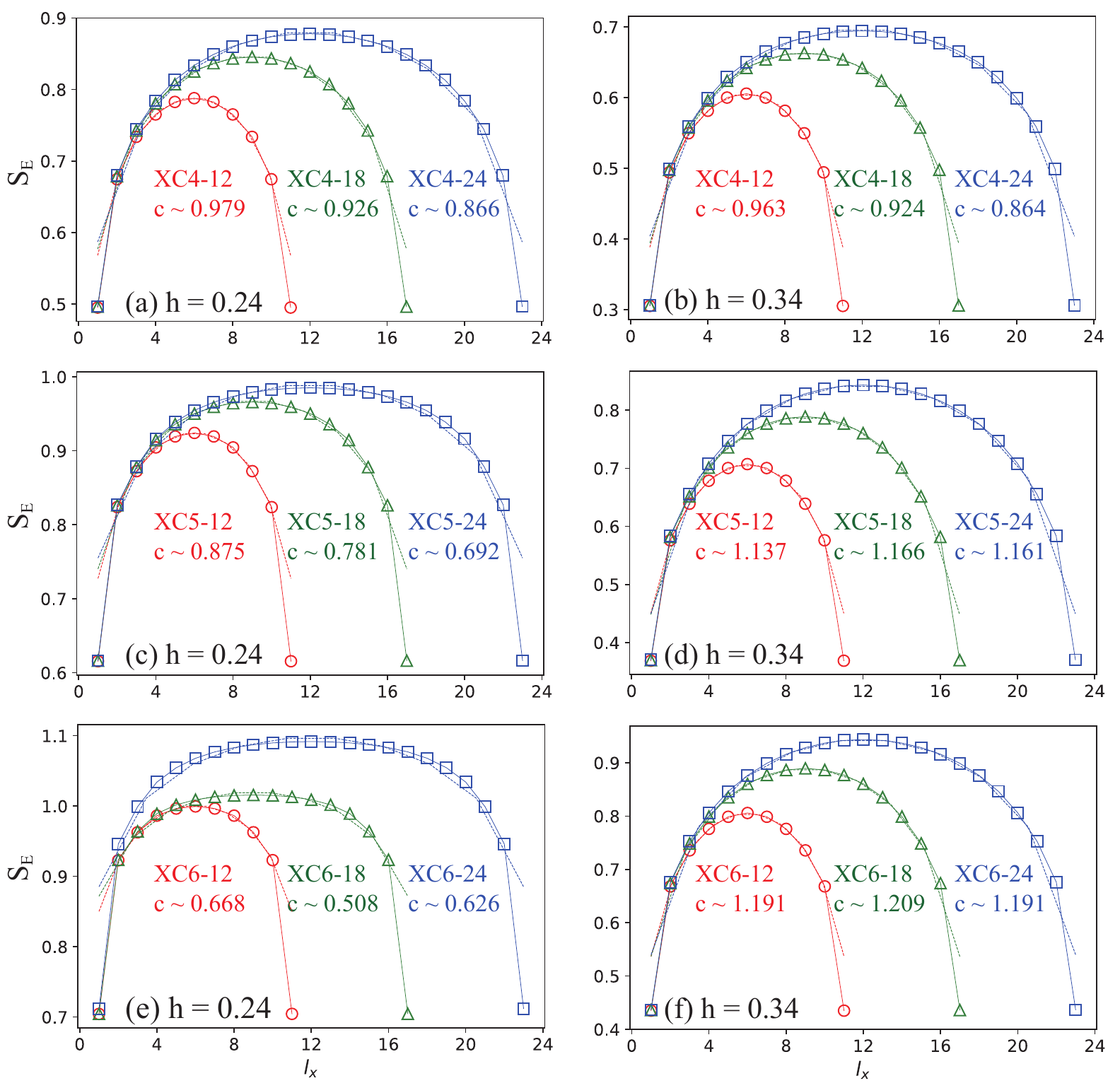}
    \caption{Entanglement entropy versus the subsystem length $l_x$ in the intermediate phase on the XC cylinders. The central charge $c$ is fitted using the formula Eq.~\eqref{eq:entropy}, $S(l_x) = \frac{c}{6} \ln{[(\frac{L_x}{\pi})\sin (\frac{\pi l_x}{L_x})]}+g$. (a), (c), (e) are the entropy for $h = 0.24$ obtained on the $L_y = 4, 5, 6$ XC cylinders, respectively. (b), (d), (f) are the similar plots for $h = 0.34$.
    }
    \label{fig:XCEE_line}
\end{figure}

\begin{figure}
    \centering
    \includegraphics[width = 1\linewidth]{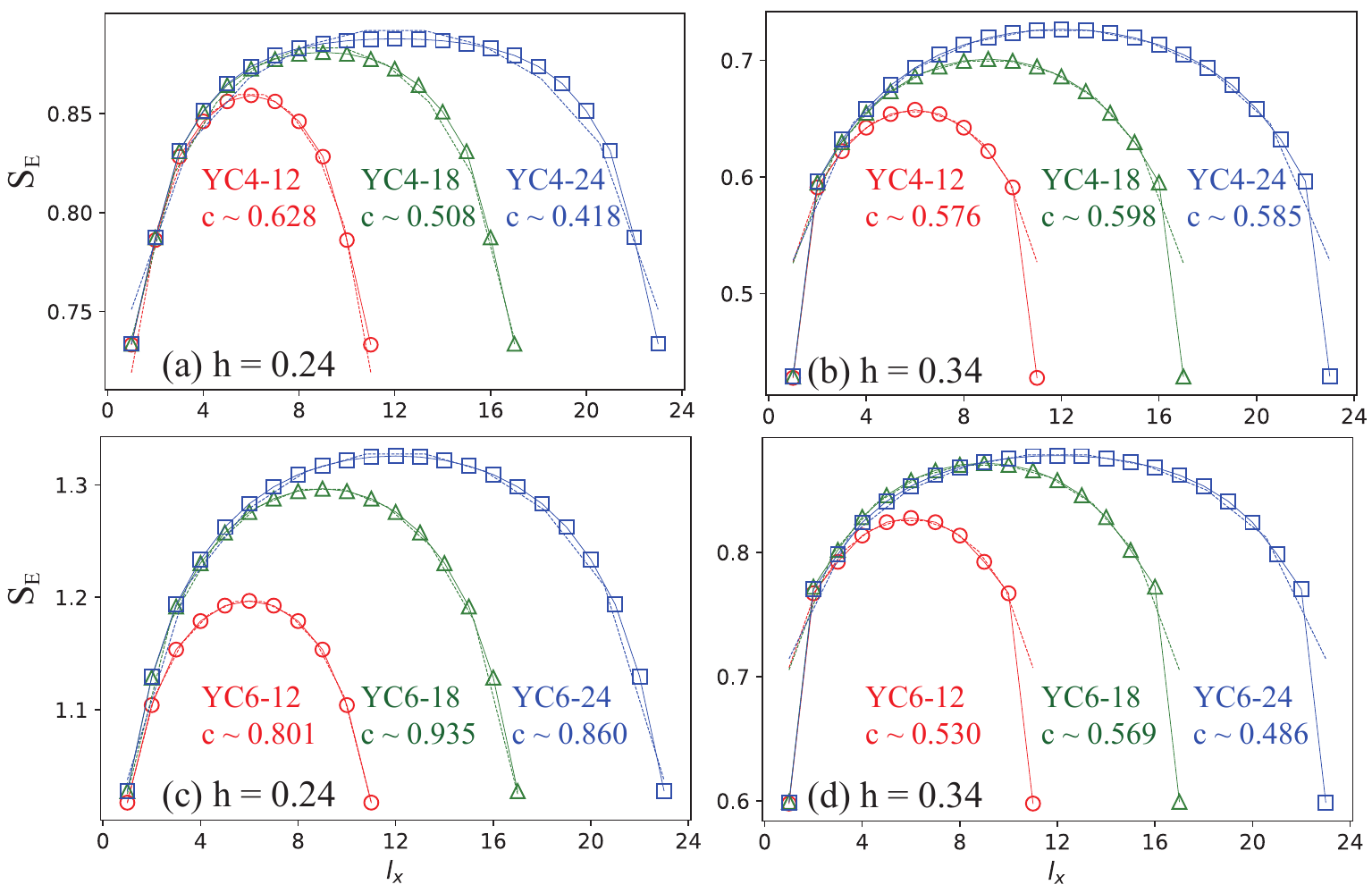}
    \caption{Entanglement entropy versus the subsystem length $l_x$ in the intermediate phase on the YC cylinders. The central charge $c$ is fitted using the formula Eq.~\eqref{eq:entropy}, $S(l_x) = \frac{c}{6} \ln{[(\frac{L_x}{\pi})\sin (\frac{\pi l_x}{L_x})]}+g$. (a) and (c) are the entropy for $h = 0.24$ obtained on the $L_y = 4, 6$ YC cylinders, respectively. (b) and (d) are the similar plots for $h = 0.34$.
    }
    \label{fig:YCEE_line}
\end{figure}

In the main text, we have shown the entanglement entropy versus the conformal distance $\tilde{l}_x$ on the $L_x = 24$ cylinders. Here, we show more results of the entropy on different system sizes.
In Figs.~\ref{fig:XCEE_line} and \ref{fig:YCEE_line}, we present the entropy obtained on the XC and YC cylinders with different $L_x$.
All the entropy data can be well fitted using Eq.~\eqref{eq:entropy}, giving the close central charge on different system lengths.

\begin{table*}[htbp]
\centering
\caption{Different variational states and the corresponding energies. The system sizes are chosen from $3\times 3\times 2$ to $9\times 9 \times 2$, and the magnetic field is $h = 0.3$. Here we use $\eta = (1-\theta)$. Bdr1 = Bdr2 = $1$ denotes the periodic boundary conditions along both the ${\bf a}_1$ and ${\bf a}_2$ directions. $E$ is the variational energy per site. While the best variational state on $3\times 3\times 2$ preserves the $C_3$ symmetry with the vanished parameters $\{ \eta^{x}_7, \eta^{z}_7, \eta \}$, the best variational states on larger sizes break the $C_3$ symmetry.}\label{tab:honeycomb_vari}
\begin{tabular}{|c|c|c|c|c|c|c|c|c|c|c|c|c|c|c|}
\hline
size & Bdr1 & Bdr2 & $E$ & $\rho_{a}$ & $\rho_{c}$ & $\rho_{d}$ & $\rho_{f}$ & $\eta_{0}$ & $\eta_{3}$ & $\eta_{5}$ & $\eta_{7}^{x}$ & $\eta_{7}^{z}$ & $\eta$ & $\lambda_{0}$  \\
\hline
$3\times 3\times 2$ & 1 & 1 & -0.3725 & 0.9397 & 0.1847 & -0.1887 & -0.0079 & -0.1778 & 0.3167 & -0.0847 & 0 & 0 & 0 & 0.0257 \\
\hline
$3\times 3\times 2$ & 1 & 1 & -0.3721 & 0.6129 & 0.1742 & -0.1291 & -0.0095 & -0.2113 & 0.3062 & -0.0586 & -0.0974 & 0.0418 & 0 & 0.0311 \\
\hline
$3\times 3\times 2$ & 1 & 1 & -0.3720 & 0.5928 & 0.1928 & -0.1276 & -0.0085 & -0.2180 & 0.3063 & -0.0653 & -0.1005 & 0.0406 & 0.0356 & 0.0258 \\
\hline\hline
$4\times 4\times 2$ & 1 & 1 & -0.3720 & 1.1047 & 0.1519 & -0.1041 & -0.0094 & -0.1865 & 0.2858 & -0.1019 & 0 & 0 & 0 & 0.0280 \\
\hline
$4\times 4\times 2$ & 1 & 1 & -0.3743 & 1.2157 & 0.1769 & -0.1264 & -0.0115 & -0.1918 & 0.3187 & -0.0938 & -0.0494 & 0.0327 & 0 & 0.0365 \\
\hline
$4\times 4\times 2$ & 1 & 1 & -0.3742 & 1.1839 & 0.1698 & -0.1280 & -0.0118 & -0.1870 & 0.3179 & -0.0887 & -0.0554 & 0.0348 & 0.0181 & 0.0394 \\
\hline\hline
$6\times 6\times 2$ & 1 & 1 & -0.3715 & 0.5972 & 0.1329 & -0.1953 & -0.0153 & -0.1585 & 0.3386 & -0.0715 & 0 & 0 & 0 & 0.0407 \\
\hline
$6\times 6\times 2$ & 1 & 1 & -0.3731 & 1.3208 & 0.1746 & -0.1183 & -0.0078 & -0.1770 & 0.3039 & -0.0945 & -0.0721 & 0.0604 & 0 & 0.0235 \\
\hline
$6\times 6\times 2$ & 1 & 1 & -0.3731 & 1.3363 & 0.2109 & -0.1219 & -0.0167 & -0.1882 & 0.3225 & -0.0966 & -0.0959 & 0.0573 & 0.0008 & 0.0234 \\
\hline\hline
$8\times 8\times 2$ & 1 & 1 & -0.3691 & 0.8860 & 0.1738 & -0.1822 & -0.0072 & -0.1793 & 0.3173 & -0.1013 & 0 & 0 & 0 & 0.0267 \\
\hline
$8\times 8\times 2$ & 1 & 1 & -0.3746 & 0.6559 & 0.1854 & -0.1317 & -0.0092 & -0.1877 & 0.3065 & -0.0740 & -0.0689 & 0.0232 & 0 & 0.0310 \\
\hline
$8\times 8\times 2$ & 1 & 1 & -0.3743 & 0.7866 & 0.1881 & -0.1335 & -0.0094 & -0.1771 & 0.3152 & -0.0821 & -0.0804 & 0.0322 & 0.0351 & 0.0238 \\
\hline\hline
$9\times 9\times 2$ & 1 & 1 & -0.3684 & 1.0719 & 0.1778 & -0.1533 & -0.0083 & -0.1836 & 0.3028 & -0.0991 & 0 & 0 & 0 & 0.0276 \\
\hline
$9\times 9\times 2$ & 1 & 1 & -0.3740 & 0.8519 & 0.1835 & -0.1335 & -0.0080 & -0.1812 & 0.3172 & -0.0834 & -0.0754 & 0.0439 & 0 & 0.0291 \\
\hline
$9\times 9\times 2$ & 1 & 1 & -0.3739 & 0.9048 & 0.1904 & -0.1323 & -0.0084 & -0.1829 & 0.3183 & -0.0816 & -0.0784 & 0.0439 & 0.0294 & 0.0264 \\
\hline
\end{tabular}
\end{table*}

\section{Nematic order in Variational Monte Carlo results}
\label{app:VMC}

Using VMC calculation, we examine the mean-field Hamiltonian Eq.~\eqref{eq:MF} in the intermediate phase.
Here we show the results at the magnetic field $h = 0.3$ as an example.
On different system sizes with $L_{\rm a_1} \times L_{\rm a_2} \times 2$ ($L_{\rm a_1}$ and $L_{\rm a_2}$ denote the numbers of unit cell along the two lattice spacing directions ${\bf a}_1$ and ${\bf a}_2$) under the periodic boundary conditions (PBC), which preserve the $C_3$ symmetry, we calculate the variational energies up to the system size $9 \times 9 \times 2$, as shown in Table~\ref{tab:honeycomb_vari}.
On the smaller $3\times 3\times 2$ system sizes, the variational ground states preserve the $C_3$ symmetry and have a large overlap with an low-lying exact state obtained by exact diagonalization. Remarkably, on the larger sizes ($\geq 4 \times 4 \times 2$), the optimized ground state always breaks the $C_3$ symmetry (see Table~\ref{tab:honeycomb_vari}).

Next, to interpret the difference results for systems with different size, we perform the `symmetrization' procedure among the projected states under different boundary conditions to obtain a $C_3$ symmetric wave functions. We diagonalize the original Hamiltonian Eq.~\eqref{eq:model} in the Hilbert space with anti-periodic boundary conditions (APBC), and the obtained results are shown in Table~\ref{tab:honeycomb_renormBdr}, where the parameters $\{ \eta^{x}_7, \eta^{z}_7, \eta \} = 0$ (with $\eta = (1-\theta)$) for the three projected states in the APBC. For the $3\times 3\times 2$ system size, the energy of the symmetrized ground state is significantly lowered. Actually we have confirmed that the overlap of the symmetrized state has and the exact ground state is close to 1, and its energy also very close to that of the exact state. However, for systems with $6\times 6 \times 2$ and $8\times 8 \times 2$ sizes, the symmetrized states are not significantly proved (as shown in Table~\ref{tab:honeycomb_renormBdr} the energy correction is very small) and their energy are still higher than those of the corresponding nematic states in Table~\ref{tab:honeycomb_vari}. This result indicate that with increasing system size, within the Hilbert space of projected `non-nematic' states there does not exist a $C_3$ symmetrized state whose energy is lower than the ones with nematic order. This result supports the $C_3$ symmetry breaking in the thermodynamic limit.

For the $C_3$ symmetry breaking nematic states, we can also perform similar symmetrization procedure in the space spaned by the three nematic states related by the $C_3$ operation. The results are shown in Table~\ref{tab:honeycomb_renormC3}. We find that the symmetrization cannot efficiently lower the energy. Furthermore eigenvalues of the fidelity matrix are all close to $1$. The reason is that the overlaps between the different nematic states as well as the hopping elements of the Hamiltonian among the three states are all vanishing small, i.e. $\langle \psi^{nematic}_i | \psi^{nematic}_j \rangle \approx \langle \psi^{nematic}_i | \hat{H} | \psi^{nematic}_j \rangle \approx 0$, which indicates a three-fold degeneracy that agrees with a $C_3$ symmetry breaking.

\begin{table}[t]
\centering
\caption{Symmetrization of the variational wavefunctions under the three anti-periodic boundary conditions. The system sizes include $3\times 3\times 2$, $6\times 6\times 2$, and $8\times 8\times 2$. $E$ is the ground-state energy of the symmetrized state. $E_{+,-}$, $E_{-,+}$, and $E_{-,-}$ are the ground state energies under the three anti-periodic boundary conditions, before the symmetrization.}\label{tab:honeycomb_renormBdr}
\begin{tabular}{|c|c|c|c|c|c|c|c|c|c|c|c|c|c|c|c|}
\hline
size & $E$ & $\sqrt{|O_{0}|^{2}}$ & $E_{+,-}$ & $E_{-,+}$ & $E_{-,-}$  \\
\hline
$3\times 3\times 2$ & -0.3781 & 0.9846 & -0.3709 & -0.3706 & -0.3708 \\
\hline
$6\times 6\times 2$ & -0.3701 & / & -0.3690 & -0.3690 & -0.3689 \\
\hline
$8\times 8\times 2$ & -0.3701 & / & -0.3696 & -0.3695 & -0.3696 \\
\hline
\end{tabular}
\end{table}

\begin{table}[t]
\centering
\caption{Symmetrization of the three nematic states. The system sizes include $6\times 6\times 2$ and $8\times 8\times 2$. $E$ is the ground-state energy of the renormalized state. $E_{nematic}$ are the ground state energies of the nematic states before the symmetrization. $\rho_1, \rho_2, \rho_3$ are the eigenvalues of the density matrix.}
\label{tab:honeycomb_renormC3}
\begin{tabular}{|c|c|c|c|c|c|c|c|c|c|c|c|c|c|c|c|c|c|c|}
\hline
size  & $E$ & $E_{nematic}$ & $\rho_{1}$ & $\rho_{2}$ & $\rho_{3}$ \\
\hline
$6\times 6\times 2$ & -0.3732 & -0.3729 & 0.9558 & 1.0028 & 1.0414 \\
\hline
$8\times 8\times 2$ & -0.3757 & -0.3757 & 0.9870 & 1.0000 & 1.0130 \\
\hline
\end{tabular}
\end{table}
\newpage

\nocite{*}

\bibliography{apssamp1}

\begin{thebibliography}{89}%
\makeatletter
\providecommand \@ifxundefined [1]{%
 \@ifx{#1\undefined}
}%
\providecommand \@ifnum [1]{%
 \ifnum #1\expandafter \@firstoftwo
 \else \expandafter \@secondoftwo
 \fi
}%
\providecommand \@ifx [1]{%
 \ifx #1\expandafter \@firstoftwo
 \else \expandafter \@secondoftwo
 \fi
}%
\providecommand \natexlab [1]{#1}%
\providecommand \enquote  [1]{``#1''}%
\providecommand \bibnamefont  [1]{#1}%
\providecommand \bibfnamefont [1]{#1}%
\providecommand \citenamefont [1]{#1}%
\providecommand \href@noop [0]{\@secondoftwo}%
\providecommand \href [0]{\begingroup \@sanitize@url \@href}%
\providecommand \@href[1]{\@@startlink{#1}\@@href}%
\providecommand \@@href[1]{\endgroup#1\@@endlink}%
\providecommand \@sanitize@url [0]{\catcode `\\12\catcode `\$12\catcode
  `\&12\catcode `\#12\catcode `\^12\catcode `\_12\catcode `\%12\relax}%
\providecommand \@@startlink[1]{}%
\providecommand \@@endlink[0]{}%
\providecommand \url  [0]{\begingroup\@sanitize@url \@url }%
\providecommand \@url [1]{\endgroup\@href {#1}{\urlprefix }}%
\providecommand \urlprefix  [0]{URL }%
\providecommand \Eprint [0]{\href }%
\providecommand \doibase [0]{https://doi.org/}%
\providecommand \selectlanguage [0]{\@gobble}%
\providecommand \bibinfo  [0]{\@secondoftwo}%
\providecommand \bibfield  [0]{\@secondoftwo}%
\providecommand \translation [1]{[#1]}%
\providecommand \BibitemOpen [0]{}%
\providecommand \bibitemStop [0]{}%
\providecommand \bibitemNoStop [0]{.\EOS\space}%
\providecommand \EOS [0]{\spacefactor3000\relax}%
\providecommand \BibitemShut  [1]{\csname bibitem#1\endcsname}%
\let\auto@bib@innerbib\@empty
\bibitem [{\citenamefont {Savary}\ and\ \citenamefont
  {Balents}(2016)}]{Savary2016}%
  \BibitemOpen
  \bibfield  {author} {\bibinfo {author} {\bibfnamefont {L.}~\bibnamefont
  {Savary}}\ and\ \bibinfo {author} {\bibfnamefont {L.}~\bibnamefont
  {Balents}},\ }\bibfield  {title} {\bibinfo {title} {Quantum spin liquids: a
  review},\ }\href {https://doi.org/10.1088/0034-4885/80/1/016502} {\bibfield
  {journal} {\bibinfo  {journal} {Reports on Progress in Physics}\ }\textbf
  {\bibinfo {volume} {80}},\ \bibinfo {pages} {016502} (\bibinfo {year}
  {2016})}\BibitemShut {NoStop}%
\bibitem [{\citenamefont {Zhou}\ \emph {et~al.}(2017)\citenamefont {Zhou},
  \citenamefont {Kanoda},\ and\ \citenamefont {Ng}}]{Zhou2017}%
  \BibitemOpen
  \bibfield  {author} {\bibinfo {author} {\bibfnamefont {Y.}~\bibnamefont
  {Zhou}}, \bibinfo {author} {\bibfnamefont {K.}~\bibnamefont {Kanoda}},\ and\
  \bibinfo {author} {\bibfnamefont {T.-K.}\ \bibnamefont {Ng}},\ }\bibfield
  {title} {\bibinfo {title} {Quantum spin liquid states},\ }\href
  {https://doi.org/10.1103/RevModPhys.89.025003} {\bibfield  {journal}
  {\bibinfo  {journal} {Rev. Mod. Phys.}\ }\textbf {\bibinfo {volume} {89}},\
  \bibinfo {pages} {025003} (\bibinfo {year} {2017})}\BibitemShut {NoStop}%
\bibitem [{\citenamefont {Wen}(1991)}]{Wen1991}%
  \BibitemOpen
  \bibfield  {author} {\bibinfo {author} {\bibfnamefont {X.~G.}\ \bibnamefont
  {Wen}},\ }\bibfield  {title} {\bibinfo {title} {Mean-field theory of
  spin-liquid states with finite energy gap and topological orders},\ }\href
  {https://doi.org/10.1103/PhysRevB.44.2664} {\bibfield  {journal} {\bibinfo
  {journal} {Phys. Rev. B}\ }\textbf {\bibinfo {volume} {44}},\ \bibinfo
  {pages} {2664} (\bibinfo {year} {1991})}\BibitemShut {NoStop}%
\bibitem [{\citenamefont {Senthil}\ and\ \citenamefont
  {Fisher}(2000)}]{Senthil2000}%
  \BibitemOpen
  \bibfield  {author} {\bibinfo {author} {\bibfnamefont {T.}~\bibnamefont
  {Senthil}}\ and\ \bibinfo {author} {\bibfnamefont {M.~P.~A.}\ \bibnamefont
  {Fisher}},\ }\bibfield  {title} {\bibinfo {title} {${Z}_{2}$ gauge theory of
  electron fractionalization in strongly correlated systems},\ }\href
  {https://doi.org/10.1103/PhysRevB.62.7850} {\bibfield  {journal} {\bibinfo
  {journal} {Phys. Rev. B}\ }\textbf {\bibinfo {volume} {62}},\ \bibinfo
  {pages} {7850} (\bibinfo {year} {2000})}\BibitemShut {NoStop}%
\bibitem [{\citenamefont {Senthil}\ and\ \citenamefont
  {Fisher}(2001)}]{Senthil2001}%
  \BibitemOpen
  \bibfield  {author} {\bibinfo {author} {\bibfnamefont {T.}~\bibnamefont
  {Senthil}}\ and\ \bibinfo {author} {\bibfnamefont {M.~P.~A.}\ \bibnamefont
  {Fisher}},\ }\bibfield  {title} {\bibinfo {title} {Fractionalization in the
  cuprates: Detecting the topological order},\ }\href
  {https://doi.org/10.1103/PhysRevLett.86.292} {\bibfield  {journal} {\bibinfo
  {journal} {Phys. Rev. Lett.}\ }\textbf {\bibinfo {volume} {86}},\ \bibinfo
  {pages} {292} (\bibinfo {year} {2001})}\BibitemShut {NoStop}%
\bibitem [{\citenamefont {Kitaev}(2003)}]{Kitaev2003}%
  \BibitemOpen
  \bibfield  {author} {\bibinfo {author} {\bibfnamefont {A.}~\bibnamefont
  {Kitaev}},\ }\bibfield  {title} {\bibinfo {title} {Fault-tolerant quantum
  computation by anyons},\ }\href
  {https://doi.org/https://doi.org/10.1016/S0003-4916(02)00018-0} {\bibfield
  {journal} {\bibinfo  {journal} {Annals of Physics}\ }\textbf {\bibinfo
  {volume} {303}},\ \bibinfo {pages} {2} (\bibinfo {year} {2003})}\BibitemShut
  {NoStop}%
\bibitem [{\citenamefont {Kitaev}(2006)}]{Kitaev2006}%
  \BibitemOpen
  \bibfield  {author} {\bibinfo {author} {\bibfnamefont {A.}~\bibnamefont
  {Kitaev}},\ }\bibfield  {title} {\bibinfo {title} {Anyons in an exactly
  solved model and beyond},\ }\href
  {https://doi.org/https://doi.org/10.1016/j.aop.2005.10.005} {\bibfield
  {journal} {\bibinfo  {journal} {Annals of Physics}\ }\textbf {\bibinfo
  {volume} {321}},\ \bibinfo {pages} {2} (\bibinfo {year} {2006})}\BibitemShut
  {NoStop}%
\bibitem [{\citenamefont {Jackeli}\ and\ \citenamefont
  {Khaliullin}(2009)}]{Jackeli2009}%
  \BibitemOpen
  \bibfield  {author} {\bibinfo {author} {\bibfnamefont {G.}~\bibnamefont
  {Jackeli}}\ and\ \bibinfo {author} {\bibfnamefont {G.}~\bibnamefont
  {Khaliullin}},\ }\bibfield  {title} {\bibinfo {title} {Mott insulators in the
  strong spin-orbit coupling limit: From {H}eisenberg to a quantum compass and
  {K}itaev models},\ }\href {https://doi.org/10.1103/PhysRevLett.102.017205}
  {\bibfield  {journal} {\bibinfo  {journal} {Phys. Rev. Lett.}\ }\textbf
  {\bibinfo {volume} {102}},\ \bibinfo {pages} {017205} (\bibinfo {year}
  {2009})}\BibitemShut {NoStop}%
\bibitem [{\citenamefont {Singh}\ \emph {et~al.}(2012)\citenamefont {Singh},
  \citenamefont {Manni}, \citenamefont {Reuther}, \citenamefont {Berlijn},
  \citenamefont {Thomale}, \citenamefont {Ku}, \citenamefont {Trebst},\ and\
  \citenamefont {Gegenwart}}]{Singh2012}%
  \BibitemOpen
  \bibfield  {author} {\bibinfo {author} {\bibfnamefont {Y.}~\bibnamefont
  {Singh}}, \bibinfo {author} {\bibfnamefont {S.}~\bibnamefont {Manni}},
  \bibinfo {author} {\bibfnamefont {J.}~\bibnamefont {Reuther}}, \bibinfo
  {author} {\bibfnamefont {T.}~\bibnamefont {Berlijn}}, \bibinfo {author}
  {\bibfnamefont {R.}~\bibnamefont {Thomale}}, \bibinfo {author} {\bibfnamefont
  {W.}~\bibnamefont {Ku}}, \bibinfo {author} {\bibfnamefont {S.}~\bibnamefont
  {Trebst}},\ and\ \bibinfo {author} {\bibfnamefont {P.}~\bibnamefont
  {Gegenwart}},\ }\bibfield  {title} {\bibinfo {title} {Relevance of the
  {H}eisenberg-{K}itaev model for the honeycomb lattice iridates
  ${A}_{2}{I}r{O}_3$},\ }\href {https://doi.org/10.1103/PhysRevLett.108.127203}
  {\bibfield  {journal} {\bibinfo  {journal} {Phys. Rev. Lett.}\ }\textbf
  {\bibinfo {volume} {108}},\ \bibinfo {pages} {127203} (\bibinfo {year}
  {2012})}\BibitemShut {NoStop}%
\bibitem [{\citenamefont {Kitagawa}\ \emph {et~al.}(2018)\citenamefont
  {Kitagawa}, \citenamefont {Takayama}, \citenamefont {Matsumoto},
  \citenamefont {Kato}, \citenamefont {Takano}, \citenamefont {Kishimoto},
  \citenamefont {Bette}, \citenamefont {Dinnebier}, \citenamefont {Jackeli},\
  and\ \citenamefont {Takagi}}]{Kitagawa2018}%
  \BibitemOpen
  \bibfield  {author} {\bibinfo {author} {\bibfnamefont {K.}~\bibnamefont
  {Kitagawa}}, \bibinfo {author} {\bibfnamefont {T.}~\bibnamefont {Takayama}},
  \bibinfo {author} {\bibfnamefont {Y.}~\bibnamefont {Matsumoto}}, \bibinfo
  {author} {\bibfnamefont {A.}~\bibnamefont {Kato}}, \bibinfo {author}
  {\bibfnamefont {R.}~\bibnamefont {Takano}}, \bibinfo {author} {\bibfnamefont
  {Y.}~\bibnamefont {Kishimoto}}, \bibinfo {author} {\bibfnamefont
  {S.}~\bibnamefont {Bette}}, \bibinfo {author} {\bibfnamefont
  {R.}~\bibnamefont {Dinnebier}}, \bibinfo {author} {\bibfnamefont
  {G.}~\bibnamefont {Jackeli}},\ and\ \bibinfo {author} {\bibfnamefont
  {H.}~\bibnamefont {Takagi}},\ }\bibfield  {title} {\bibinfo {title} {A
  spin-orbital-entangled quantum liquid on a honeycomb lattice},\ }\href
  {https://doi.org/10.1038/nature25482} {\bibfield  {journal} {\bibinfo
  {journal} {Nature}\ }\textbf {\bibinfo {volume} {554}},\ \bibinfo {pages}
  {341} (\bibinfo {year} {2018})}\BibitemShut {NoStop}%
\bibitem [{\citenamefont {Sears}\ \emph {et~al.}(2015)\citenamefont {Sears},
  \citenamefont {Songvilay}, \citenamefont {Plumb}, \citenamefont {Clancy},
  \citenamefont {Qiu}, \citenamefont {Zhao}, \citenamefont {Parshall},\ and\
  \citenamefont {Kim}}]{Sears2015}%
  \BibitemOpen
  \bibfield  {author} {\bibinfo {author} {\bibfnamefont {J.~A.}\ \bibnamefont
  {Sears}}, \bibinfo {author} {\bibfnamefont {M.}~\bibnamefont {Songvilay}},
  \bibinfo {author} {\bibfnamefont {K.~W.}\ \bibnamefont {Plumb}}, \bibinfo
  {author} {\bibfnamefont {J.~P.}\ \bibnamefont {Clancy}}, \bibinfo {author}
  {\bibfnamefont {Y.}~\bibnamefont {Qiu}}, \bibinfo {author} {\bibfnamefont
  {Y.}~\bibnamefont {Zhao}}, \bibinfo {author} {\bibfnamefont {D.}~\bibnamefont
  {Parshall}},\ and\ \bibinfo {author} {\bibfnamefont {Y.-J.}\ \bibnamefont
  {Kim}},\ }\bibfield  {title} {\bibinfo {title} {Magnetic order in
  $\alpha$-{R}u{C}l$_3$: A honeycomb-lattice quantum magnet with strong
  spin-orbit coupling},\ }\href {https://doi.org/10.1103/PhysRevB.91.144420}
  {\bibfield  {journal} {\bibinfo  {journal} {Phys. Rev. B}\ }\textbf {\bibinfo
  {volume} {91}},\ \bibinfo {pages} {144420} (\bibinfo {year}
  {2015})}\BibitemShut {NoStop}%
\bibitem [{\citenamefont {Johnson}\ \emph {et~al.}(2015)\citenamefont
  {Johnson}, \citenamefont {Williams}, \citenamefont {Haghighirad},
  \citenamefont {Singleton}, \citenamefont {Zapf}, \citenamefont {Manuel},
  \citenamefont {Mazin}, \citenamefont {Li}, \citenamefont {Jeschke},
  \citenamefont {Valent\'{\i}},\ and\ \citenamefont {Coldea}}]{Johnson2015}%
  \BibitemOpen
  \bibfield  {author} {\bibinfo {author} {\bibfnamefont {R.~D.}\ \bibnamefont
  {Johnson}}, \bibinfo {author} {\bibfnamefont {S.~C.}\ \bibnamefont
  {Williams}}, \bibinfo {author} {\bibfnamefont {A.~A.}\ \bibnamefont
  {Haghighirad}}, \bibinfo {author} {\bibfnamefont {J.}~\bibnamefont
  {Singleton}}, \bibinfo {author} {\bibfnamefont {V.}~\bibnamefont {Zapf}},
  \bibinfo {author} {\bibfnamefont {P.}~\bibnamefont {Manuel}}, \bibinfo
  {author} {\bibfnamefont {I.~I.}\ \bibnamefont {Mazin}}, \bibinfo {author}
  {\bibfnamefont {Y.}~\bibnamefont {Li}}, \bibinfo {author} {\bibfnamefont
  {H.~O.}\ \bibnamefont {Jeschke}}, \bibinfo {author} {\bibfnamefont
  {R.}~\bibnamefont {Valent\'{\i}}},\ and\ \bibinfo {author} {\bibfnamefont
  {R.}~\bibnamefont {Coldea}},\ }\bibfield  {title} {\bibinfo {title}
  {Monoclinic crystal structure of $\alpha$-{R}u{C}l$_3$ and the zigzag
  antiferromagnetic ground state},\ }\href
  {https://doi.org/10.1103/PhysRevB.92.235119} {\bibfield  {journal} {\bibinfo
  {journal} {Phys. Rev. B}\ }\textbf {\bibinfo {volume} {92}},\ \bibinfo
  {pages} {235119} (\bibinfo {year} {2015})}\BibitemShut {NoStop}%
\bibitem [{\citenamefont {Banerjee}\ \emph {et~al.}(2017)\citenamefont
  {Banerjee}, \citenamefont {Yan}, \citenamefont {Knolle}, \citenamefont
  {Bridges}, \citenamefont {Stone}, \citenamefont {Lumsden}, \citenamefont
  {Mandrus}, \citenamefont {Tennant}, \citenamefont {Moessner},\ and\
  \citenamefont {Nagler}}]{Banerjee2017}%
  \BibitemOpen
  \bibfield  {author} {\bibinfo {author} {\bibfnamefont {A.}~\bibnamefont
  {Banerjee}}, \bibinfo {author} {\bibfnamefont {J.}~\bibnamefont {Yan}},
  \bibinfo {author} {\bibfnamefont {J.}~\bibnamefont {Knolle}}, \bibinfo
  {author} {\bibfnamefont {C.~A.}\ \bibnamefont {Bridges}}, \bibinfo {author}
  {\bibfnamefont {M.~B.}\ \bibnamefont {Stone}}, \bibinfo {author}
  {\bibfnamefont {M.~D.}\ \bibnamefont {Lumsden}}, \bibinfo {author}
  {\bibfnamefont {D.~G.}\ \bibnamefont {Mandrus}}, \bibinfo {author}
  {\bibfnamefont {D.~A.}\ \bibnamefont {Tennant}}, \bibinfo {author}
  {\bibfnamefont {R.}~\bibnamefont {Moessner}},\ and\ \bibinfo {author}
  {\bibfnamefont {S.~E.}\ \bibnamefont {Nagler}},\ }\bibfield  {title}
  {\bibinfo {title} {Neutron scattering in the proximate quantum spin liquid
  $\alpha$-{R}u{C}l$_3$},\ }\href {https://doi.org/10.1126/science.aah6015}
  {\bibfield  {journal} {\bibinfo  {journal} {Science}\ }\textbf {\bibinfo
  {volume} {356}},\ \bibinfo {pages} {1055} (\bibinfo {year}
  {2017})}\BibitemShut {NoStop}%
\bibitem [{\citenamefont {Do}\ \emph {et~al.}(2017)\citenamefont {Do},
  \citenamefont {Park}, \citenamefont {Yoshitake}, \citenamefont {Nasu},
  \citenamefont {Motome}, \citenamefont {Kwon}, \citenamefont {Adroja},
  \citenamefont {Voneshen}, \citenamefont {Kim}, \citenamefont {Jang},
  \citenamefont {Park}, \citenamefont {Choi},\ and\ \citenamefont
  {Ji}}]{Do2017}%
  \BibitemOpen
  \bibfield  {author} {\bibinfo {author} {\bibfnamefont {S.-H.}\ \bibnamefont
  {Do}}, \bibinfo {author} {\bibfnamefont {S.-Y.}\ \bibnamefont {Park}},
  \bibinfo {author} {\bibfnamefont {J.}~\bibnamefont {Yoshitake}}, \bibinfo
  {author} {\bibfnamefont {J.}~\bibnamefont {Nasu}}, \bibinfo {author}
  {\bibfnamefont {Y.}~\bibnamefont {Motome}}, \bibinfo {author} {\bibfnamefont
  {Y.}~\bibnamefont {Kwon}}, \bibinfo {author} {\bibfnamefont {D.~T.}\
  \bibnamefont {Adroja}}, \bibinfo {author} {\bibfnamefont {D.~J.}\
  \bibnamefont {Voneshen}}, \bibinfo {author} {\bibfnamefont {K.}~\bibnamefont
  {Kim}}, \bibinfo {author} {\bibfnamefont {T.-H.}\ \bibnamefont {Jang}},
  \bibinfo {author} {\bibfnamefont {J.-H.}\ \bibnamefont {Park}}, \bibinfo
  {author} {\bibfnamefont {K.-Y.}\ \bibnamefont {Choi}},\ and\ \bibinfo
  {author} {\bibfnamefont {S.}~\bibnamefont {Ji}},\ }\bibfield  {title}
  {\bibinfo {title} {{M}ajorana fermions in the {K}itaev quantum spin system
  $\alpha$-{R}u{C}l$_3$},\ }\href {https://doi.org/10.1038/nphys4264}
  {\bibfield  {journal} {\bibinfo  {journal} {Nature Physics}\ }\textbf
  {\bibinfo {volume} {13}},\ \bibinfo {pages} {1079} (\bibinfo {year}
  {2017})}\BibitemShut {NoStop}%
\bibitem [{\citenamefont {{Lin}}\ \emph {et~al.}(2021)\citenamefont {{Lin}},
  \citenamefont {{Jeong}}, \citenamefont {{Kim}}, \citenamefont {{Wang}},
  \citenamefont {{Huang}}, \citenamefont {{Masuda}}, \citenamefont {{Asai}},
  \citenamefont {{Itoh}}, \citenamefont {{G{\"u}nther}}, \citenamefont
  {{Russina}}, \citenamefont {{Lu}}, \citenamefont {{Sheng}}, \citenamefont
  {{Wang}}, \citenamefont {{Wang}}, \citenamefont {{Wang}}, \citenamefont
  {{Ren}}, \citenamefont {{Xi}}, \citenamefont {{Tong}}, \citenamefont
  {{Ling}}, \citenamefont {{Liu}}, \citenamefont {{Wu}}, \citenamefont {{Mei}},
  \citenamefont {{Qu}}, \citenamefont {{Zhou}}, \citenamefont {{Wang}},
  \citenamefont {{Park}}, \citenamefont {{Wan}},\ and\ \citenamefont
  {{Ma}}}]{Lin2021NC}%
  \BibitemOpen
  \bibfield  {author} {\bibinfo {author} {\bibfnamefont {G.}~\bibnamefont
  {{Lin}}}, \bibinfo {author} {\bibfnamefont {J.}~\bibnamefont {{Jeong}}},
  \bibinfo {author} {\bibfnamefont {C.}~\bibnamefont {{Kim}}}, \bibinfo
  {author} {\bibfnamefont {Y.}~\bibnamefont {{Wang}}}, \bibinfo {author}
  {\bibfnamefont {Q.}~\bibnamefont {{Huang}}}, \bibinfo {author} {\bibfnamefont
  {T.}~\bibnamefont {{Masuda}}}, \bibinfo {author} {\bibfnamefont
  {S.}~\bibnamefont {{Asai}}}, \bibinfo {author} {\bibfnamefont
  {S.}~\bibnamefont {{Itoh}}}, \bibinfo {author} {\bibfnamefont
  {G.}~\bibnamefont {{G{\"u}nther}}}, \bibinfo {author} {\bibfnamefont
  {M.}~\bibnamefont {{Russina}}}, \bibinfo {author} {\bibfnamefont
  {Z.}~\bibnamefont {{Lu}}}, \bibinfo {author} {\bibfnamefont {J.}~\bibnamefont
  {{Sheng}}}, \bibinfo {author} {\bibfnamefont {L.}~\bibnamefont {{Wang}}},
  \bibinfo {author} {\bibfnamefont {J.}~\bibnamefont {{Wang}}}, \bibinfo
  {author} {\bibfnamefont {G.}~\bibnamefont {{Wang}}}, \bibinfo {author}
  {\bibfnamefont {Q.}~\bibnamefont {{Ren}}}, \bibinfo {author} {\bibfnamefont
  {C.}~\bibnamefont {{Xi}}}, \bibinfo {author} {\bibfnamefont {W.}~\bibnamefont
  {{Tong}}}, \bibinfo {author} {\bibfnamefont {L.}~\bibnamefont {{Ling}}},
  \bibinfo {author} {\bibfnamefont {Z.}~\bibnamefont {{Liu}}}, \bibinfo
  {author} {\bibfnamefont {L.}~\bibnamefont {{Wu}}}, \bibinfo {author}
  {\bibfnamefont {J.}~\bibnamefont {{Mei}}}, \bibinfo {author} {\bibfnamefont
  {Z.}~\bibnamefont {{Qu}}}, \bibinfo {author} {\bibfnamefont {H.}~\bibnamefont
  {{Zhou}}}, \bibinfo {author} {\bibfnamefont {X.}~\bibnamefont {{Wang}}},
  \bibinfo {author} {\bibfnamefont {J.-G.}\ \bibnamefont {{Park}}}, \bibinfo
  {author} {\bibfnamefont {Y.}~\bibnamefont {{Wan}}},\ and\ \bibinfo {author}
  {\bibfnamefont {J.}~\bibnamefont {{Ma}}},\ }\bibfield  {title} {\bibinfo
  {title} {{Field-induced quantum spin disordered state in spin-1/2 honeycomb
  magnet Na$_{2}$Co$_{2}$TeO$_{6}$}},\ }\href
  {https://doi.org/10.1038/s41467-021-25567-7} {\bibfield  {journal} {\bibinfo
  {journal} {Nature Communications}\ }\textbf {\bibinfo {volume} {12}},\
  \bibinfo {eid} {5559} (\bibinfo {year} {2021})}\BibitemShut {NoStop}%
\bibitem [{\citenamefont {Yao}\ \emph {et~al.}(2022)\citenamefont {Yao},
  \citenamefont {Iida}, \citenamefont {Kamazawa},\ and\ \citenamefont
  {Li}}]{Yao2022PRL}%
  \BibitemOpen
  \bibfield  {author} {\bibinfo {author} {\bibfnamefont {W.}~\bibnamefont
  {Yao}}, \bibinfo {author} {\bibfnamefont {K.}~\bibnamefont {Iida}}, \bibinfo
  {author} {\bibfnamefont {K.}~\bibnamefont {Kamazawa}},\ and\ \bibinfo
  {author} {\bibfnamefont {Y.}~\bibnamefont {Li}},\ }\bibfield  {title}
  {\bibinfo {title} {Excitations in the ordered and paramagnetic states of
  honeycomb magnet {Na$_2$Co$_2$TeO$_6$}},\ }\href
  {https://doi.org/10.1103/PhysRevLett.129.147202} {\bibfield  {journal}
  {\bibinfo  {journal} {Phys. Rev. Lett.}\ }\textbf {\bibinfo {volume} {129}},\
  \bibinfo {pages} {147202} (\bibinfo {year} {2022})}\BibitemShut {NoStop}%
\bibitem [{\citenamefont {Sano}\ \emph {et~al.}(2018)\citenamefont {Sano},
  \citenamefont {Kato},\ and\ \citenamefont {Motome}}]{Sano2018}%
  \BibitemOpen
  \bibfield  {author} {\bibinfo {author} {\bibfnamefont {R.}~\bibnamefont
  {Sano}}, \bibinfo {author} {\bibfnamefont {Y.}~\bibnamefont {Kato}},\ and\
  \bibinfo {author} {\bibfnamefont {Y.}~\bibnamefont {Motome}},\ }\bibfield
  {title} {\bibinfo {title} {{K}itaev-{H}eisenberg {H}amiltonian for high-spin
  ${d}^{7}$ {M}ott insulators},\ }\href
  {https://doi.org/10.1103/PhysRevB.97.014408} {\bibfield  {journal} {\bibinfo
  {journal} {Phys. Rev. B}\ }\textbf {\bibinfo {volume} {97}},\ \bibinfo
  {pages} {014408} (\bibinfo {year} {2018})}\BibitemShut {NoStop}%
\bibitem [{\citenamefont {Songvilay}\ \emph {et~al.}(2020)\citenamefont
  {Songvilay}, \citenamefont {Robert}, \citenamefont {Petit}, \citenamefont
  {Rodriguez-Rivera}, \citenamefont {Ratcliff}, \citenamefont {Damay},
  \citenamefont {Bal\'edent}, \citenamefont {Jim\'enez-Ruiz}, \citenamefont
  {Lejay}, \citenamefont {Pachoud}, \citenamefont {Hadj-Azzem}, \citenamefont
  {Simonet},\ and\ \citenamefont {Stock}}]{Songvilay2020}%
  \BibitemOpen
  \bibfield  {author} {\bibinfo {author} {\bibfnamefont {M.}~\bibnamefont
  {Songvilay}}, \bibinfo {author} {\bibfnamefont {J.}~\bibnamefont {Robert}},
  \bibinfo {author} {\bibfnamefont {S.}~\bibnamefont {Petit}}, \bibinfo
  {author} {\bibfnamefont {J.~A.}\ \bibnamefont {Rodriguez-Rivera}}, \bibinfo
  {author} {\bibfnamefont {W.~D.}\ \bibnamefont {Ratcliff}}, \bibinfo {author}
  {\bibfnamefont {F.}~\bibnamefont {Damay}}, \bibinfo {author} {\bibfnamefont
  {V.}~\bibnamefont {Bal\'edent}}, \bibinfo {author} {\bibfnamefont
  {M.}~\bibnamefont {Jim\'enez-Ruiz}}, \bibinfo {author} {\bibfnamefont
  {P.}~\bibnamefont {Lejay}}, \bibinfo {author} {\bibfnamefont
  {E.}~\bibnamefont {Pachoud}}, \bibinfo {author} {\bibfnamefont
  {A.}~\bibnamefont {Hadj-Azzem}}, \bibinfo {author} {\bibfnamefont
  {V.}~\bibnamefont {Simonet}},\ and\ \bibinfo {author} {\bibfnamefont
  {C.}~\bibnamefont {Stock}},\ }\bibfield  {title} {\bibinfo {title} {Kitaev
  interactions in the {C}o honeycomb antiferromagnets
  {N}a$_3${C}o$_2${S}b{O}$_6$ and {N}a$_2${C}o$_2${T}e{O}$_6$},\ }\href
  {https://doi.org/10.1103/PhysRevB.102.224429} {\bibfield  {journal} {\bibinfo
   {journal} {Phys. Rev. B}\ }\textbf {\bibinfo {volume} {102}},\ \bibinfo
  {pages} {224429} (\bibinfo {year} {2020})}\BibitemShut {NoStop}%
\bibitem [{\citenamefont {Zhong}\ \emph {et~al.}(2020)\citenamefont {Zhong},
  \citenamefont {Gao}, \citenamefont {Ong},\ and\ \citenamefont
  {Cava}}]{Zhong2020}%
  \BibitemOpen
  \bibfield  {author} {\bibinfo {author} {\bibfnamefont {R.}~\bibnamefont
  {Zhong}}, \bibinfo {author} {\bibfnamefont {T.}~\bibnamefont {Gao}}, \bibinfo
  {author} {\bibfnamefont {N.~P.}\ \bibnamefont {Ong}},\ and\ \bibinfo {author}
  {\bibfnamefont {R.~J.}\ \bibnamefont {Cava}},\ }\bibfield  {title} {\bibinfo
  {title} {Weak-field induced nonmagnetic state in a {C}o-based honeycomb},\
  }\href {https://doi.org/10.1126/sciadv.aay6953} {\bibfield  {journal}
  {\bibinfo  {journal} {Science Advances}\ }\textbf {\bibinfo {volume} {6}},\
  \bibinfo {pages} {eaay6953} (\bibinfo {year} {2020})}\BibitemShut {NoStop}%
\bibitem [{\citenamefont {{Zhang}}\ \emph {et~al.}(2023)\citenamefont
  {{Zhang}}, \citenamefont {{Xu}}, \citenamefont {{Halloran}}, \citenamefont
  {{Zhong}}, \citenamefont {{Broholm}}, \citenamefont {{Cava}}, \citenamefont
  {{Drichko}},\ and\ \citenamefont {{Armitage}}}]{Zhang2023}%
  \BibitemOpen
  \bibfield  {author} {\bibinfo {author} {\bibfnamefont {X.}~\bibnamefont
  {{Zhang}}}, \bibinfo {author} {\bibfnamefont {Y.}~\bibnamefont {{Xu}}},
  \bibinfo {author} {\bibfnamefont {T.}~\bibnamefont {{Halloran}}}, \bibinfo
  {author} {\bibfnamefont {R.}~\bibnamefont {{Zhong}}}, \bibinfo {author}
  {\bibfnamefont {C.}~\bibnamefont {{Broholm}}}, \bibinfo {author}
  {\bibfnamefont {R.~J.}\ \bibnamefont {{Cava}}}, \bibinfo {author}
  {\bibfnamefont {N.}~\bibnamefont {{Drichko}}},\ and\ \bibinfo {author}
  {\bibfnamefont {N.~P.}\ \bibnamefont {{Armitage}}},\ }\bibfield  {title}
  {\bibinfo {title} {{A magnetic continuum in the cobalt-based honeycomb magnet
  {B}a{C}o$_{2}$({As}{O}$_{4}$)$_{2}$}},\ }\href
  {https://doi.org/10.1038/s41563-022-01403-1} {\bibfield  {journal} {\bibinfo
  {journal} {Nature Materials}\ }\textbf {\bibinfo {volume} {22}},\ \bibinfo
  {pages} {58} (\bibinfo {year} {2023})}\BibitemShut {NoStop}%
\bibitem [{\citenamefont {Sears}\ \emph {et~al.}(2017)\citenamefont {Sears},
  \citenamefont {Zhao}, \citenamefont {Xu}, \citenamefont {Lynn},\ and\
  \citenamefont {Kim}}]{Sears2017}%
  \BibitemOpen
  \bibfield  {author} {\bibinfo {author} {\bibfnamefont {J.~A.}\ \bibnamefont
  {Sears}}, \bibinfo {author} {\bibfnamefont {Y.}~\bibnamefont {Zhao}},
  \bibinfo {author} {\bibfnamefont {Z.}~\bibnamefont {Xu}}, \bibinfo {author}
  {\bibfnamefont {J.~W.}\ \bibnamefont {Lynn}},\ and\ \bibinfo {author}
  {\bibfnamefont {Y.-J.}\ \bibnamefont {Kim}},\ }\bibfield  {title} {\bibinfo
  {title} {Phase diagram of $\alpha$-{R}u{C}l$_3$ in an in-plane magnetic
  field},\ }\href {https://doi.org/10.1103/PhysRevB.95.180411} {\bibfield
  {journal} {\bibinfo  {journal} {Phys. Rev. B}\ }\textbf {\bibinfo {volume}
  {95}},\ \bibinfo {pages} {180411} (\bibinfo {year} {2017})}\BibitemShut
  {NoStop}%
\bibitem [{\citenamefont {Zheng}\ \emph {et~al.}(2017)\citenamefont {Zheng},
  \citenamefont {Ran}, \citenamefont {Li}, \citenamefont {Wang}, \citenamefont
  {Wang}, \citenamefont {Liu}, \citenamefont {Liu}, \citenamefont {Normand},
  \citenamefont {Wen},\ and\ \citenamefont {Yu}}]{Zheng2017}%
  \BibitemOpen
  \bibfield  {author} {\bibinfo {author} {\bibfnamefont {J.}~\bibnamefont
  {Zheng}}, \bibinfo {author} {\bibfnamefont {K.}~\bibnamefont {Ran}}, \bibinfo
  {author} {\bibfnamefont {T.}~\bibnamefont {Li}}, \bibinfo {author}
  {\bibfnamefont {J.}~\bibnamefont {Wang}}, \bibinfo {author} {\bibfnamefont
  {P.}~\bibnamefont {Wang}}, \bibinfo {author} {\bibfnamefont {B.}~\bibnamefont
  {Liu}}, \bibinfo {author} {\bibfnamefont {Z.-X.}\ \bibnamefont {Liu}},
  \bibinfo {author} {\bibfnamefont {B.}~\bibnamefont {Normand}}, \bibinfo
  {author} {\bibfnamefont {J.}~\bibnamefont {Wen}},\ and\ \bibinfo {author}
  {\bibfnamefont {W.}~\bibnamefont {Yu}},\ }\bibfield  {title} {\bibinfo
  {title} {Gapless spin excitations in the field-induced quantum spin liquid
  phase of $\alpha$-{R}u{C}l$_3$},\ }\href
  {https://doi.org/10.1103/PhysRevLett.119.227208} {\bibfield  {journal}
  {\bibinfo  {journal} {Phys. Rev. Lett.}\ }\textbf {\bibinfo {volume} {119}},\
  \bibinfo {pages} {227208} (\bibinfo {year} {2017})}\BibitemShut {NoStop}%
\bibitem [{\citenamefont {Banerjee}\ \emph {et~al.}(2018)\citenamefont
  {Banerjee}, \citenamefont {Lampen-Kelley}, \citenamefont {Knolle},
  \citenamefont {Balz}, \citenamefont {Aczel}, \citenamefont {Winn},
  \citenamefont {Liu}, \citenamefont {Pajerowski}, \citenamefont {Yan},
  \citenamefont {Bridges}, \citenamefont {Savici}, \citenamefont {Chakoumakos},
  \citenamefont {Lumsden}, \citenamefont {Tennant}, \citenamefont {Moessner},
  \citenamefont {Mandrus},\ and\ \citenamefont {Nagler}}]{Banerjee2018}%
  \BibitemOpen
  \bibfield  {author} {\bibinfo {author} {\bibfnamefont {A.}~\bibnamefont
  {Banerjee}}, \bibinfo {author} {\bibfnamefont {P.}~\bibnamefont
  {Lampen-Kelley}}, \bibinfo {author} {\bibfnamefont {J.}~\bibnamefont
  {Knolle}}, \bibinfo {author} {\bibfnamefont {C.}~\bibnamefont {Balz}},
  \bibinfo {author} {\bibfnamefont {A.~A.}\ \bibnamefont {Aczel}}, \bibinfo
  {author} {\bibfnamefont {B.}~\bibnamefont {Winn}}, \bibinfo {author}
  {\bibfnamefont {Y.}~\bibnamefont {Liu}}, \bibinfo {author} {\bibfnamefont
  {D.}~\bibnamefont {Pajerowski}}, \bibinfo {author} {\bibfnamefont
  {J.}~\bibnamefont {Yan}}, \bibinfo {author} {\bibfnamefont {C.~A.}\
  \bibnamefont {Bridges}}, \bibinfo {author} {\bibfnamefont {A.~T.}\
  \bibnamefont {Savici}}, \bibinfo {author} {\bibfnamefont {B.~C.}\
  \bibnamefont {Chakoumakos}}, \bibinfo {author} {\bibfnamefont {M.~D.}\
  \bibnamefont {Lumsden}}, \bibinfo {author} {\bibfnamefont {D.~A.}\
  \bibnamefont {Tennant}}, \bibinfo {author} {\bibfnamefont {R.}~\bibnamefont
  {Moessner}}, \bibinfo {author} {\bibfnamefont {D.~G.}\ \bibnamefont
  {Mandrus}},\ and\ \bibinfo {author} {\bibfnamefont {S.~E.}\ \bibnamefont
  {Nagler}},\ }\bibfield  {title} {\bibinfo {title} {Excitations in the
  field-induced quantum spin liquid state of $\alpha$-{R}u{C}l$_3$},\ }\href
  {https://doi.org/10.1038/s41535-018-0079-2} {\bibfield  {journal} {\bibinfo
  {journal} {npj Quantum Materials}\ }\textbf {\bibinfo {volume} {3}},\
  \bibinfo {pages} {8} (\bibinfo {year} {2018})}\BibitemShut {NoStop}%
\bibitem [{\citenamefont {Banerjee}\ \emph {et~al.}(2016)\citenamefont
  {Banerjee}, \citenamefont {Bridges}, \citenamefont {Yan}, \citenamefont
  {Aczel}, \citenamefont {Li}, \citenamefont {Stone}, \citenamefont {Granroth},
  \citenamefont {Lumsden}, \citenamefont {Yiu}, \citenamefont {Knolle},
  \citenamefont {Bhattacharjee}, \citenamefont {Kovrizhin}, \citenamefont
  {Moessner}, \citenamefont {Tennant}, \citenamefont {Mandrus},\ and\
  \citenamefont {Nagler}}]{Banerjee2016}%
  \BibitemOpen
  \bibfield  {author} {\bibinfo {author} {\bibfnamefont {A.}~\bibnamefont
  {Banerjee}}, \bibinfo {author} {\bibfnamefont {C.~A.}\ \bibnamefont
  {Bridges}}, \bibinfo {author} {\bibfnamefont {J.-Q.}\ \bibnamefont {Yan}},
  \bibinfo {author} {\bibfnamefont {A.~A.}\ \bibnamefont {Aczel}}, \bibinfo
  {author} {\bibfnamefont {L.}~\bibnamefont {Li}}, \bibinfo {author}
  {\bibfnamefont {M.~B.}\ \bibnamefont {Stone}}, \bibinfo {author}
  {\bibfnamefont {G.~E.}\ \bibnamefont {Granroth}}, \bibinfo {author}
  {\bibfnamefont {M.~D.}\ \bibnamefont {Lumsden}}, \bibinfo {author}
  {\bibfnamefont {Y.}~\bibnamefont {Yiu}}, \bibinfo {author} {\bibfnamefont
  {J.}~\bibnamefont {Knolle}}, \bibinfo {author} {\bibfnamefont
  {S.}~\bibnamefont {Bhattacharjee}}, \bibinfo {author} {\bibfnamefont {D.~L.}\
  \bibnamefont {Kovrizhin}}, \bibinfo {author} {\bibfnamefont {R.}~\bibnamefont
  {Moessner}}, \bibinfo {author} {\bibfnamefont {D.~A.}\ \bibnamefont
  {Tennant}}, \bibinfo {author} {\bibfnamefont {D.~G.}\ \bibnamefont
  {Mandrus}},\ and\ \bibinfo {author} {\bibfnamefont {S.~E.}\ \bibnamefont
  {Nagler}},\ }\bibfield  {title} {\bibinfo {title} {Proximate {K}itaev quantum
  spin liquid behaviour in a honeycomb magnet},\ }\href
  {https://doi.org/10.1038/nmat4604} {\bibfield  {journal} {\bibinfo  {journal}
  {Nature Materials}\ }\textbf {\bibinfo {volume} {15}},\ \bibinfo {pages}
  {733} (\bibinfo {year} {2016})}\BibitemShut {NoStop}%
\bibitem [{\citenamefont {Kasahara}\ \emph {et~al.}(2018)\citenamefont
  {Kasahara}, \citenamefont {Ohnishi}, \citenamefont {Mizukami}, \citenamefont
  {Tanaka}, \citenamefont {Ma}, \citenamefont {Sugii}, \citenamefont {Kurita},
  \citenamefont {Tanaka}, \citenamefont {Nasu}, \citenamefont {Motome},
  \citenamefont {Shibauchi},\ and\ \citenamefont {Matsuda}}]{Kasahara2018}%
  \BibitemOpen
  \bibfield  {author} {\bibinfo {author} {\bibfnamefont {Y.}~\bibnamefont
  {Kasahara}}, \bibinfo {author} {\bibfnamefont {T.}~\bibnamefont {Ohnishi}},
  \bibinfo {author} {\bibfnamefont {Y.}~\bibnamefont {Mizukami}}, \bibinfo
  {author} {\bibfnamefont {O.}~\bibnamefont {Tanaka}}, \bibinfo {author}
  {\bibfnamefont {S.}~\bibnamefont {Ma}}, \bibinfo {author} {\bibfnamefont
  {K.}~\bibnamefont {Sugii}}, \bibinfo {author} {\bibfnamefont
  {N.}~\bibnamefont {Kurita}}, \bibinfo {author} {\bibfnamefont
  {H.}~\bibnamefont {Tanaka}}, \bibinfo {author} {\bibfnamefont
  {J.}~\bibnamefont {Nasu}}, \bibinfo {author} {\bibfnamefont {Y.}~\bibnamefont
  {Motome}}, \bibinfo {author} {\bibfnamefont {T.}~\bibnamefont {Shibauchi}},\
  and\ \bibinfo {author} {\bibfnamefont {Y.}~\bibnamefont {Matsuda}},\
  }\bibfield  {title} {\bibinfo {title} {Majorana quantization and half-integer
  thermal quantum {H}all effect in a {K}itaev spin liquid},\ }\href
  {https://doi.org/10.1038/s41586-018-0274-0} {\bibfield  {journal} {\bibinfo
  {journal} {Nature}\ }\textbf {\bibinfo {volume} {559}},\ \bibinfo {pages}
  {227} (\bibinfo {year} {2018})}\BibitemShut {NoStop}%
\bibitem [{\citenamefont {Yokoi}\ \emph {et~al.}(2021)\citenamefont {Yokoi},
  \citenamefont {Ma}, \citenamefont {Kasahara}, \citenamefont {Kasahara},
  \citenamefont {Shibauchi}, \citenamefont {Kurita}, \citenamefont {Tanaka},
  \citenamefont {Nasu}, \citenamefont {Motome}, \citenamefont {Hickey},
  \citenamefont {Trebst},\ and\ \citenamefont {Matsuda}}]{Yokoi2021}%
  \BibitemOpen
  \bibfield  {author} {\bibinfo {author} {\bibfnamefont {T.}~\bibnamefont
  {Yokoi}}, \bibinfo {author} {\bibfnamefont {S.}~\bibnamefont {Ma}}, \bibinfo
  {author} {\bibfnamefont {Y.}~\bibnamefont {Kasahara}}, \bibinfo {author}
  {\bibfnamefont {S.}~\bibnamefont {Kasahara}}, \bibinfo {author}
  {\bibfnamefont {T.}~\bibnamefont {Shibauchi}}, \bibinfo {author}
  {\bibfnamefont {N.}~\bibnamefont {Kurita}}, \bibinfo {author} {\bibfnamefont
  {H.}~\bibnamefont {Tanaka}}, \bibinfo {author} {\bibfnamefont
  {J.}~\bibnamefont {Nasu}}, \bibinfo {author} {\bibfnamefont {Y.}~\bibnamefont
  {Motome}}, \bibinfo {author} {\bibfnamefont {C.}~\bibnamefont {Hickey}},
  \bibinfo {author} {\bibfnamefont {S.}~\bibnamefont {Trebst}},\ and\ \bibinfo
  {author} {\bibfnamefont {Y.}~\bibnamefont {Matsuda}},\ }\bibfield  {title}
  {\bibinfo {title} {Half-integer quantized anomalous thermal {H}all effect in
  the {K}itaev material candidate $\alpha$-{R}u{C}l$_3$},\ }\href
  {https://doi.org/10.1126/science.aay5551} {\bibfield  {journal} {\bibinfo
  {journal} {Science}\ }\textbf {\bibinfo {volume} {373}},\ \bibinfo {pages}
  {568} (\bibinfo {year} {2021})}\BibitemShut {NoStop}%
\bibitem [{\citenamefont {Bruin}\ \emph {et~al.}(2022)\citenamefont {Bruin},
  \citenamefont {Claus}, \citenamefont {Matsumoto}, \citenamefont {Kurita},
  \citenamefont {Tanaka},\ and\ \citenamefont {Takagi}}]{Bruin2022}%
  \BibitemOpen
  \bibfield  {author} {\bibinfo {author} {\bibfnamefont {J.~A.~N.}\
  \bibnamefont {Bruin}}, \bibinfo {author} {\bibfnamefont {R.~R.}\ \bibnamefont
  {Claus}}, \bibinfo {author} {\bibfnamefont {Y.}~\bibnamefont {Matsumoto}},
  \bibinfo {author} {\bibfnamefont {N.}~\bibnamefont {Kurita}}, \bibinfo
  {author} {\bibfnamefont {H.}~\bibnamefont {Tanaka}},\ and\ \bibinfo {author}
  {\bibfnamefont {H.}~\bibnamefont {Takagi}},\ }\bibfield  {title} {\bibinfo
  {title} {Robustness of the thermal {H}all effect close to half-quantization
  in $\alpha$-{R}u{C}l$_3$},\ }\href
  {https://doi.org/10.1038/s41567-021-01501-y} {\bibfield  {journal} {\bibinfo
  {journal} {Nature Physics}\ }\textbf {\bibinfo {volume} {18}},\ \bibinfo
  {pages} {401} (\bibinfo {year} {2022})}\BibitemShut {NoStop}%
\bibitem [{\citenamefont {Vinkler-Aviv}\ and\ \citenamefont
  {Rosch}(2018)}]{Aviv2018}%
  \BibitemOpen
  \bibfield  {author} {\bibinfo {author} {\bibfnamefont {Y.}~\bibnamefont
  {Vinkler-Aviv}}\ and\ \bibinfo {author} {\bibfnamefont {A.}~\bibnamefont
  {Rosch}},\ }\bibfield  {title} {\bibinfo {title} {Approximately quantized
  thermal {H}all effect of chiral liquids coupled to phonons},\ }\href
  {https://doi.org/10.1103/PhysRevX.8.031032} {\bibfield  {journal} {\bibinfo
  {journal} {Phys. Rev. X}\ }\textbf {\bibinfo {volume} {8}},\ \bibinfo {pages}
  {031032} (\bibinfo {year} {2018})}\BibitemShut {NoStop}%
\bibitem [{\citenamefont {Ye}\ \emph {et~al.}(2018)\citenamefont {Ye},
  \citenamefont {Hal\'asz}, \citenamefont {Savary},\ and\ \citenamefont
  {Balents}}]{Ye2018}%
  \BibitemOpen
  \bibfield  {author} {\bibinfo {author} {\bibfnamefont {M.}~\bibnamefont
  {Ye}}, \bibinfo {author} {\bibfnamefont {G.~B.}\ \bibnamefont {Hal\'asz}},
  \bibinfo {author} {\bibfnamefont {L.}~\bibnamefont {Savary}},\ and\ \bibinfo
  {author} {\bibfnamefont {L.}~\bibnamefont {Balents}},\ }\bibfield  {title}
  {\bibinfo {title} {Quantization of the thermal {H}all conductivity at small
  {H}all angles},\ }\href {https://doi.org/10.1103/PhysRevLett.121.147201}
  {\bibfield  {journal} {\bibinfo  {journal} {Phys. Rev. Lett.}\ }\textbf
  {\bibinfo {volume} {121}},\ \bibinfo {pages} {147201} (\bibinfo {year}
  {2018})}\BibitemShut {NoStop}%
\bibitem [{\citenamefont {Czajka}\ \emph {et~al.}(2021)\citenamefont {Czajka},
  \citenamefont {Gao}, \citenamefont {Hirschberger}, \citenamefont
  {Lampen-Kelley}, \citenamefont {Banerjee}, \citenamefont {Yan}, \citenamefont
  {Mandrus}, \citenamefont {Nagler},\ and\ \citenamefont {Ong}}]{Czajka2021}%
  \BibitemOpen
  \bibfield  {author} {\bibinfo {author} {\bibfnamefont {P.}~\bibnamefont
  {Czajka}}, \bibinfo {author} {\bibfnamefont {T.}~\bibnamefont {Gao}},
  \bibinfo {author} {\bibfnamefont {M.}~\bibnamefont {Hirschberger}}, \bibinfo
  {author} {\bibfnamefont {P.}~\bibnamefont {Lampen-Kelley}}, \bibinfo {author}
  {\bibfnamefont {A.}~\bibnamefont {Banerjee}}, \bibinfo {author}
  {\bibfnamefont {J.}~\bibnamefont {Yan}}, \bibinfo {author} {\bibfnamefont
  {D.~G.}\ \bibnamefont {Mandrus}}, \bibinfo {author} {\bibfnamefont {S.~E.}\
  \bibnamefont {Nagler}},\ and\ \bibinfo {author} {\bibfnamefont {N.~P.}\
  \bibnamefont {Ong}},\ }\bibfield  {title} {\bibinfo {title} {Oscillations of
  the thermal conductivity in the spin-liquid state of $\alpha$-{R}u{C}l$_3$},\
  }\href {https://doi.org/10.1038/s41567-021-01243-x} {\bibfield  {journal}
  {\bibinfo  {journal} {Nature Physics}\ }\textbf {\bibinfo {volume} {17}},\
  \bibinfo {pages} {915} (\bibinfo {year} {2021})}\BibitemShut {NoStop}%
\bibitem [{\citenamefont {Janssen}\ and\ \citenamefont
  {Vojta}(2019)}]{Janssen2019}%
  \BibitemOpen
  \bibfield  {author} {\bibinfo {author} {\bibfnamefont {L.}~\bibnamefont
  {Janssen}}\ and\ \bibinfo {author} {\bibfnamefont {M.}~\bibnamefont
  {Vojta}},\ }\bibfield  {title} {\bibinfo {title} {{H}eisenberg–{K}itaev
  physics in magnetic fields},\ }\href
  {https://doi.org/10.1088/1361-648X/ab283e} {\bibfield  {journal} {\bibinfo
  {journal} {Journal of Physics: Condensed Matter}\ }\textbf {\bibinfo {volume}
  {31}},\ \bibinfo {pages} {423002} (\bibinfo {year} {2019})}\BibitemShut
  {NoStop}%
\bibitem [{\citenamefont {Baek}\ \emph {et~al.}(2017)\citenamefont {Baek},
  \citenamefont {Do}, \citenamefont {Choi}, \citenamefont {Kwon}, \citenamefont
  {Wolter}, \citenamefont {Nishimoto}, \citenamefont {van~den Brink},\ and\
  \citenamefont {B\"uchner}}]{Baek2017}%
  \BibitemOpen
  \bibfield  {author} {\bibinfo {author} {\bibfnamefont {S.-H.}\ \bibnamefont
  {Baek}}, \bibinfo {author} {\bibfnamefont {S.-H.}\ \bibnamefont {Do}},
  \bibinfo {author} {\bibfnamefont {K.-Y.}\ \bibnamefont {Choi}}, \bibinfo
  {author} {\bibfnamefont {Y.~S.}\ \bibnamefont {Kwon}}, \bibinfo {author}
  {\bibfnamefont {A.~U.~B.}\ \bibnamefont {Wolter}}, \bibinfo {author}
  {\bibfnamefont {S.}~\bibnamefont {Nishimoto}}, \bibinfo {author}
  {\bibfnamefont {J.}~\bibnamefont {van~den Brink}},\ and\ \bibinfo {author}
  {\bibfnamefont {B.}~\bibnamefont {B\"uchner}},\ }\bibfield  {title} {\bibinfo
  {title} {Evidence for a field-induced quantum spin liquid in
  $\alpha$-{R}u{C}l$_3$},\ }\href
  {https://doi.org/10.1103/PhysRevLett.119.037201} {\bibfield  {journal}
  {\bibinfo  {journal} {Phys. Rev. Lett.}\ }\textbf {\bibinfo {volume} {119}},\
  \bibinfo {pages} {037201} (\bibinfo {year} {2017})}\BibitemShut {NoStop}%
\bibitem [{\citenamefont {Wolter}\ \emph {et~al.}(2017)\citenamefont {Wolter},
  \citenamefont {Corredor}, \citenamefont {Janssen}, \citenamefont {Nenkov},
  \citenamefont {Sch\"onecker}, \citenamefont {Do}, \citenamefont {Choi},
  \citenamefont {Albrecht}, \citenamefont {Hunger}, \citenamefont {Doert},
  \citenamefont {Vojta},\ and\ \citenamefont {B\"uchner}}]{Wolter2017}%
  \BibitemOpen
  \bibfield  {author} {\bibinfo {author} {\bibfnamefont {A.~U.~B.}\
  \bibnamefont {Wolter}}, \bibinfo {author} {\bibfnamefont {L.~T.}\
  \bibnamefont {Corredor}}, \bibinfo {author} {\bibfnamefont {L.}~\bibnamefont
  {Janssen}}, \bibinfo {author} {\bibfnamefont {K.}~\bibnamefont {Nenkov}},
  \bibinfo {author} {\bibfnamefont {S.}~\bibnamefont {Sch\"onecker}}, \bibinfo
  {author} {\bibfnamefont {S.-H.}\ \bibnamefont {Do}}, \bibinfo {author}
  {\bibfnamefont {K.-Y.}\ \bibnamefont {Choi}}, \bibinfo {author}
  {\bibfnamefont {R.}~\bibnamefont {Albrecht}}, \bibinfo {author}
  {\bibfnamefont {J.}~\bibnamefont {Hunger}}, \bibinfo {author} {\bibfnamefont
  {T.}~\bibnamefont {Doert}}, \bibinfo {author} {\bibfnamefont
  {M.}~\bibnamefont {Vojta}},\ and\ \bibinfo {author} {\bibfnamefont
  {B.}~\bibnamefont {B\"uchner}},\ }\bibfield  {title} {\bibinfo {title}
  {Field-induced quantum criticality in the {K}itaev system
  $\alpha$-{R}u{C}l$_3$},\ }\href {https://doi.org/10.1103/PhysRevB.96.041405}
  {\bibfield  {journal} {\bibinfo  {journal} {Phys. Rev. B}\ }\textbf {\bibinfo
  {volume} {96}},\ \bibinfo {pages} {041405} (\bibinfo {year}
  {2017})}\BibitemShut {NoStop}%
\bibitem [{\citenamefont {Gass}\ \emph {et~al.}(2020)\citenamefont {Gass},
  \citenamefont {C\^onsoli}, \citenamefont {Kocsis}, \citenamefont {Corredor},
  \citenamefont {Lampen-Kelley}, \citenamefont {Mandrus}, \citenamefont
  {Nagler}, \citenamefont {Janssen}, \citenamefont {Vojta}, \citenamefont
  {B\"uchner},\ and\ \citenamefont {Wolter}}]{Gass2020}%
  \BibitemOpen
  \bibfield  {author} {\bibinfo {author} {\bibfnamefont {S.}~\bibnamefont
  {Gass}}, \bibinfo {author} {\bibfnamefont {P.~M.}\ \bibnamefont {C\^onsoli}},
  \bibinfo {author} {\bibfnamefont {V.}~\bibnamefont {Kocsis}}, \bibinfo
  {author} {\bibfnamefont {L.~T.}\ \bibnamefont {Corredor}}, \bibinfo {author}
  {\bibfnamefont {P.}~\bibnamefont {Lampen-Kelley}}, \bibinfo {author}
  {\bibfnamefont {D.~G.}\ \bibnamefont {Mandrus}}, \bibinfo {author}
  {\bibfnamefont {S.~E.}\ \bibnamefont {Nagler}}, \bibinfo {author}
  {\bibfnamefont {L.}~\bibnamefont {Janssen}}, \bibinfo {author} {\bibfnamefont
  {M.}~\bibnamefont {Vojta}}, \bibinfo {author} {\bibfnamefont
  {B.}~\bibnamefont {B\"uchner}},\ and\ \bibinfo {author} {\bibfnamefont
  {A.~U.~B.}\ \bibnamefont {Wolter}},\ }\bibfield  {title} {\bibinfo {title}
  {Field-induced transitions in the {K}itaev material $\alpha$-{R}u{C}l$_3$
  probed by thermal expansion and magnetostriction},\ }\href
  {https://doi.org/10.1103/PhysRevB.101.245158} {\bibfield  {journal} {\bibinfo
   {journal} {Phys. Rev. B}\ }\textbf {\bibinfo {volume} {101}},\ \bibinfo
  {pages} {245158} (\bibinfo {year} {2020})}\BibitemShut {NoStop}%
\bibitem [{\citenamefont {Hentrich}\ \emph {et~al.}(2018)\citenamefont
  {Hentrich}, \citenamefont {Wolter}, \citenamefont {Zotos}, \citenamefont
  {Brenig}, \citenamefont {Nowak}, \citenamefont {Isaeva}, \citenamefont
  {Doert}, \citenamefont {Banerjee}, \citenamefont {Lampen-Kelley},
  \citenamefont {Mandrus}, \citenamefont {Nagler}, \citenamefont {Sears},
  \citenamefont {Kim}, \citenamefont {B\"uchner},\ and\ \citenamefont
  {Hess}}]{Hentrich2018}%
  \BibitemOpen
  \bibfield  {author} {\bibinfo {author} {\bibfnamefont {R.}~\bibnamefont
  {Hentrich}}, \bibinfo {author} {\bibfnamefont {A.~U.~B.}\ \bibnamefont
  {Wolter}}, \bibinfo {author} {\bibfnamefont {X.}~\bibnamefont {Zotos}},
  \bibinfo {author} {\bibfnamefont {W.}~\bibnamefont {Brenig}}, \bibinfo
  {author} {\bibfnamefont {D.}~\bibnamefont {Nowak}}, \bibinfo {author}
  {\bibfnamefont {A.}~\bibnamefont {Isaeva}}, \bibinfo {author} {\bibfnamefont
  {T.}~\bibnamefont {Doert}}, \bibinfo {author} {\bibfnamefont
  {A.}~\bibnamefont {Banerjee}}, \bibinfo {author} {\bibfnamefont
  {P.}~\bibnamefont {Lampen-Kelley}}, \bibinfo {author} {\bibfnamefont {D.~G.}\
  \bibnamefont {Mandrus}}, \bibinfo {author} {\bibfnamefont {S.~E.}\
  \bibnamefont {Nagler}}, \bibinfo {author} {\bibfnamefont {J.}~\bibnamefont
  {Sears}}, \bibinfo {author} {\bibfnamefont {Y.-J.}\ \bibnamefont {Kim}},
  \bibinfo {author} {\bibfnamefont {B.}~\bibnamefont {B\"uchner}},\ and\
  \bibinfo {author} {\bibfnamefont {C.}~\bibnamefont {Hess}},\ }\bibfield
  {title} {\bibinfo {title} {Unusual phonon heat transport in
  $\alpha$-{R}u{C}l$_3$: Strong spin-phonon scattering and field-induced spin
  gap},\ }\href {https://doi.org/10.1103/PhysRevLett.120.117204} {\bibfield
  {journal} {\bibinfo  {journal} {Phys. Rev. Lett.}\ }\textbf {\bibinfo
  {volume} {120}},\ \bibinfo {pages} {117204} (\bibinfo {year}
  {2018})}\BibitemShut {NoStop}%
\bibitem [{\citenamefont {Hentrich}\ \emph {et~al.}(2019)\citenamefont
  {Hentrich}, \citenamefont {Roslova}, \citenamefont {Isaeva}, \citenamefont
  {Doert}, \citenamefont {Brenig}, \citenamefont {B\"uchner},\ and\
  \citenamefont {Hess}}]{Hentrich2019}%
  \BibitemOpen
  \bibfield  {author} {\bibinfo {author} {\bibfnamefont {R.}~\bibnamefont
  {Hentrich}}, \bibinfo {author} {\bibfnamefont {M.}~\bibnamefont {Roslova}},
  \bibinfo {author} {\bibfnamefont {A.}~\bibnamefont {Isaeva}}, \bibinfo
  {author} {\bibfnamefont {T.}~\bibnamefont {Doert}}, \bibinfo {author}
  {\bibfnamefont {W.}~\bibnamefont {Brenig}}, \bibinfo {author} {\bibfnamefont
  {B.}~\bibnamefont {B\"uchner}},\ and\ \bibinfo {author} {\bibfnamefont
  {C.}~\bibnamefont {Hess}},\ }\bibfield  {title} {\bibinfo {title} {Large
  thermal {H}all effect in $\alpha$-{R}u{C}l$_3$: Evidence for heat transport
  by {K}itaev-{H}eisenberg paramagnons},\ }\href
  {https://doi.org/10.1103/PhysRevB.99.085136} {\bibfield  {journal} {\bibinfo
  {journal} {Phys. Rev. B}\ }\textbf {\bibinfo {volume} {99}},\ \bibinfo
  {pages} {085136} (\bibinfo {year} {2019})}\BibitemShut {NoStop}%
\bibitem [{\citenamefont {Hentrich}\ \emph {et~al.}(2020)\citenamefont
  {Hentrich}, \citenamefont {Hong}, \citenamefont {Gillig}, \citenamefont
  {Caglieris}, \citenamefont {\ifmmode~\check{C}\else \v{C}\fi{}ulo},
  \citenamefont {Shahrokhvand}, \citenamefont {Zeitler}, \citenamefont
  {Roslova}, \citenamefont {Isaeva}, \citenamefont {Doert}, \citenamefont
  {Janssen}, \citenamefont {Vojta}, \citenamefont {B\"uchner},\ and\
  \citenamefont {Hess}}]{Hentrich2020}%
  \BibitemOpen
  \bibfield  {author} {\bibinfo {author} {\bibfnamefont {R.}~\bibnamefont
  {Hentrich}}, \bibinfo {author} {\bibfnamefont {X.}~\bibnamefont {Hong}},
  \bibinfo {author} {\bibfnamefont {M.}~\bibnamefont {Gillig}}, \bibinfo
  {author} {\bibfnamefont {F.}~\bibnamefont {Caglieris}}, \bibinfo {author}
  {\bibfnamefont {M.}~\bibnamefont {\ifmmode~\check{C}\else \v{C}\fi{}ulo}},
  \bibinfo {author} {\bibfnamefont {M.}~\bibnamefont {Shahrokhvand}}, \bibinfo
  {author} {\bibfnamefont {U.}~\bibnamefont {Zeitler}}, \bibinfo {author}
  {\bibfnamefont {M.}~\bibnamefont {Roslova}}, \bibinfo {author} {\bibfnamefont
  {A.}~\bibnamefont {Isaeva}}, \bibinfo {author} {\bibfnamefont
  {T.}~\bibnamefont {Doert}}, \bibinfo {author} {\bibfnamefont
  {L.}~\bibnamefont {Janssen}}, \bibinfo {author} {\bibfnamefont
  {M.}~\bibnamefont {Vojta}}, \bibinfo {author} {\bibfnamefont
  {B.}~\bibnamefont {B\"uchner}},\ and\ \bibinfo {author} {\bibfnamefont
  {C.}~\bibnamefont {Hess}},\ }\bibfield  {title} {\bibinfo {title} {High-field
  thermal transport properties of the {K}itaev quantum magnet
  $\alpha$-{R}u{C}l$_3$: Evidence for low-energy excitations beyond the
  critical field},\ }\href {https://doi.org/10.1103/PhysRevB.102.235155}
  {\bibfield  {journal} {\bibinfo  {journal} {Phys. Rev. B}\ }\textbf {\bibinfo
  {volume} {102}},\ \bibinfo {pages} {235155} (\bibinfo {year}
  {2020})}\BibitemShut {NoStop}%
\bibitem [{\citenamefont {Balz}\ \emph {et~al.}(2019)\citenamefont {Balz},
  \citenamefont {Lampen-Kelley}, \citenamefont {Banerjee}, \citenamefont {Yan},
  \citenamefont {Lu}, \citenamefont {Hu}, \citenamefont {Yadav}, \citenamefont
  {Takano}, \citenamefont {Liu}, \citenamefont {Tennant}, \citenamefont
  {Lumsden}, \citenamefont {Mandrus},\ and\ \citenamefont {Nagler}}]{Balz2019}%
  \BibitemOpen
  \bibfield  {author} {\bibinfo {author} {\bibfnamefont {C.}~\bibnamefont
  {Balz}}, \bibinfo {author} {\bibfnamefont {P.}~\bibnamefont {Lampen-Kelley}},
  \bibinfo {author} {\bibfnamefont {A.}~\bibnamefont {Banerjee}}, \bibinfo
  {author} {\bibfnamefont {J.}~\bibnamefont {Yan}}, \bibinfo {author}
  {\bibfnamefont {Z.}~\bibnamefont {Lu}}, \bibinfo {author} {\bibfnamefont
  {X.}~\bibnamefont {Hu}}, \bibinfo {author} {\bibfnamefont {S.~M.}\
  \bibnamefont {Yadav}}, \bibinfo {author} {\bibfnamefont {Y.}~\bibnamefont
  {Takano}}, \bibinfo {author} {\bibfnamefont {Y.}~\bibnamefont {Liu}},
  \bibinfo {author} {\bibfnamefont {D.~A.}\ \bibnamefont {Tennant}}, \bibinfo
  {author} {\bibfnamefont {M.~D.}\ \bibnamefont {Lumsden}}, \bibinfo {author}
  {\bibfnamefont {D.}~\bibnamefont {Mandrus}},\ and\ \bibinfo {author}
  {\bibfnamefont {S.~E.}\ \bibnamefont {Nagler}},\ }\bibfield  {title}
  {\bibinfo {title} {Finite field regime for a quantum spin liquid in
  $\alpha$-{R}u{C}l$_3$},\ }\href {https://doi.org/10.1103/PhysRevB.100.060405}
  {\bibfield  {journal} {\bibinfo  {journal} {Phys. Rev. B}\ }\textbf {\bibinfo
  {volume} {100}},\ \bibinfo {pages} {060405} (\bibinfo {year}
  {2019})}\BibitemShut {NoStop}%
\bibitem [{\citenamefont {Balz}\ \emph {et~al.}(2021)\citenamefont {Balz},
  \citenamefont {Janssen}, \citenamefont {Lampen-Kelley}, \citenamefont
  {Banerjee}, \citenamefont {Liu}, \citenamefont {Yan}, \citenamefont
  {Mandrus}, \citenamefont {Vojta},\ and\ \citenamefont {Nagler}}]{Balz2021}%
  \BibitemOpen
  \bibfield  {author} {\bibinfo {author} {\bibfnamefont {C.}~\bibnamefont
  {Balz}}, \bibinfo {author} {\bibfnamefont {L.}~\bibnamefont {Janssen}},
  \bibinfo {author} {\bibfnamefont {P.}~\bibnamefont {Lampen-Kelley}}, \bibinfo
  {author} {\bibfnamefont {A.}~\bibnamefont {Banerjee}}, \bibinfo {author}
  {\bibfnamefont {Y.~H.}\ \bibnamefont {Liu}}, \bibinfo {author} {\bibfnamefont
  {J.-Q.}\ \bibnamefont {Yan}}, \bibinfo {author} {\bibfnamefont {D.~G.}\
  \bibnamefont {Mandrus}}, \bibinfo {author} {\bibfnamefont {M.}~\bibnamefont
  {Vojta}},\ and\ \bibinfo {author} {\bibfnamefont {S.~E.}\ \bibnamefont
  {Nagler}},\ }\bibfield  {title} {\bibinfo {title} {Field-induced intermediate
  ordered phase and anisotropic interlayer interactions in
  $\alpha$-{R}u{C}l$_3$},\ }\href {https://doi.org/10.1103/PhysRevB.103.174417}
  {\bibfield  {journal} {\bibinfo  {journal} {Phys. Rev. B}\ }\textbf {\bibinfo
  {volume} {103}},\ \bibinfo {pages} {174417} (\bibinfo {year}
  {2021})}\BibitemShut {NoStop}%
\bibitem [{\citenamefont {Bachus}\ \emph {et~al.}(2020)\citenamefont {Bachus},
  \citenamefont {Kaib}, \citenamefont {Tokiwa}, \citenamefont {Jesche},
  \citenamefont {Tsurkan}, \citenamefont {Loidl}, \citenamefont {Winter},
  \citenamefont {Tsirlin}, \citenamefont {Valent\'{\i}},\ and\ \citenamefont
  {Gegenwart}}]{Bachus2020}%
  \BibitemOpen
  \bibfield  {author} {\bibinfo {author} {\bibfnamefont {S.}~\bibnamefont
  {Bachus}}, \bibinfo {author} {\bibfnamefont {D.~A.~S.}\ \bibnamefont {Kaib}},
  \bibinfo {author} {\bibfnamefont {Y.}~\bibnamefont {Tokiwa}}, \bibinfo
  {author} {\bibfnamefont {A.}~\bibnamefont {Jesche}}, \bibinfo {author}
  {\bibfnamefont {V.}~\bibnamefont {Tsurkan}}, \bibinfo {author} {\bibfnamefont
  {A.}~\bibnamefont {Loidl}}, \bibinfo {author} {\bibfnamefont {S.~M.}\
  \bibnamefont {Winter}}, \bibinfo {author} {\bibfnamefont {A.~A.}\
  \bibnamefont {Tsirlin}}, \bibinfo {author} {\bibfnamefont {R.}~\bibnamefont
  {Valent\'{\i}}},\ and\ \bibinfo {author} {\bibfnamefont {P.}~\bibnamefont
  {Gegenwart}},\ }\bibfield  {title} {\bibinfo {title} {Thermodynamic
  perspective on field-induced behavior of $\alpha$-{R}u{C}l$_3$},\ }\href
  {https://doi.org/10.1103/PhysRevLett.125.097203} {\bibfield  {journal}
  {\bibinfo  {journal} {Phys. Rev. Lett.}\ }\textbf {\bibinfo {volume} {125}},\
  \bibinfo {pages} {097203} (\bibinfo {year} {2020})}\BibitemShut {NoStop}%
\bibitem [{\citenamefont {Bachus}\ \emph {et~al.}(2021)\citenamefont {Bachus},
  \citenamefont {Kaib}, \citenamefont {Jesche}, \citenamefont {Tsurkan},
  \citenamefont {Loidl}, \citenamefont {Winter}, \citenamefont {Tsirlin},
  \citenamefont {Valent\'{\i}},\ and\ \citenamefont {Gegenwart}}]{Bachus2021}%
  \BibitemOpen
  \bibfield  {author} {\bibinfo {author} {\bibfnamefont {S.}~\bibnamefont
  {Bachus}}, \bibinfo {author} {\bibfnamefont {D.~A.~S.}\ \bibnamefont {Kaib}},
  \bibinfo {author} {\bibfnamefont {A.}~\bibnamefont {Jesche}}, \bibinfo
  {author} {\bibfnamefont {V.}~\bibnamefont {Tsurkan}}, \bibinfo {author}
  {\bibfnamefont {A.}~\bibnamefont {Loidl}}, \bibinfo {author} {\bibfnamefont
  {S.~M.}\ \bibnamefont {Winter}}, \bibinfo {author} {\bibfnamefont {A.~A.}\
  \bibnamefont {Tsirlin}}, \bibinfo {author} {\bibfnamefont {R.}~\bibnamefont
  {Valent\'{\i}}},\ and\ \bibinfo {author} {\bibfnamefont {P.}~\bibnamefont
  {Gegenwart}},\ }\bibfield  {title} {\bibinfo {title} {Angle-dependent
  thermodynamics of $\alpha$-{R}u{C}l$_3$},\ }\href
  {https://doi.org/10.1103/PhysRevB.103.054440} {\bibfield  {journal} {\bibinfo
   {journal} {Phys. Rev. B}\ }\textbf {\bibinfo {volume} {103}},\ \bibinfo
  {pages} {054440} (\bibinfo {year} {2021})}\BibitemShut {NoStop}%
\bibitem [{\citenamefont {Leahy}\ \emph {et~al.}(2017)\citenamefont {Leahy},
  \citenamefont {Pocs}, \citenamefont {Siegfried}, \citenamefont {Graf},
  \citenamefont {Do}, \citenamefont {Choi}, \citenamefont {Normand},\ and\
  \citenamefont {Lee}}]{Leahy2017}%
  \BibitemOpen
  \bibfield  {author} {\bibinfo {author} {\bibfnamefont {I.~A.}\ \bibnamefont
  {Leahy}}, \bibinfo {author} {\bibfnamefont {C.~A.}\ \bibnamefont {Pocs}},
  \bibinfo {author} {\bibfnamefont {P.~E.}\ \bibnamefont {Siegfried}}, \bibinfo
  {author} {\bibfnamefont {D.}~\bibnamefont {Graf}}, \bibinfo {author}
  {\bibfnamefont {S.-H.}\ \bibnamefont {Do}}, \bibinfo {author} {\bibfnamefont
  {K.-Y.}\ \bibnamefont {Choi}}, \bibinfo {author} {\bibfnamefont
  {B.}~\bibnamefont {Normand}},\ and\ \bibinfo {author} {\bibfnamefont
  {M.}~\bibnamefont {Lee}},\ }\bibfield  {title} {\bibinfo {title} {Anomalous
  thermal conductivity and magnetic torque response in the honeycomb magnet
  $\alpha$-{R}u{C}l$_3$},\ }\href
  {https://doi.org/10.1103/PhysRevLett.118.187203} {\bibfield  {journal}
  {\bibinfo  {journal} {Phys. Rev. Lett.}\ }\textbf {\bibinfo {volume} {118}},\
  \bibinfo {pages} {187203} (\bibinfo {year} {2017})}\BibitemShut {NoStop}%
\bibitem [{\citenamefont {Jan{\v{s}}a}\ \emph {et~al.}(2018)\citenamefont
  {Jan{\v{s}}a}, \citenamefont {Zorko}, \citenamefont {Gomil{\v{s}}ek},
  \citenamefont {Pregelj}, \citenamefont {Kr{\"a}mer}, \citenamefont {Biner},
  \citenamefont {Biffin}, \citenamefont {R{\"u}egg},\ and\ \citenamefont
  {Klanj{\v{s}}ek}}]{Jansa2018}%
  \BibitemOpen
  \bibfield  {author} {\bibinfo {author} {\bibfnamefont {N.}~\bibnamefont
  {Jan{\v{s}}a}}, \bibinfo {author} {\bibfnamefont {A.}~\bibnamefont {Zorko}},
  \bibinfo {author} {\bibfnamefont {M.}~\bibnamefont {Gomil{\v{s}}ek}},
  \bibinfo {author} {\bibfnamefont {M.}~\bibnamefont {Pregelj}}, \bibinfo
  {author} {\bibfnamefont {K.~W.}\ \bibnamefont {Kr{\"a}mer}}, \bibinfo
  {author} {\bibfnamefont {D.}~\bibnamefont {Biner}}, \bibinfo {author}
  {\bibfnamefont {A.}~\bibnamefont {Biffin}}, \bibinfo {author} {\bibfnamefont
  {C.}~\bibnamefont {R{\"u}egg}},\ and\ \bibinfo {author} {\bibfnamefont
  {M.}~\bibnamefont {Klanj{\v{s}}ek}},\ }\bibfield  {title} {\bibinfo {title}
  {Observation of two types of fractional excitation in the {K}itaev honeycomb
  magnet},\ }\href {https://doi.org/10.1038/s41567-018-0129-5} {\bibfield
  {journal} {\bibinfo  {journal} {Nature Physics}\ }\textbf {\bibinfo {volume}
  {14}},\ \bibinfo {pages} {786} (\bibinfo {year} {2018})}\BibitemShut
  {NoStop}%
\bibitem [{\citenamefont {Winter}\ \emph {et~al.}(2018)\citenamefont {Winter},
  \citenamefont {Riedl}, \citenamefont {Kaib}, \citenamefont {Coldea},\ and\
  \citenamefont {Valent\'{\i}}}]{Winter2018}%
  \BibitemOpen
  \bibfield  {author} {\bibinfo {author} {\bibfnamefont {S.~M.}\ \bibnamefont
  {Winter}}, \bibinfo {author} {\bibfnamefont {K.}~\bibnamefont {Riedl}},
  \bibinfo {author} {\bibfnamefont {D.}~\bibnamefont {Kaib}}, \bibinfo {author}
  {\bibfnamefont {R.}~\bibnamefont {Coldea}},\ and\ \bibinfo {author}
  {\bibfnamefont {R.}~\bibnamefont {Valent\'{\i}}},\ }\bibfield  {title}
  {\bibinfo {title} {Probing $\alpha$-{R}u{C}l$_3$ beyond magnetic order:
  Effects of temperature and magnetic field},\ }\href
  {https://doi.org/10.1103/PhysRevLett.120.077203} {\bibfield  {journal}
  {\bibinfo  {journal} {Phys. Rev. Lett.}\ }\textbf {\bibinfo {volume} {120}},\
  \bibinfo {pages} {077203} (\bibinfo {year} {2018})}\BibitemShut {NoStop}%
\bibitem [{\citenamefont {Kaib}\ \emph {et~al.}(2019)\citenamefont {Kaib},
  \citenamefont {Winter},\ and\ \citenamefont {Valent\'{\i}}}]{Kaib2019}%
  \BibitemOpen
  \bibfield  {author} {\bibinfo {author} {\bibfnamefont {D.~A.~S.}\
  \bibnamefont {Kaib}}, \bibinfo {author} {\bibfnamefont {S.~M.}\ \bibnamefont
  {Winter}},\ and\ \bibinfo {author} {\bibfnamefont {R.}~\bibnamefont
  {Valent\'{\i}}},\ }\bibfield  {title} {\bibinfo {title} {Kitaev honeycomb
  models in magnetic fields: Dynamical response and dual models},\ }\href
  {https://doi.org/10.1103/PhysRevB.100.144445} {\bibfield  {journal} {\bibinfo
   {journal} {Phys. Rev. B}\ }\textbf {\bibinfo {volume} {100}},\ \bibinfo
  {pages} {144445} (\bibinfo {year} {2019})}\BibitemShut {NoStop}%
\bibitem [{\citenamefont {Chern}\ \emph {et~al.}(2021)\citenamefont {Chern},
  \citenamefont {Zhang},\ and\ \citenamefont {Kim}}]{Chern2021}%
  \BibitemOpen
  \bibfield  {author} {\bibinfo {author} {\bibfnamefont {L.~E.}\ \bibnamefont
  {Chern}}, \bibinfo {author} {\bibfnamefont {E.~Z.}\ \bibnamefont {Zhang}},\
  and\ \bibinfo {author} {\bibfnamefont {Y.~B.}\ \bibnamefont {Kim}},\
  }\bibfield  {title} {\bibinfo {title} {Sign structure of thermal {H}all
  conductivity and topological magnons for in-plane field polarized {K}itaev
  magnets},\ }\href {https://doi.org/10.1103/PhysRevLett.126.147201} {\bibfield
   {journal} {\bibinfo  {journal} {Phys. Rev. Lett.}\ }\textbf {\bibinfo
  {volume} {126}},\ \bibinfo {pages} {147201} (\bibinfo {year}
  {2021})}\BibitemShut {NoStop}%
\bibitem [{\citenamefont {Winter}\ \emph {et~al.}(2017)\citenamefont {Winter},
  \citenamefont {Riedl}, \citenamefont {Maksimov}, \citenamefont {Chernyshev},
  \citenamefont {Honecker},\ and\ \citenamefont {Valent{\'i}}}]{Winter2017nc}%
  \BibitemOpen
  \bibfield  {author} {\bibinfo {author} {\bibfnamefont {S.~M.}\ \bibnamefont
  {Winter}}, \bibinfo {author} {\bibfnamefont {K.}~\bibnamefont {Riedl}},
  \bibinfo {author} {\bibfnamefont {P.~A.}\ \bibnamefont {Maksimov}}, \bibinfo
  {author} {\bibfnamefont {A.~L.}\ \bibnamefont {Chernyshev}}, \bibinfo
  {author} {\bibfnamefont {A.}~\bibnamefont {Honecker}},\ and\ \bibinfo
  {author} {\bibfnamefont {R.}~\bibnamefont {Valent{\'i}}},\ }\bibfield
  {title} {\bibinfo {title} {Breakdown of magnons in a strongly spin-orbital
  coupled magnet},\ }\href {https://doi.org/10.1038/s41467-017-01177-0}
  {\bibfield  {journal} {\bibinfo  {journal} {Nature Communications}\ }\textbf
  {\bibinfo {volume} {8}},\ \bibinfo {pages} {1152} (\bibinfo {year}
  {2017})}\BibitemShut {NoStop}%
\bibitem [{\citenamefont {Cookmeyer}\ and\ \citenamefont
  {Moore}(2018)}]{Cookmeyer2018}%
  \BibitemOpen
  \bibfield  {author} {\bibinfo {author} {\bibfnamefont {T.}~\bibnamefont
  {Cookmeyer}}\ and\ \bibinfo {author} {\bibfnamefont {J.~E.}\ \bibnamefont
  {Moore}},\ }\bibfield  {title} {\bibinfo {title} {Spin-wave analysis of the
  low-temperature thermal {H}all effect in the candidate {K}itaev spin liquid
  $\alpha$-{R}u{C}l$_3$},\ }\href {https://doi.org/10.1103/PhysRevB.98.060412}
  {\bibfield  {journal} {\bibinfo  {journal} {Phys. Rev. B}\ }\textbf {\bibinfo
  {volume} {98}},\ \bibinfo {pages} {060412} (\bibinfo {year}
  {2018})}\BibitemShut {NoStop}%
\bibitem [{\citenamefont {Kim}\ and\ \citenamefont {Kee}(2016)}]{Kim2016}%
  \BibitemOpen
  \bibfield  {author} {\bibinfo {author} {\bibfnamefont {H.-S.}\ \bibnamefont
  {Kim}}\ and\ \bibinfo {author} {\bibfnamefont {H.-Y.}\ \bibnamefont {Kee}},\
  }\bibfield  {title} {\bibinfo {title} {Crystal structure and magnetism in
  $\alpha$-{R}u{C}l$_3$: An ab initio study},\ }\href
  {https://doi.org/10.1103/PhysRevB.93.155143} {\bibfield  {journal} {\bibinfo
  {journal} {Phys. Rev. B}\ }\textbf {\bibinfo {volume} {93}},\ \bibinfo
  {pages} {155143} (\bibinfo {year} {2016})}\BibitemShut {NoStop}%
\bibitem [{\citenamefont {Suzuki}\ and\ \citenamefont
  {Suga}(2018)}]{Suzuki2019}%
  \BibitemOpen
  \bibfield  {author} {\bibinfo {author} {\bibfnamefont {T.}~\bibnamefont
  {Suzuki}}\ and\ \bibinfo {author} {\bibfnamefont {S.-i.}\ \bibnamefont
  {Suga}},\ }\bibfield  {title} {\bibinfo {title} {Effective model with strong
  {K}itaev interactions for $\alpha$-{R}u{C}l$_3$},\ }\href
  {https://doi.org/10.1103/PhysRevB.97.134424} {\bibfield  {journal} {\bibinfo
  {journal} {Phys. Rev. B}\ }\textbf {\bibinfo {volume} {97}},\ \bibinfo
  {pages} {134424} (\bibinfo {year} {2018})}\BibitemShut {NoStop}%
\bibitem [{\citenamefont {Ran}\ \emph {et~al.}(2017)\citenamefont {Ran},
  \citenamefont {Wang}, \citenamefont {Wang}, \citenamefont {Dong},
  \citenamefont {Ren}, \citenamefont {Bao}, \citenamefont {Li}, \citenamefont
  {Ma}, \citenamefont {Gan}, \citenamefont {Zhang}, \citenamefont {Park},
  \citenamefont {Deng}, \citenamefont {Danilkin}, \citenamefont {Yu},
  \citenamefont {Li},\ and\ \citenamefont {Wen}}]{Ran2017}%
  \BibitemOpen
  \bibfield  {author} {\bibinfo {author} {\bibfnamefont {K.}~\bibnamefont
  {Ran}}, \bibinfo {author} {\bibfnamefont {J.}~\bibnamefont {Wang}}, \bibinfo
  {author} {\bibfnamefont {W.}~\bibnamefont {Wang}}, \bibinfo {author}
  {\bibfnamefont {Z.-Y.}\ \bibnamefont {Dong}}, \bibinfo {author}
  {\bibfnamefont {X.}~\bibnamefont {Ren}}, \bibinfo {author} {\bibfnamefont
  {S.}~\bibnamefont {Bao}}, \bibinfo {author} {\bibfnamefont {S.}~\bibnamefont
  {Li}}, \bibinfo {author} {\bibfnamefont {Z.}~\bibnamefont {Ma}}, \bibinfo
  {author} {\bibfnamefont {Y.}~\bibnamefont {Gan}}, \bibinfo {author}
  {\bibfnamefont {Y.}~\bibnamefont {Zhang}}, \bibinfo {author} {\bibfnamefont
  {J.~T.}\ \bibnamefont {Park}}, \bibinfo {author} {\bibfnamefont
  {G.}~\bibnamefont {Deng}}, \bibinfo {author} {\bibfnamefont {S.}~\bibnamefont
  {Danilkin}}, \bibinfo {author} {\bibfnamefont {S.-L.}\ \bibnamefont {Yu}},
  \bibinfo {author} {\bibfnamefont {J.-X.}\ \bibnamefont {Li}},\ and\ \bibinfo
  {author} {\bibfnamefont {J.}~\bibnamefont {Wen}},\ }\bibfield  {title}
  {\bibinfo {title} {Spin-wave excitations evidencing the {K}itaev interaction
  in single crystalline $\alpha$-{R}u{C}l$_3$},\ }\href
  {https://doi.org/10.1103/PhysRevLett.118.107203} {\bibfield  {journal}
  {\bibinfo  {journal} {Phys. Rev. Lett.}\ }\textbf {\bibinfo {volume} {118}},\
  \bibinfo {pages} {107203} (\bibinfo {year} {2017})}\BibitemShut {NoStop}%
\bibitem [{\citenamefont {Ozel}\ \emph {et~al.}(2019)\citenamefont {Ozel},
  \citenamefont {Belvin}, \citenamefont {Baldini}, \citenamefont {Kimchi},
  \citenamefont {Do}, \citenamefont {Choi},\ and\ \citenamefont
  {Gedik}}]{Ozel2019}%
  \BibitemOpen
  \bibfield  {author} {\bibinfo {author} {\bibfnamefont {I.~O.}\ \bibnamefont
  {Ozel}}, \bibinfo {author} {\bibfnamefont {C.~A.}\ \bibnamefont {Belvin}},
  \bibinfo {author} {\bibfnamefont {E.}~\bibnamefont {Baldini}}, \bibinfo
  {author} {\bibfnamefont {I.}~\bibnamefont {Kimchi}}, \bibinfo {author}
  {\bibfnamefont {S.}~\bibnamefont {Do}}, \bibinfo {author} {\bibfnamefont
  {K.-Y.}\ \bibnamefont {Choi}},\ and\ \bibinfo {author} {\bibfnamefont
  {N.}~\bibnamefont {Gedik}},\ }\bibfield  {title} {\bibinfo {title} {Magnetic
  field-dependent low-energy magnon dynamics in $\alpha$-{R}u{C}l$_3$},\ }\href
  {https://doi.org/10.1103/PhysRevB.100.085108} {\bibfield  {journal} {\bibinfo
   {journal} {Phys. Rev. B}\ }\textbf {\bibinfo {volume} {100}},\ \bibinfo
  {pages} {085108} (\bibinfo {year} {2019})}\BibitemShut {NoStop}%
\bibitem [{\citenamefont {Kim}\ \emph {et~al.}(2015)\citenamefont {Kim},
  \citenamefont {V.}, \citenamefont {Catuneanu},\ and\ \citenamefont
  {Kee}}]{Kim2015}%
  \BibitemOpen
  \bibfield  {author} {\bibinfo {author} {\bibfnamefont {H.-S.}\ \bibnamefont
  {Kim}}, \bibinfo {author} {\bibfnamefont {V.~S.}\ \bibnamefont {V.}},
  \bibinfo {author} {\bibfnamefont {A.}~\bibnamefont {Catuneanu}},\ and\
  \bibinfo {author} {\bibfnamefont {H.-Y.}\ \bibnamefont {Kee}},\ }\bibfield
  {title} {\bibinfo {title} {Kitaev magnetism in honeycomb {R}u{C}l$_3$ with
  intermediate spin-orbit coupling},\ }\href
  {https://doi.org/10.1103/PhysRevB.91.241110} {\bibfield  {journal} {\bibinfo
  {journal} {Phys. Rev. B}\ }\textbf {\bibinfo {volume} {91}},\ \bibinfo
  {pages} {241110} (\bibinfo {year} {2015})}\BibitemShut {NoStop}%
\bibitem [{\citenamefont {Zhu}\ \emph {et~al.}(2018)\citenamefont {Zhu},
  \citenamefont {Kimchi}, \citenamefont {Sheng},\ and\ \citenamefont
  {Fu}}]{Zhu2018}%
  \BibitemOpen
  \bibfield  {author} {\bibinfo {author} {\bibfnamefont {Z.}~\bibnamefont
  {Zhu}}, \bibinfo {author} {\bibfnamefont {I.}~\bibnamefont {Kimchi}},
  \bibinfo {author} {\bibfnamefont {D.~N.}\ \bibnamefont {Sheng}},\ and\
  \bibinfo {author} {\bibfnamefont {L.}~\bibnamefont {Fu}},\ }\bibfield
  {title} {\bibinfo {title} {Robust non-abelian spin liquid and a possible
  intermediate phase in the antiferromagnetic {K}itaev model with magnetic
  field},\ }\href {https://doi.org/10.1103/PhysRevB.97.241110} {\bibfield
  {journal} {\bibinfo  {journal} {Phys. Rev. B}\ }\textbf {\bibinfo {volume}
  {97}},\ \bibinfo {pages} {241110} (\bibinfo {year} {2018})}\BibitemShut
  {NoStop}%
\bibitem [{\citenamefont {Gohlke}\ \emph
  {et~al.}(2018{\natexlab{a}})\citenamefont {Gohlke}, \citenamefont
  {Moessner},\ and\ \citenamefont {Pollmann}}]{Gohlke2018dynamic}%
  \BibitemOpen
  \bibfield  {author} {\bibinfo {author} {\bibfnamefont {M.}~\bibnamefont
  {Gohlke}}, \bibinfo {author} {\bibfnamefont {R.}~\bibnamefont {Moessner}},\
  and\ \bibinfo {author} {\bibfnamefont {F.}~\bibnamefont {Pollmann}},\
  }\bibfield  {title} {\bibinfo {title} {Dynamical and topological properties
  of the {K}itaev model in a [111] magnetic field},\ }\href
  {https://doi.org/10.1103/PhysRevB.98.014418} {\bibfield  {journal} {\bibinfo
  {journal} {Phys. Rev. B}\ }\textbf {\bibinfo {volume} {98}},\ \bibinfo
  {pages} {014418} (\bibinfo {year} {2018}{\natexlab{a}})}\BibitemShut
  {NoStop}%
\bibitem [{\citenamefont {{Jiang}}\ \emph {et~al.}(2018)\citenamefont
  {{Jiang}}, \citenamefont {{Wang}}, \citenamefont {{Huang}},\ and\
  \citenamefont {{Lu}}}]{Jiang2018}%
  \BibitemOpen
  \bibfield  {author} {\bibinfo {author} {\bibfnamefont {H.-C.}\ \bibnamefont
  {{Jiang}}}, \bibinfo {author} {\bibfnamefont {C.-Y.}\ \bibnamefont {{Wang}}},
  \bibinfo {author} {\bibfnamefont {B.}~\bibnamefont {{Huang}}},\ and\ \bibinfo
  {author} {\bibfnamefont {Y.-M.}\ \bibnamefont {{Lu}}},\ }\bibfield  {title}
  {\bibinfo {title} {Field induced quantum spin liquid with spinon {Fermi}
  surfaces in the {Kitaev} model},\ }\href {https://arxiv.org/abs/1809.08247}
  {\bibfield  {journal} {\bibinfo  {journal} {arXiv:1809.08247}\ } (\bibinfo
  {year} {2018})}\BibitemShut {NoStop}%
\bibitem [{\citenamefont {Patel}\ and\ \citenamefont
  {Trivedi}(2019)}]{Patel2019}%
  \BibitemOpen
  \bibfield  {author} {\bibinfo {author} {\bibfnamefont {N.~D.}\ \bibnamefont
  {Patel}}\ and\ \bibinfo {author} {\bibfnamefont {N.}~\bibnamefont
  {Trivedi}},\ }\bibfield  {title} {\bibinfo {title} {Magnetic field-induced
  intermediate quantum spin liquid with a spinon {F}ermi surface},\ }\href
  {https://doi.org/10.1073/pnas.1821406116} {\bibfield  {journal} {\bibinfo
  {journal} {Proceedings of the National Academy of Sciences}\ }\textbf
  {\bibinfo {volume} {116}},\ \bibinfo {pages} {12199} (\bibinfo {year}
  {2019})}\BibitemShut {NoStop}%
\bibitem [{\citenamefont {Hickey}\ and\ \citenamefont
  {Trebst}(2019)}]{Hickey2019}%
  \BibitemOpen
  \bibfield  {author} {\bibinfo {author} {\bibfnamefont {C.}~\bibnamefont
  {Hickey}}\ and\ \bibinfo {author} {\bibfnamefont {S.}~\bibnamefont
  {Trebst}},\ }\bibfield  {title} {\bibinfo {title} {Emergence of a
  field-driven {U}(1) spin liquid in the {K}itaev honeycomb model},\ }\href
  {https://doi.org/10.1038/s41467-019-08459-9} {\bibfield  {journal} {\bibinfo
  {journal} {Nature Communications}\ }\textbf {\bibinfo {volume} {10}},\
  \bibinfo {pages} {530} (\bibinfo {year} {2019})}\BibitemShut {NoStop}%
\bibitem [{\citenamefont {Ronquillo}\ \emph {et~al.}(2019)\citenamefont
  {Ronquillo}, \citenamefont {Vengal},\ and\ \citenamefont
  {Trivedi}}]{Ronquillo2019}%
  \BibitemOpen
  \bibfield  {author} {\bibinfo {author} {\bibfnamefont {D.~C.}\ \bibnamefont
  {Ronquillo}}, \bibinfo {author} {\bibfnamefont {A.}~\bibnamefont {Vengal}},\
  and\ \bibinfo {author} {\bibfnamefont {N.}~\bibnamefont {Trivedi}},\
  }\bibfield  {title} {\bibinfo {title} {Signatures of magnetic-field-driven
  quantum phase transitions in the entanglement entropy and spin dynamics of
  the {K}itaev honeycomb model},\ }\href
  {https://doi.org/10.1103/PhysRevB.99.140413} {\bibfield  {journal} {\bibinfo
  {journal} {Phys. Rev. B}\ }\textbf {\bibinfo {volume} {99}},\ \bibinfo
  {pages} {140413} (\bibinfo {year} {2019})}\BibitemShut {NoStop}%
\bibitem [{\citenamefont {Zou}\ and\ \citenamefont {He}(2020)}]{Zou2020}%
  \BibitemOpen
  \bibfield  {author} {\bibinfo {author} {\bibfnamefont {L.}~\bibnamefont
  {Zou}}\ and\ \bibinfo {author} {\bibfnamefont {Y.-C.}\ \bibnamefont {He}},\
  }\bibfield  {title} {\bibinfo {title} {Field-induced
  {QCD}$_3$-{C}hern-{S}imons quantum criticalities in {K}itaev materials},\
  }\href {https://doi.org/10.1103/PhysRevResearch.2.013072} {\bibfield
  {journal} {\bibinfo  {journal} {Phys. Rev. Res.}\ }\textbf {\bibinfo {volume}
  {2}},\ \bibinfo {pages} {013072} (\bibinfo {year} {2020})}\BibitemShut
  {NoStop}%
\bibitem [{\citenamefont {Gohlke}\ \emph
  {et~al.}(2018{\natexlab{b}})\citenamefont {Gohlke}, \citenamefont {Wachtel},
  \citenamefont {Yamaji}, \citenamefont {Pollmann},\ and\ \citenamefont
  {Kim}}]{Gohlke2018}%
  \BibitemOpen
  \bibfield  {author} {\bibinfo {author} {\bibfnamefont {M.}~\bibnamefont
  {Gohlke}}, \bibinfo {author} {\bibfnamefont {G.}~\bibnamefont {Wachtel}},
  \bibinfo {author} {\bibfnamefont {Y.}~\bibnamefont {Yamaji}}, \bibinfo
  {author} {\bibfnamefont {F.}~\bibnamefont {Pollmann}},\ and\ \bibinfo
  {author} {\bibfnamefont {Y.~B.}\ \bibnamefont {Kim}},\ }\bibfield  {title}
  {\bibinfo {title} {Quantum spin liquid signatures in {K}itaev-like frustrated
  magnets},\ }\href {https://doi.org/10.1103/PhysRevB.97.075126} {\bibfield
  {journal} {\bibinfo  {journal} {Phys. Rev. B}\ }\textbf {\bibinfo {volume}
  {97}},\ \bibinfo {pages} {075126} (\bibinfo {year}
  {2018}{\natexlab{b}})}\BibitemShut {NoStop}%
\bibitem [{\citenamefont {Liu}\ and\ \citenamefont {Normand}(2018)}]{Liu2018}%
  \BibitemOpen
  \bibfield  {author} {\bibinfo {author} {\bibfnamefont {Z.-X.}\ \bibnamefont
  {Liu}}\ and\ \bibinfo {author} {\bibfnamefont {B.}~\bibnamefont {Normand}},\
  }\bibfield  {title} {\bibinfo {title} {Dirac and chiral quantum spin liquids
  on the honeycomb lattice in a magnetic field},\ }\href
  {https://doi.org/10.1103/PhysRevLett.120.187201} {\bibfield  {journal}
  {\bibinfo  {journal} {Phys. Rev. Lett.}\ }\textbf {\bibinfo {volume} {120}},\
  \bibinfo {pages} {187201} (\bibinfo {year} {2018})}\BibitemShut {NoStop}%
\bibitem [{\citenamefont {Yamada}\ and\ \citenamefont
  {Fujimoto}(2021)}]{Yamada2021}%
  \BibitemOpen
  \bibfield  {author} {\bibinfo {author} {\bibfnamefont {M.~G.}\ \bibnamefont
  {Yamada}}\ and\ \bibinfo {author} {\bibfnamefont {S.}~\bibnamefont
  {Fujimoto}},\ }\bibfield  {title} {\bibinfo {title} {Quantum liquid crystals
  in the finite-field {K}-{$\Gamma$} model for $\alpha$-{R}u{C}l$_3$},\ }\href
  {https://arxiv.org/abs/2107.03045} {\bibfield  {journal} {\bibinfo  {journal}
  {arXiv:2107.03045}\ } (\bibinfo {year} {2021})}\BibitemShut {NoStop}%
\bibitem [{\citenamefont {Wang}\ \emph {et~al.}(2020)\citenamefont {Wang},
  \citenamefont {Zhao}, \citenamefont {Wang},\ and\ \citenamefont
  {Liu}}]{Wang2020}%
  \BibitemOpen
  \bibfield  {author} {\bibinfo {author} {\bibfnamefont {J.}~\bibnamefont
  {Wang}}, \bibinfo {author} {\bibfnamefont {Q.}~\bibnamefont {Zhao}}, \bibinfo
  {author} {\bibfnamefont {X.}~\bibnamefont {Wang}},\ and\ \bibinfo {author}
  {\bibfnamefont {Z.-X.}\ \bibnamefont {Liu}},\ }\bibfield  {title} {\bibinfo
  {title} {Multinode quantum spin liquids on the honeycomb lattice},\ }\href
  {https://doi.org/10.1103/PhysRevB.102.144427} {\bibfield  {journal} {\bibinfo
   {journal} {Phys. Rev. B}\ }\textbf {\bibinfo {volume} {102}},\ \bibinfo
  {pages} {144427} (\bibinfo {year} {2020})}\BibitemShut {NoStop}%
\bibitem [{\citenamefont {Gordon}\ \emph {et~al.}(2019)\citenamefont {Gordon},
  \citenamefont {Catuneanu}, \citenamefont {S{\o}rensen},\ and\ \citenamefont
  {Kee}}]{Gordon2019}%
  \BibitemOpen
  \bibfield  {author} {\bibinfo {author} {\bibfnamefont {J.~S.}\ \bibnamefont
  {Gordon}}, \bibinfo {author} {\bibfnamefont {A.}~\bibnamefont {Catuneanu}},
  \bibinfo {author} {\bibfnamefont {E.~S.}\ \bibnamefont {S{\o}rensen}},\ and\
  \bibinfo {author} {\bibfnamefont {H.-Y.}\ \bibnamefont {Kee}},\ }\bibfield
  {title} {\bibinfo {title} {Theory of the field-revealed {K}itaev spin
  liquid},\ }\href {https://doi.org/10.1038/s41467-019-10405-8} {\bibfield
  {journal} {\bibinfo  {journal} {Nature Communications}\ }\textbf {\bibinfo
  {volume} {10}},\ \bibinfo {pages} {2470} (\bibinfo {year}
  {2019})}\BibitemShut {NoStop}%
\bibitem [{\citenamefont {Lee}\ \emph {et~al.}(2020)\citenamefont {Lee},
  \citenamefont {Kaneko}, \citenamefont {Chern}, \citenamefont {Okubo},
  \citenamefont {Yamaji}, \citenamefont {Kawashima},\ and\ \citenamefont
  {Kim}}]{Lee2020}%
  \BibitemOpen
  \bibfield  {author} {\bibinfo {author} {\bibfnamefont {H.-Y.}\ \bibnamefont
  {Lee}}, \bibinfo {author} {\bibfnamefont {R.}~\bibnamefont {Kaneko}},
  \bibinfo {author} {\bibfnamefont {L.~E.}\ \bibnamefont {Chern}}, \bibinfo
  {author} {\bibfnamefont {T.}~\bibnamefont {Okubo}}, \bibinfo {author}
  {\bibfnamefont {Y.}~\bibnamefont {Yamaji}}, \bibinfo {author} {\bibfnamefont
  {N.}~\bibnamefont {Kawashima}},\ and\ \bibinfo {author} {\bibfnamefont
  {Y.~B.}\ \bibnamefont {Kim}},\ }\bibfield  {title} {\bibinfo {title}
  {Magnetic field induced quantum phases in a tensor network study of {K}itaev
  magnets},\ }\href {https://doi.org/10.1038/s41467-020-15320-x} {\bibfield
  {journal} {\bibinfo  {journal} {Nature Communications}\ }\textbf {\bibinfo
  {volume} {11}},\ \bibinfo {pages} {1639} (\bibinfo {year}
  {2020})}\BibitemShut {NoStop}%
\bibitem [{\citenamefont {Gohlke}\ \emph {et~al.}(2020)\citenamefont {Gohlke},
  \citenamefont {Chern}, \citenamefont {Kee},\ and\ \citenamefont
  {Kim}}]{Gohlke2020}%
  \BibitemOpen
  \bibfield  {author} {\bibinfo {author} {\bibfnamefont {M.}~\bibnamefont
  {Gohlke}}, \bibinfo {author} {\bibfnamefont {L.~E.}\ \bibnamefont {Chern}},
  \bibinfo {author} {\bibfnamefont {H.-Y.}\ \bibnamefont {Kee}},\ and\ \bibinfo
  {author} {\bibfnamefont {Y.~B.}\ \bibnamefont {Kim}},\ }\bibfield  {title}
  {\bibinfo {title} {Emergence of nematic paramagnet via quantum
  order-by-disorder and pseudo-{G}oldstone modes in {K}itaev magnets},\ }\href
  {https://doi.org/10.1103/PhysRevResearch.2.043023} {\bibfield  {journal}
  {\bibinfo  {journal} {Phys. Rev. Research}\ }\textbf {\bibinfo {volume}
  {2}},\ \bibinfo {pages} {043023} (\bibinfo {year} {2020})}\BibitemShut
  {NoStop}%
\bibitem [{\citenamefont {Li}\ \emph {et~al.}(2021)\citenamefont {Li},
  \citenamefont {Zhang}, \citenamefont {Wang}, \citenamefont {Wu},
  \citenamefont {Gao}, \citenamefont {Qu}, \citenamefont {Liu}, \citenamefont
  {Gong},\ and\ \citenamefont {Li}}]{Li2021}%
  \BibitemOpen
  \bibfield  {author} {\bibinfo {author} {\bibfnamefont {H.}~\bibnamefont
  {Li}}, \bibinfo {author} {\bibfnamefont {H.-K.}\ \bibnamefont {Zhang}},
  \bibinfo {author} {\bibfnamefont {J.}~\bibnamefont {Wang}}, \bibinfo {author}
  {\bibfnamefont {H.-Q.}\ \bibnamefont {Wu}}, \bibinfo {author} {\bibfnamefont
  {Y.}~\bibnamefont {Gao}}, \bibinfo {author} {\bibfnamefont {D.-W.}\
  \bibnamefont {Qu}}, \bibinfo {author} {\bibfnamefont {Z.-X.}\ \bibnamefont
  {Liu}}, \bibinfo {author} {\bibfnamefont {S.-S.}\ \bibnamefont {Gong}},\ and\
  \bibinfo {author} {\bibfnamefont {W.}~\bibnamefont {Li}},\ }\bibfield
  {title} {\bibinfo {title} {Identification of magnetic interactions and
  high-field quantum spin liquid in $\alpha$-{R}u{C}l$_3$},\ }\href
  {https://doi.org/10.1038/s41467-021-24257-8} {\bibfield  {journal} {\bibinfo
  {journal} {Nature Communications}\ }\textbf {\bibinfo {volume} {12}},\
  \bibinfo {pages} {4007} (\bibinfo {year} {2021})}\BibitemShut {NoStop}%
\bibitem [{\citenamefont {{Zhou}}\ \emph {et~al.}()\citenamefont {{Zhou}},
  \citenamefont {{Li}}, \citenamefont {{Matsuda}}, \citenamefont {{Matsuo}},
  \citenamefont {{Li}}, \citenamefont {{Kurita}}, \citenamefont {{Kindo}},\
  and\ \citenamefont {{Tanaka}}}]{Zhou2022arXiv}%
  \BibitemOpen
  \bibfield  {author} {\bibinfo {author} {\bibfnamefont {X.-G.}\ \bibnamefont
  {{Zhou}}}, \bibinfo {author} {\bibfnamefont {H.}~\bibnamefont {{Li}}},
  \bibinfo {author} {\bibfnamefont {Y.~H.}\ \bibnamefont {{Matsuda}}}, \bibinfo
  {author} {\bibfnamefont {A.}~\bibnamefont {{Matsuo}}}, \bibinfo {author}
  {\bibfnamefont {W.}~\bibnamefont {{Li}}}, \bibinfo {author} {\bibfnamefont
  {N.}~\bibnamefont {{Kurita}}}, \bibinfo {author} {\bibfnamefont
  {K.}~\bibnamefont {{Kindo}}},\ and\ \bibinfo {author} {\bibfnamefont
  {H.}~\bibnamefont {{Tanaka}}},\ }\bibfield  {title} {\bibinfo {title}
  {{Intermediate Quantum Spin Liquid Phase in the Kitaev Material
  $\alpha$-RuCl$_3$ under High Magnetic Fields up to 100 T}},\ }\href@noop {}
  {\ }\Eprint {https://arxiv.org/abs/2201.04597 (2022)} {arXiv:2201.04597
  (2022)} \BibitemShut {NoStop}%
\bibitem [{\citenamefont {White}(1992)}]{White1992}%
  \BibitemOpen
  \bibfield  {author} {\bibinfo {author} {\bibfnamefont {S.~R.}\ \bibnamefont
  {White}},\ }\bibfield  {title} {\bibinfo {title} {Density matrix formulation
  for quantum renormalization groups},\ }\href
  {https://doi.org/10.1103/PhysRevLett.69.2863} {\bibfield  {journal} {\bibinfo
   {journal} {Phys. Rev. Lett.}\ }\textbf {\bibinfo {volume} {69}},\ \bibinfo
  {pages} {2863} (\bibinfo {year} {1992})}\BibitemShut {NoStop}%
\bibitem [{\citenamefont {McCulloch}(2008)}]{Mcculloch2008}%
  \BibitemOpen
  \bibfield  {author} {\bibinfo {author} {\bibfnamefont {I.~P.}\ \bibnamefont
  {McCulloch}},\ }\href {https://arxiv.org/abs/0804.2509} {\bibinfo {title}
  {Infinite size density matrix renormalization group, revisited}} (\bibinfo
  {year} {2008})\BibitemShut {NoStop}%
\bibitem [{\citenamefont {Hauschild}\ and\ \citenamefont
  {Pollmann}(2018)}]{Hauschild2018}%
  \BibitemOpen
  \bibfield  {author} {\bibinfo {author} {\bibfnamefont {J.}~\bibnamefont
  {Hauschild}}\ and\ \bibinfo {author} {\bibfnamefont {F.}~\bibnamefont
  {Pollmann}},\ }\bibfield  {title} {\bibinfo {title} {Efficient numerical
  simulations with tensor networks: Tensor network python (tenpy)},\ }\href
  {https://doi.org/10.21468/SciPostPhysLectNotes.5} {\bibfield  {journal}
  {\bibinfo  {journal} {SciPost Phys. Lect. Notes}\ ,\ \bibinfo {pages} {5}}
  (\bibinfo {year} {2018})}\BibitemShut {NoStop}%
\bibitem [{\citenamefont {Chen}\ \emph {et~al.}(2018)\citenamefont {Chen},
  \citenamefont {Chen}, \citenamefont {Chen}, \citenamefont {Li},\ and\
  \citenamefont {Weichselbaum}}]{Chen2018}%
  \BibitemOpen
  \bibfield  {author} {\bibinfo {author} {\bibfnamefont {B.-B.}\ \bibnamefont
  {Chen}}, \bibinfo {author} {\bibfnamefont {L.}~\bibnamefont {Chen}}, \bibinfo
  {author} {\bibfnamefont {Z.}~\bibnamefont {Chen}}, \bibinfo {author}
  {\bibfnamefont {W.}~\bibnamefont {Li}},\ and\ \bibinfo {author}
  {\bibfnamefont {A.}~\bibnamefont {Weichselbaum}},\ }\bibfield  {title}
  {\bibinfo {title} {Exponential thermal tensor network approach for quantum
  lattice models},\ }\href {https://doi.org/10.1103/PhysRevX.8.031082}
  {\bibfield  {journal} {\bibinfo  {journal} {Phys. Rev. X}\ }\textbf {\bibinfo
  {volume} {8}},\ \bibinfo {pages} {031082} (\bibinfo {year}
  {2018})}\BibitemShut {NoStop}%
\bibitem [{\citenamefont {Li}\ \emph {et~al.}()\citenamefont {Li},
  \citenamefont {Gao}, \citenamefont {He}, \citenamefont {Qi}, \citenamefont
  {Chen},\ and\ \citenamefont {Li}}]{Li2022}%
  \BibitemOpen
  \bibfield  {author} {\bibinfo {author} {\bibfnamefont {Q.}~\bibnamefont
  {Li}}, \bibinfo {author} {\bibfnamefont {Y.}~\bibnamefont {Gao}}, \bibinfo
  {author} {\bibfnamefont {Y.-Y.}\ \bibnamefont {He}}, \bibinfo {author}
  {\bibfnamefont {Y.}~\bibnamefont {Qi}}, \bibinfo {author} {\bibfnamefont
  {B.-B.}\ \bibnamefont {Chen}},\ and\ \bibinfo {author} {\bibfnamefont
  {W.}~\bibnamefont {Li}},\ }\bibfield  {title} {\bibinfo {title} {Tangent
  space approach for thermal tensor network simulations of 2{D} {H}ubbard
  model},\ }\href {https://arxiv.org/abs/2212.11973} {\bibinfo  {journal}
  {arXiv: 2212.11973 (2022)}\ }\BibitemShut {NoStop}%
\bibitem [{\citenamefont {Gong}\ \emph {et~al.}(2014)\citenamefont {Gong},
  \citenamefont {Zhu},\ and\ \citenamefont {Sheng}}]{gong2014}%
  \BibitemOpen
\bibfield  {journal} {  }\bibfield  {author} {\bibinfo {author} {\bibfnamefont
  {S.-S.}\ \bibnamefont {Gong}}, \bibinfo {author} {\bibfnamefont
  {W.}~\bibnamefont {Zhu}},\ and\ \bibinfo {author} {\bibfnamefont {D.~N.}\
  \bibnamefont {Sheng}},\ }\bibfield  {title} {\bibinfo {title} {Emergent
  chiral spin liquid: Fractional quantum {H}all effect in a kagome {H}eisenberg
  model},\ }\href {https://doi.org/10.1038/srep06317} {\bibfield  {journal}
  {\bibinfo  {journal} {Scientific Reports}\ }\textbf {\bibinfo {volume} {4}},\
  \bibinfo {pages} {6317} (\bibinfo {year} {2014})}\BibitemShut {NoStop}%
\bibitem [{\citenamefont {Luo}\ \emph {et~al.}(2022)\citenamefont {Luo},
  \citenamefont {Stavropoulos}, \citenamefont {Gordon},\ and\ \citenamefont
  {Kee}}]{Luo2022}%
  \BibitemOpen
  \bibfield  {author} {\bibinfo {author} {\bibfnamefont {Q.}~\bibnamefont
  {Luo}}, \bibinfo {author} {\bibfnamefont {P.~P.}\ \bibnamefont
  {Stavropoulos}}, \bibinfo {author} {\bibfnamefont {J.~S.}\ \bibnamefont
  {Gordon}},\ and\ \bibinfo {author} {\bibfnamefont {H.-Y.}\ \bibnamefont
  {Kee}},\ }\bibfield  {title} {\bibinfo {title} {Spontaneous chiral-spin
  ordering in spin-orbit coupled honeycomb magnets},\ }\href
  {https://doi.org/10.1103/PhysRevResearch.4.013062} {\bibfield  {journal}
  {\bibinfo  {journal} {Phys. Rev. Research}\ }\textbf {\bibinfo {volume}
  {4}},\ \bibinfo {pages} {013062} (\bibinfo {year} {2022})}\BibitemShut
  {NoStop}%
\bibitem [{\citenamefont {Luo}\ and\ \citenamefont {Kee}(2022)}]{Luo2022prb}%
  \BibitemOpen
  \bibfield  {author} {\bibinfo {author} {\bibfnamefont {Q.}~\bibnamefont
  {Luo}}\ and\ \bibinfo {author} {\bibfnamefont {H.-Y.}\ \bibnamefont {Kee}},\
  }\bibfield  {title} {\bibinfo {title} {Interplay of magnetic field and
  trigonal distortion in the honeycomb $\mathrm{\ensuremath{\Gamma}}$ model:
  Occurrence of a spin-flop phase},\ }\href
  {https://doi.org/10.1103/PhysRevB.105.174435} {\bibfield  {journal} {\bibinfo
   {journal} {Phys. Rev. B}\ }\textbf {\bibinfo {volume} {105}},\ \bibinfo
  {pages} {174435} (\bibinfo {year} {2022})}\BibitemShut {NoStop}%
\bibitem [{\citenamefont {S{\o}rensen}\ \emph {et~al.}(2021)\citenamefont
  {S{\o}rensen}, \citenamefont {Catuneanu}, \citenamefont {Gordon},\ and\
  \citenamefont {Kee}}]{Sorensen2021}%
  \BibitemOpen
  \bibfield  {author} {\bibinfo {author} {\bibfnamefont {E.~S.}\ \bibnamefont
  {S{\o}rensen}}, \bibinfo {author} {\bibfnamefont {A.}~\bibnamefont
  {Catuneanu}}, \bibinfo {author} {\bibfnamefont {J.~S.}\ \bibnamefont
  {Gordon}},\ and\ \bibinfo {author} {\bibfnamefont {H.-Y.}\ \bibnamefont
  {Kee}},\ }\bibfield  {title} {\bibinfo {title} {Heart of entanglement:
  Chiral, nematic, and incommensurate phases in the {K}itaev-{$\Gamma$} ladder
  in a field},\ }\bibfield  {journal} {\bibinfo  {journal} {Phys. Rev. X}\
  }\textbf {\bibinfo {volume} {11}},\ \href {https://doi.org/Phys. Rev. X
  11.011013} {Phys. Rev. X 11.011013} (\bibinfo {year} {2021})\BibitemShut
  {NoStop}%
\bibitem [{\citenamefont {Hu}\ \emph {et~al.}(2019)\citenamefont {Hu},
  \citenamefont {Gong}, \citenamefont {Lai}, \citenamefont {Hu}, \citenamefont
  {Si},\ and\ \citenamefont {Nevidomskyy}}]{hu2019}%
  \BibitemOpen
  \bibfield  {author} {\bibinfo {author} {\bibfnamefont {W.-J.}\ \bibnamefont
  {Hu}}, \bibinfo {author} {\bibfnamefont {S.-S.}\ \bibnamefont {Gong}},
  \bibinfo {author} {\bibfnamefont {H.-H.}\ \bibnamefont {Lai}}, \bibinfo
  {author} {\bibfnamefont {H.}~\bibnamefont {Hu}}, \bibinfo {author}
  {\bibfnamefont {Q.}~\bibnamefont {Si}},\ and\ \bibinfo {author}
  {\bibfnamefont {A.~H.}\ \bibnamefont {Nevidomskyy}},\ }\bibfield  {title}
  {\bibinfo {title} {Nematic spin liquid phase in a frustrated spin-1 system on
  the square lattice},\ }\href {https://doi.org/10.1103/PhysRevB.100.165142}
  {\bibfield  {journal} {\bibinfo  {journal} {Phys. Rev. B}\ }\textbf {\bibinfo
  {volume} {100}},\ \bibinfo {pages} {165142} (\bibinfo {year}
  {2019})}\BibitemShut {NoStop}%
\bibitem [{\citenamefont {Hu}\ \emph {et~al.}(2020{\natexlab{a}})\citenamefont
  {Hu}, \citenamefont {Gong}, \citenamefont {Lai}, \citenamefont {Si},\ and\
  \citenamefont {Dagotto}}]{hu2020prb}%
  \BibitemOpen
  \bibfield  {author} {\bibinfo {author} {\bibfnamefont {W.-J.}\ \bibnamefont
  {Hu}}, \bibinfo {author} {\bibfnamefont {S.-S.}\ \bibnamefont {Gong}},
  \bibinfo {author} {\bibfnamefont {H.-H.}\ \bibnamefont {Lai}}, \bibinfo
  {author} {\bibfnamefont {Q.}~\bibnamefont {Si}},\ and\ \bibinfo {author}
  {\bibfnamefont {E.}~\bibnamefont {Dagotto}},\ }\bibfield  {title} {\bibinfo
  {title} {Density matrix renormalization group study of nematicity in two
  dimensions: Application to a spin-1 bilinear-biquadratic model on the square
  lattice},\ }\href {https://doi.org/10.1103/PhysRevB.101.014421} {\bibfield
  {journal} {\bibinfo  {journal} {Phys. Rev. B}\ }\textbf {\bibinfo {volume}
  {101}},\ \bibinfo {pages} {014421} (\bibinfo {year}
  {2020}{\natexlab{a}})}\BibitemShut {NoStop}%
\bibitem [{\citenamefont {Hu}\ \emph {et~al.}(2020{\natexlab{b}})\citenamefont
  {Hu}, \citenamefont {Lai}, \citenamefont {Gong}, \citenamefont {Yu},
  \citenamefont {Dagotto},\ and\ \citenamefont {Si}}]{hu2020prr}%
  \BibitemOpen
  \bibfield  {author} {\bibinfo {author} {\bibfnamefont {W.-J.}\ \bibnamefont
  {Hu}}, \bibinfo {author} {\bibfnamefont {H.-H.}\ \bibnamefont {Lai}},
  \bibinfo {author} {\bibfnamefont {S.-S.}\ \bibnamefont {Gong}}, \bibinfo
  {author} {\bibfnamefont {R.}~\bibnamefont {Yu}}, \bibinfo {author}
  {\bibfnamefont {E.}~\bibnamefont {Dagotto}},\ and\ \bibinfo {author}
  {\bibfnamefont {Q.}~\bibnamefont {Si}},\ }\bibfield  {title} {\bibinfo
  {title} {Quantum transitions of nematic phases in a spin-1
  bilinear-biquadratic model and their implications for {F}e{S}e},\ }\href
  {https://doi.org/10.1103/PhysRevResearch.2.023359} {\bibfield  {journal}
  {\bibinfo  {journal} {Phys. Rev. Research}\ }\textbf {\bibinfo {volume}
  {2}},\ \bibinfo {pages} {023359} (\bibinfo {year}
  {2020}{\natexlab{b}})}\BibitemShut {NoStop}%
\bibitem [{\citenamefont {Calabrese}\ and\ \citenamefont
  {Cardy}(2004)}]{Calabrese2004}%
  \BibitemOpen
  \bibfield  {author} {\bibinfo {author} {\bibfnamefont {P.}~\bibnamefont
  {Calabrese}}\ and\ \bibinfo {author} {\bibfnamefont {J.}~\bibnamefont
  {Cardy}},\ }\bibfield  {title} {\bibinfo {title} {Entanglement entropy and
  quantum field theory: A non-technical introduction},\ }\href
  {https://doi.org/10.1088/1742-5468/2004/06/p06002} {\bibfield  {journal}
  {\bibinfo  {journal} {Journal of Statistical Mechanics: Theory and
  Experiment}\ }\textbf {\bibinfo {volume} {2004}},\ \bibinfo {pages} {P06002}
  (\bibinfo {year} {2004})}\BibitemShut {NoStop}%
\bibitem [{\citenamefont {Wang}\ \emph {et~al.}(2019)\citenamefont {Wang},
  \citenamefont {Normand},\ and\ \citenamefont {Liu}}]{Wang2019}%
  \BibitemOpen
  \bibfield  {author} {\bibinfo {author} {\bibfnamefont {J.}~\bibnamefont
  {Wang}}, \bibinfo {author} {\bibfnamefont {B.}~\bibnamefont {Normand}},\ and\
  \bibinfo {author} {\bibfnamefont {Z.-X.}\ \bibnamefont {Liu}},\ }\bibfield
  {title} {\bibinfo {title} {One proximate {K}itaev spin liquid in the
  {$K$-$J$-$\Gamma$} model on the honeycomb lattice},\ }\href
  {https://doi.org/10.1103/PhysRevLett.123.197201} {\bibfield  {journal}
  {\bibinfo  {journal} {Phys. Rev. Lett.}\ }\textbf {\bibinfo {volume} {123}},\
  \bibinfo {pages} {197201} (\bibinfo {year} {2019})}\BibitemShut {NoStop}%
\bibitem [{\citenamefont {Wen}(2002)}]{Wen2002}%
  \BibitemOpen
  \bibfield  {author} {\bibinfo {author} {\bibfnamefont {X.-G.}\ \bibnamefont
  {Wen}},\ }\bibfield  {title} {\bibinfo {title} {Quantum orders and symmetric
  spin liquids},\ }\href {https://doi.org/10.1103/PhysRevB.65.165113}
  {\bibfield  {journal} {\bibinfo  {journal} {Phys. Rev. B}\ }\textbf {\bibinfo
  {volume} {65}},\ \bibinfo {pages} {165113} (\bibinfo {year}
  {2002})}\BibitemShut {NoStop}%
\bibitem [{\citenamefont {Song}\ \emph {et~al.}(2016)\citenamefont {Song},
  \citenamefont {You},\ and\ \citenamefont {Balents}}]{Song2016}%
  \BibitemOpen
  \bibfield  {author} {\bibinfo {author} {\bibfnamefont {X.-Y.}\ \bibnamefont
  {Song}}, \bibinfo {author} {\bibfnamefont {Y.-Z.}\ \bibnamefont {You}},\ and\
  \bibinfo {author} {\bibfnamefont {L.}~\bibnamefont {Balents}},\ }\bibfield
  {title} {\bibinfo {title} {Low-energy spin dynamics of the honeycomb spin
  liquid beyond the {K}itaev limit},\ }\bibfield  {journal} {\bibinfo
  {journal} {Physical Review Letters}\ }\textbf {\bibinfo {volume} {117}},\
  \href {https://doi.org/Phys. Rev. Lett. 117.037209} {Phys. Rev. Lett.
  117.037209} (\bibinfo {year} {2016})\BibitemShut {NoStop}%
\bibitem [{\citenamefont {You}\ \emph {et~al.}(2012)\citenamefont {You},
  \citenamefont {Kimchi},\ and\ \citenamefont {Vishwanath}}]{You2012}%
  \BibitemOpen
  \bibfield  {author} {\bibinfo {author} {\bibfnamefont {Y.-Z.}\ \bibnamefont
  {You}}, \bibinfo {author} {\bibfnamefont {I.}~\bibnamefont {Kimchi}},\ and\
  \bibinfo {author} {\bibfnamefont {A.}~\bibnamefont {Vishwanath}},\ }\bibfield
   {title} {\bibinfo {title} {Doping a spin-orbit {M}ott insulator: Topological
  superconductivity from the {K}itaev-{H}eisenberg model and possible
  application to {(Na${}_{2}$/Li${}_{2}$)IrO${}_{3}$}},\ }\href
  {https://doi.org/10.1103/PhysRevB.86.085145} {\bibfield  {journal} {\bibinfo
  {journal} {Phys. Rev. B}\ }\textbf {\bibinfo {volume} {86}},\ \bibinfo
  {pages} {085145} (\bibinfo {year} {2012})}\BibitemShut {NoStop}%
\bibitem [{\citenamefont {Meier}\ \emph {et~al.}(2011)\citenamefont {Meier},
  \citenamefont {Greiser}, \citenamefont {Haase}, \citenamefont
  {Herrmannsdörfer}, \citenamefont {Wolff-Fabris},\ and\ \citenamefont
  {Wosnitza}}]{Meier2016}%
  \BibitemOpen
  \bibfield  {author} {\bibinfo {author} {\bibfnamefont {B.}~\bibnamefont
  {Meier}}, \bibinfo {author} {\bibfnamefont {S.}~\bibnamefont {Greiser}},
  \bibinfo {author} {\bibfnamefont {J.}~\bibnamefont {Haase}}, \bibinfo
  {author} {\bibfnamefont {T.}~\bibnamefont {Herrmannsdörfer}}, \bibinfo
  {author} {\bibfnamefont {F.}~\bibnamefont {Wolff-Fabris}},\ and\ \bibinfo
  {author} {\bibfnamefont {J.}~\bibnamefont {Wosnitza}},\ }\bibfield  {title}
  {\bibinfo {title} {{NMR} signal averaging in 62 {T} pulsed fields},\ }\href
  {https://doi.org/https://doi.org/10.1016/j.jmr.2011.02.007} {\bibfield
  {journal} {\bibinfo  {journal} {Journal of Magnetic Resonance}\ }\textbf
  {\bibinfo {volume} {210}},\ \bibinfo {pages} {1} (\bibinfo {year}
  {2011})}\BibitemShut {NoStop}%
\bibitem [{\citenamefont {Akaki}\ \emph {et~al.}(2018)\citenamefont {Akaki},
  \citenamefont {Kanaida},\ and\ \citenamefont {Hagiwara}}]{Akaki2018}%
  \BibitemOpen
  \bibfield  {author} {\bibinfo {author} {\bibfnamefont {M.}~\bibnamefont
  {Akaki}}, \bibinfo {author} {\bibfnamefont {K.}~\bibnamefont {Kanaida}},\
  and\ \bibinfo {author} {\bibfnamefont {M.}~\bibnamefont {Hagiwara}},\
  }\bibfield  {title} {\bibinfo {title} {Magnetoelectric properties of
  {Ca$_2$CoSi$_2$O$_7$} studied by high-field electron spin resonance},\ }\href
  {https://doi.org/10.1088/1742-6596/969/1/012102} {\bibfield  {journal}
  {\bibinfo  {journal} {Journal of Physics: Conference Series}\ }\textbf
  {\bibinfo {volume} {969}},\ \bibinfo {pages} {012102} (\bibinfo {year}
  {2018})}\BibitemShut {NoStop}%
\bibitem [{\citenamefont {Yao}\ and\ \citenamefont {Li}(2020)}]{Yao2020}%
  \BibitemOpen
  \bibfield  {author} {\bibinfo {author} {\bibfnamefont {W.}~\bibnamefont
  {Yao}}\ and\ \bibinfo {author} {\bibfnamefont {Y.}~\bibnamefont {Li}},\
  }\bibfield  {title} {\bibinfo {title} {Ferrimagnetism and anisotropic phase
  tunability by magnetic fields in
  ${\mathrm{na}}_{2}{\mathrm{co}}_{2}{\mathrm{teo}}_{6}$},\ }\href
  {https://doi.org/10.1103/PhysRevB.101.085120} {\bibfield  {journal} {\bibinfo
   {journal} {Phys. Rev. B}\ }\textbf {\bibinfo {volume} {101}},\ \bibinfo
  {pages} {085120} (\bibinfo {year} {2020})}\BibitemShut {NoStop}%
\end{thebibliography}%


%
\end{document}